\documentclass[12pt]{article}

\usepackage{ifpdf}

\usepackage{amsfonts,amssymb,amstext,amsmath}
\usepackage[T1]{fontenc}
\usepackage{dsfont}
\usepackage[vcentermath]{youngtab}

\usepackage[parsep]{collref}

\usepackage{subfigure}
\ifpdf
\usepackage[pdftex]{graphicx}
\usepackage[usenames,dvipsnames]{color}
\else
\usepackage[dvips]{graphicx}
\usepackage[usenames,dvips]{color}
\fi

\ifpdf
\usepackage[pdftex,a4paper,unicode]{hyperref}
\else
\usepackage[hypertex,a4paper,unicode]{hyperref}
\fi

\ifpdf

\hypersetup{
pdfstartview= FitH,
pdfview= FitH,
pdfmenubar= True
}

\hypersetup{
bookmarksopen= true,
bookmarksopenlevel= 2,
bookmarksnumbered= true
}

\hypersetup{
colorlinks= false,
linkcolor= blue,
citecolor= magenta,
}

\fi

\newcommand{\bibname}{References}

\newcommand{\bea}{\begin{eqnarray}}
\newcommand{\eea}{\end{eqnarray}}
\newcommand{\nn}{\nonumber}

\renewcommand{\l}    {\left}
\renewcommand{\r}    {\right}
  \newcommand{\wh}   {\widehat}
  \newcommand{\wt}   {\widetilde}
\renewcommand{\Re}   {{\mathrm{  Re}}}
\renewcommand{\Im}   {{\mathrm{  Im}}}
  \newcommand{\Tr}   {{\mathrm{  Tr}}}

  \newcommand{\CYM}  {{\mathrm{ CYM}}}

  \newcommand{\R}    {{\mathcal{  R}}}
  \newcommand{\1}    {{\mathds{   1}}}
  \newcommand{\vecx} {{\mathbf{   x}}}

  \newcommand{\Nhalf}{{\lfloor{N/2}\rfloor}}

\newcommand{\blue}[1]{\textcolor{blue}{#1}}

\newcommand{\comm}[1]{
\mbox{\sl\footnotesize\blue{#1}}~\blue{\rightarrow}
}

\newcommand{\maintitle}{
Monte Carlo Algorithms for Reduced Lattices, Mixed Actions,\\and Double-Trace Deformations
}

\newcommand{\authors}{
H\'{e}lvio Vairinhos
}

\newcommand{\addresses}{
Centre for Computational Physics, 
Department of Physics,\\
University of Coimbra,
3004-516 Coimbra,
Portugal
}

\title{\maintitle}
\author{\authors}
\date{\today}

\ifpdf\hypersetup{
pdftitle={\maintitle},
pdfauthor={\authors},
}\fi

\begin{document}

\begin{titlepage}
	\vspace*{0.60in}
	\begin{center}
		{\large\bf\maintitle\\}
		\vspace*{0.65in}
		{\authors\\
		\vspace*{.25in}
		\addresses}
	\end{center}
	\vspace*{0.50in}
	\begin{center}
		{\bf Abstract}
	\end{center}
	We construct efficient Monte Carlo updating algorithms for two classes of pure $SU(N)$ lattice gauge actions with non-linear dependence on the link variables. Our construction generalises the method of auxiliary variables used by Fabricius and Haan in the framework of Eguchi-Kawai models. We first review the original Fabricius-Haan method of constructing a pseudo-heatbath algorithm for fully reduced models, and discuss its  extension to lattices with any number of reduced directions. We then use a similar method to construct updating algorithms for generic $SU(N)$ mixed Wilson actions. We construct explicit examples of algorithms for Wilson actions whose plaquettes are in an irreducible representation of $SU(N)$ with $N$-ality $k \leq 3$. We also construct updating algorithms for the lattice version of centre-stabilised $SU(N)$ Yang-Mills theories defined on $\mathds{R}^{d-1} \times S^1$, including the case of a fully reduced compact direction. We simulate the new algorithms and show that they are, in general, significantly more efficient than their Metropolis counterparts.

\end{titlepage}
\setcounter{page}{1}
\newpage
\pagestyle{plain}
\vspace*{0.30in}
\tableofcontents

\section
{Introduction
\label{sec:intro}}

Simulations of pure lattice gauge theories normally involve the generation of Markov chains of gauge field configurations with the help of a local Monte Carlo algorithm. Such configurations (link variables on hypercubic lattices) are generated with respect to a probability distribution characterised by the Boltzmann weight of the Euclidean partition function. Configurations are generated locally, i.e. the link variables are updated individually while keeping the remaining links fixed. A new configuration results from a sequence of local updates over the whole lattice (one sweep). It is thus necessary to know the form of the Boltzmann probability distribution restricted to only one link. The purpose of the local Monte Carlo (MC) algorithm is to sample this distribution function efficiently. However, save in some special cases, the direct sampling of the probability distributions of individual links is often impracticable. Therefore, it is essential to find a good algorithm that samples such distributions efficiently, albeit indirectly. The ideal MC algorithm should be ergodic, fast, and produce decorrelated data. 

One straightforward possibility is to generate new link variables using a local Metropolis algorithm \cite{Metropolis}. Metropolis has the advantages of having an universal scope (it can sample any probability distribution) and of being very easy to implement. However, it is usually slow, decorrelates poorly, and may fail to be ergodic.  Also, it requires the acceptance rates of new link proposals to be tuned at optimal values (which must be at around 50\%, in order to avoid `too fast' or `too slow' an exploration of the configuration space). Therefore, despite the advantages, Metropolis is not a very efficient algorithm for simulating lattice gauge theories. In many situations, however, it is the only known algorithm. Fortunately, for pure $SU(N)$ lattice gauge theories with the standard Wilson action it is possible to construct a more efficient alternative to the Metropolis algorithm. 
\\

Consider the standard $SU(N)$ Wilson action (up to an additive constant) with plaquettes in the fundamental representation:
\bea
\label{eq:F:Action}
S_F (\beta_F; [U])
&=&
-\frac{\beta_F}{N} \sum_p \Re\Tr\l\{ U_p \r\}
\eea
Here $p$ labels positively oriented plaquettes in the $d$--dimensional hypercubic lattice $\Lambda$, $U_p$ is the plaquette operator, defined as the path-ordered product of all link variables $U_{\mu,x} \in SU(N)$ in the boundary of $p$,
\bea
\label{eq:F:Plaquette}
U_p
&\equiv&
U_{\mu,x} U_{\nu,x + \hat\mu} 
U_{\mu,x + \hat\nu}^\dag U_{\nu,x}^\dag
\eea
($\hat\mu$ denotes the unit lattice vector in the $\mu$--direction), $\beta_F$ is the bare lattice coupling, and $\Tr$ denotes the character of the fundamental representation of $SU(N)$. The probability distribution of link configurations $[U]$ is Boltzmann:
\bea
\label{eq:F:Distrib}
dP(\beta_F; [U])
&=&
\mu_{\rm H}[U]~ \exp\l( -S_F (\beta_F; [U]) \r) / Z_F(\beta_F)
\eea
where $\mu_{\rm H}$ denotes the product of the $SU(N)$--invariant Haar measures of all link variables,
\bea
\label{eq:F:Haar}
\mu_{\rm H} [U]
&\equiv&
\prod_{x \in \Lambda} \prod_{\mu=1}^d dU_{\mu,x}
\eea
and $Z_F$ is the Euclidean partition function of the lattice gauge theory:
\bea
\label{eq:F:Z}
Z_F(\beta_F)
&=&
\int \mu_{\rm H}[U]~ \exp\l( -S_F (\beta_F; [U]) \r)
\eea
Because the Wilson action \eqref{eq:F:Action} is linear with respect to each link variable, the probability distribution of individual links is given by:
\bea
\label{eq:F:Distrib:Single}
dP(U_{\mu,x}) 
&\propto&
dU_{\mu,x}~ \exp
\l( 
\Re\Tr\l\{ 
V_{\mu,x}^\dag U_{\mu,x} 
\r\} 
\r)
\eea
where $V_{\mu,x}$ is the sum of all `staples' that couple to $U_{\mu,x}$ multiplied by the lattice coupling:
\bea
\label{eq:F:Staples}
V_{\mu,x} 
&=&
\frac{\beta_F}{N} \sum_{{\nu=1}\atop{(\nu\neq\mu)}}^d 
\l(~
U_{\nu,x} U_{\mu,x + \hat\nu} 
U_{\nu,x + \hat\mu}^\dag
~+~ 
U_{\nu,x - \hat\nu}^\dag U_{\mu,x - \hat\nu} 
U_{\nu,x - \hat\nu + \hat\mu}
~\r)
\eea
It is important to note that $V_{\mu,x}$ does not (and should not) depend on $U_{\mu,x}$.
\\

In the $SU(2)$ case, it is possible to perform a very efficient sampling of the probability distribution \eqref{eq:F:Distrib:Single} \cite{CreutzSU2, FabriciusHaan, KennedyPendletonSU2}. It relies on the fact that any sum $v$ of $SU(2)$ matrices, e.g. \eqref{eq:F:Staples}, is always proportional to an $SU(2)$ matrix. Because of this property, the probability distribution of individual $SU(2)$ matrices is of the form:
\bea
\label{eq:F:Distrib:Single:SU2}
dP(u) 
&\propto& 
du~ \exp\l( \xi~ \Re\Tr
\l\{\l( 
v \xi^{-1} \r)^\dag u 
\r\}\r)
\eea
where $u, v \xi^{-1} \in SU(2)$, $\xi = \surd\det(v)$, and $du$ is the $SU(2)$--invariant Haar measure. The $SU(2)$--projected sum of `staples' $v\xi^{-1}$ can be absorbed by the Haar measure. Consequently, the probability distribution \eqref{eq:F:Distrib:Single:SU2} only depends on $a_0 \equiv \Tr\l\{u\r\} \in [-1,1]$: 
\bea
dP(a_0) 
&\propto&
da_0~ \l( 1-a_0^2 \r)^{\frac{1}{2}} \exp\l( 2\xi a_0 \r)
\eea
which can be sampled very efficiently. Since $SU(2)$ is isomorphic to $S^3$, $a_0$ corresponds to a `latitude' parameter. The traceless part of $u$ is then uniformly distributed over the $S^2$ `equator', i.e. with a probability distribution given by the solid angle $d^2\Omega$, which can be sampled trivially. The algorithms \cite{CreutzSU2, FabriciusHaan, KennedyPendletonSU2} that generate random $SU(2)$ matrices with the probability distribution \eqref{eq:F:Distrib:Single:SU2} are the heatbath algorithms. 

There are two other very useful $SU(2)$ algorithms for lattice gauge theory simulations: the overrelaxation and cooling algorithms. The overrelaxation algorithm \cite{BrownWoch,CreutzOverrelaxation} performs `large' local changes to the gauge field configurations that keep the Boltzmann weight invariant. Because the sum $v$ of $SU(2)$ matrices is always proportional to an $SU(2)$ matrix, an exact overrelaxation update of $u$ is obtained with the proposal:
\bea
\label{eq:Overrelax:SU2}
u ~\mapsto~ u' 
&=&
\l(v\xi^{-1}\r) \cdot
u^\dag 
\cdot \l(v\xi^{-1}\r)
~\in SU(2)
\eea
Using overrelaxation updates together with heatbath updates improves the efficiency of the overall MC algorithm, because \eqref{eq:Overrelax:SU2} corresponds to take a large `step' on $SU(2)$ as compared with the typical heatbath `steps'. This allows a broader exploration of the configuration space around the local minimum probed by the simulation. On the other hand, the cooling algorithm \cite{TeperCooling, IlgenfritzCooling} generates gauge field configurations that minimise the lattice action locally. Because the sum $v$ of $SU(2)$ matrices is always proportional to an $SU(2)$ matrix, an exact cooling update of $u$ is obtained with the proposal:
\bea
\label{eq:Cooling:SU2}
u ~\mapsto~ u'
&=&
\l(v\xi^{-1}\r)
~\in SU(2)
\eea
The cooling of lattice gauge fields suppresses their quantum fluctuations. This is useful for the search of topological structures in the vacuum of the gauge theory, e.g. instantons.
\\

For gauge groups larger than $SU(2)$, however, the situation is different: the sum of $SU(N)$ matrices is not proportional to an $SU(N)$ matrix, in general. Therefore, none of the algorithms described above can be directly extrapolated to $N > 2$. A way to circumvent this obstacle is to construct algorithms that only update the $SU(2)$ subgroups of $SU(N)$, and not the whole group \cite{CabibboMarinari}. Due to the lattice action \eqref{eq:F:Distrib:Single} being linear with respect to individual links $U_{\mu,x}$, it is then possible to single out each of its $N(N-1)/2$ $SU(2)$ subgroup elements $u \subset U_{\mu,x}$, and obtain a distribution for $u$ that is of the form \eqref{eq:F:Distrib:Single:SU2}. In this way, the $SU(2)$ algorithms described above can still be used. The update of all (or a subset\footnote{
The subset of $SU(2)$ subgroups must be such that the remaining subgroups do not generate a left ideal \cite{CabibboMarinari}. The minimal number of $SU(2)$ subgroups that must be updated is therefore $N-1$, i.e. those of the form $u_{i,i+1}$ ($u_{ij}$ represents the $2 \times 2$ submatrix whose diagonal elements lie on the positions $i$ and $j$ of the diagonal of the $N \times N$ matrix it belongs to). Despite the large difference between the minimal number and the total number of $SU(2)$ subgroups, which becomes very significant for large $N$, it is recommended to update all of them. Updating too small a number of $SU(2)$ subgroups may result in a poor performance. In our simulations, we always update all the $N(N - 1)/2$ $SU(2)$ subgroups of $SU(N)$.}
of all) $SU(2)$ submatrices of a link variable results in a new link that is compatible with the probability distribution \eqref{eq:F:Distrib:Single}. The algorithm that generates random $SU(N)$ matrices is known as Cabibbo-Marinari pseudo-heatbath algorithm \cite{CabibboMarinari}. Even though it doesn't sample \eqref{eq:F:Distrib:Single} directly, it performs much better than the Metropolis algorithm in all aspects: it provides faster equilibration times, smaller autocorrelations, and it doesn't require any tuning. Because of all this, it has naturally become the standard algorithm for the generation of equilibrium configurations in pure lattice gauge theories.\footnote{
An exact heatbath algorithm has been suggested for $SU(3)$ by Pietarinen \cite{Pietarinen}. However, it is hard to implement, and so the Cabibbo-Marinari pseudo-heatbath strategy has also been adopted as the standard heatbath algorithm for $SU(3)$.}

In the same way, the overrelaxation and cooling of $SU(N)$ gauge configurations can be done by restricting such updates to the $SU(2)$ subgroups. 
Recently, algorithms for the overrelaxation of the full $SU(N)$ group, and not just of its $SU(2)$ subgroups, have been suggested \cite{KiskisNarayananNeuberger, ForcrandJahn}. They are based on the singular value decomposition of $V_{\mu,x}$, and are shown to perform better than the overrelaxation of $SU(2)$ subgroups. In the same line of thought, a cooling algorithm for the full $SU(N)$ group can easily be constructed using the singular value decomposition of $V_{\mu,x}$.
\\

However, the $SU(N)$ updating algorithms discussed above only apply if the probability distribution of individual link variables is of the form \eqref{eq:F:Distrib:Single}, i.e. if the lattice action is linear with respect to each link. If not, the contribution to the partition function of individual $SU(2)$ subgroups is not of the form \eqref{eq:F:Distrib:Single:SU2}, and so the $SU(2)$ algorithms discussed above cannot be used. For the same reasons, the full $SU(N)$ overrelaxation/cooling algorithms cannot be used, too. In sum, if a lattice action is not linear with respect to the link variables, and no other efficient algorithms are known, then using a Metropolis algorithm is the only way to simulate such a theory.

In this paper, we construct MC updating algorithms for some $SU(N)$ lattice gauge actions whose dependence on the link variables is nonlinear. Our strategy is based on a generalisation of the Fabricius-Haan method of auxiliary variables \cite{FabriciusHaan}. We perform numerical simulations with the new algorithms and show that they are, in most cases, significantly more efficient then their Metropolis alternatives. In Section \ref{sec:red}, we review the Fabricius-Haan method of constructing a pseudo-heatbath algorithm for the Eguchi-Kawai model. We then generalise their method, and apply it to the case of hypercubic lattices with any number of fully reduced directions. In Section \ref{sec:mix}, we apply similar methods to pure lattice gauge theories with a mixed Wilson action. We explicitly construct MC updating algorithms for Wilson actions whose plaquettes are in irreducible representations of $SU(N)$ with $N$--ality $k \leq 3$. We then study the numerical performance of these algorithms against Metropolis. In Section \ref{sec:cym}, we deal with the lattice regularisation of centre-stabilised $SU(N)$ Yang-Mills theories defined on $\mathds{R}^3 \times S^1$. We explicitly construct MC an updating algorithm for these theories and study its numerical performance against Metropolis. We conclude our paper with a discussion on the advantages, efficiency and applicability of the new algorithms. We summarise in Appendix \ref{app:MC} all the algorithms proposed in this paper.

\section
{Reduced lattices
\label{sec:red}}

A situation where the probability distribution of individual links is not of the form \eqref{eq:F:Distrib:Single} occurs already in pure lattice gauge theory with the standard Wilson action. If one or more lattice directions of a hypercubic lattice are reduced (i.e. have length $N_\mu = 1$ in lattice units, for some direction $\mu$), the Wilson action \eqref{eq:F:Action} becomes quadratic with respect to the link variables orthogonal to the reduced directions. Because of the this, gauge field configurations on reduced lattices would have to be generated with a Metropolis algorithm. However, Fabricius and Haan \cite{FabriciusHaan} were able to construct an efficient pseudo-heatbath algorithm for the case of a fully reduced ($1^d$) lattice, also known as Eguchi-Kawai (EK) model \cite{EguchiKawai}.\footnote{
In fact, the pseudo-heatbath algorithm of Fabricius and Haan was originally constructed for $1^d$ lattices with twisted boundary conditions, also known as twisted Eguchi-Kawai (TEK) models \cite{TEK}. We omit the twists factors in the lattice action because they are not relevant to our discussion.}
They used a method of auxiliary variables, which we review below.

\subsection
{Fabricius-Haan method
\label{sec:red:FH}}

The EK model \cite{EguchiKawai} is the original proposal of a matrix model for the large $N$ limit of pure $SU(N)$ lattice Yang-Mills theories. Initially, it was believed that their planar sectors would coincide, but the spontaneous breaking of an important symmetry eliminated that hope \cite{QEK}. Since then, a number of alternative reduced models have been suggested \cite{QEK, TEK, UnsalYaffe}. The goal is to find a zero-volume model of the full gauge theory in the large $N$ limit, in which the set of global symmetries relevant for the large $N$ equivalence stay intact for all values of the coupling. This is hard to achieve, however, because those symmetries are sensitive to the volume.

The action of the EK model is simply the fundamental Wilson action \eqref{eq:F:Action} on a $1^d$ lattice:
\bea
\label{eq:EK:Action}
S_{\rm EK}(\beta_F; [U])
&=& 
-\frac{\beta_F}{N} \sum_{\mu<\nu}^d 
\Re\Tr\l\{
U_\mu U_\nu U_\mu^\dag U_\nu^\dag 
\r\}
\eea
Notice that the total absence of spacetime degrees of freedom in \eqref{eq:EK:Action} makes the plaquette operators quadratic on the link variables. Therefore, the probability distribution of individual links cannot be put in the form of \eqref{eq:F:Distrib:Single}, because the sum of `staples' \eqref{eq:F:Staples} would not be independent of $U_{\mu}$. Consequently, none of the efficient updating algorithms discussed in Section \ref{sec:intro} can be applied directly, and the Metropolis algorithm seems to be the only alternative to simulate the EK model.

Fabricius and Haan \cite{FabriciusHaan} circumvented this no-go by adding auxiliary degrees of freedom to the EK model. The new lattice variables are complex $N \times N$ matrices associated with the plaquettes, $\wt Q_{\mu\nu} \equiv \wt Q_{\nu\mu}, \forall \mu<\nu$. They are given a free dynamics completely decoupled from the gauge field. In other words, the spurious degrees of freedom $\wt Q_{\mu\nu}$ are random matrices with the normal distribution, whose only effect is to multiply the partition function of the EK model by a constant factor:
\bea
\label{eq:EK:Z*Gauss}
Z_{\rm EK}(\beta_F)
\propto
\int \mu_{\rm H}[U]~ 
\exp\l( 
-S_{\rm EK}( \beta_F; [U] ) 
\r) 
\times 
\underbrace{ 
\int [d\wt Q^\dag d\wt Q]~ \exp\l( -\frac{1}{2}\sum_{\mu<\nu}^d \Tr\l\{ \wt Q_{\mu\nu}^\dag \wt Q_{\mu\nu} \r\}
\r) 
}_{\rm constant}
\eea
$\mu_{\rm H}[U]$ is the product of $SU(N)$--invariant Haar measures of all $d$ link variables, and $[d\wt Q^\dag d\wt Q]$ is the product of standard flat measures for the auxiliary fields,
\bea
\label{eq:EK:Gauss:mu(flat)}
[d\wt Q^\dag d\wt Q] 
&\equiv& 
\prod_{\mu<\nu}^d \prod_{a,b=1}^N 
d\Re ( {\wt Q_{\mu\nu}} )_{ab}~ 
d\Im ( {\wt Q_{\mu\nu}} )_{ab}
\eea
To simplify the notation, we use $\mu_\sigma$ to denote the Gaussian measure with variance $\sigma^2$ of a complex variable $z \in \mathds{C}$,
\bea
\label{eq:Gauss:mu(z)}
\mu_\sigma(z)
&\equiv&
dz^\ast dz~ (2\pi\sigma^2)^{-1}
\exp\l( -\frac{1}{2\sigma^2} \l| z \r|^2 \r)
\eea
or more generally, of a complex $N \times N$ matrix $A \in M(N,\mathds{C})$,
\bea
\label{eq:Gauss:mu(A)}
\mu_\sigma(A)
&\equiv&
dA^\dag dA~ (2\pi\sigma^2)^{-N^2}
\exp\l( -\frac{1}{2\sigma^2} \Tr\l\{ A^\dag A \r\} \r)
\eea
In this notation, the Gaussian integral multiplying the EK partition function is:
\bea
\label{eq:EK:Gauss:Integral}
\int \mu_1[\wt Q] &=& 1
\eea
where $ \mu_1[\wt Q] \equiv \prod_{\mu<\nu}^d  \mu_1 (\wt Q_{\mu\nu}) $.

The second step in the Fabricius-Haan method consists in performing a particular change of variables $(\wt Q, U) \mapsto (Q, U)$ given by:
\bea
\label{eq:EK:ChangeVars}
\wt Q_{\mu\nu} 
&=& 
\l(\frac{\beta_F}{N}\r)^{\frac{1}{2}} 
\l( 
Q_{\mu\nu} - U_\mu U_\nu - U_\nu U_\mu 
\r)
\eea
This transformation keeps the integration measure \eqref{eq:EK:Gauss:mu(flat)} invariant, up to a constant factor. It also cancels out the quadratic terms in the EK action, replacing them with Gaussian and linear terms.  To see this, consider the effect of this change of variables on the exponent of the Gaussian term in \eqref{eq:EK:Z*Gauss}:
\bea
\label{eq:EK:Gauss:Terms(HS)}
-\frac{1}{2} \sum_{\mu<\nu}^d \Tr\l\{ {\wt Q_{\mu\nu}}^\dag \wt Q_{\mu\nu} \r\}
&\stackrel{ \eqref{eq:EK:ChangeVars} }{=}&
\nn\\ 
\comm{Gaussian} &&-\frac{\beta_F}{2N} \sum_{\mu<\nu}^d \Tr\l\{ Q_{\mu\nu}^\dag Q_{\mu\nu} \r\} 
\nn\\
\comm{linear} &&+\frac{\beta_F}{N} \sum_{\mu\neq\nu}^d \Re\Tr\l\{ Q_{\mu\nu}^\dag U_\mu U_\nu \r\} 
\nn\\
\comm{cancels out} && +\frac{\beta_F}{N} \sum_{\mu<\nu}^d \Re\Tr\l\{ U_\mu U_\nu U_\mu^\dag U_\nu^\dag \r\}
\eea
The change of variables produces Gaussian terms for the $Q_{\mu\nu}$, linear terms on the link variables, and quadratic terms that have the same functional form as the terms in \eqref{eq:EK:Action}, but with opposite sign. Therefore, the quadratic terms in \eqref{eq:EK:Z*Gauss} are cancelled out and only the Gaussian and linear terms survive. The partition function of the EK model with auxiliary variables then becomes:
\bea
\label{eq:EK:Z*Gauss(HS)}%
Z_{\rm EK}(\beta_F)
&=&
\int 
\mu_{\rm H}[U]~ 
\mu_\sigma[Q]~ 
\exp\l( 
\frac{\beta_F}{N} \sum_{\mu\neq\nu}^d \Re\Tr\l\{ Q_{\mu\nu}^\dag U_\mu U_\nu \r\}
\r) 
\eea
with $\sigma^2 = N/\beta_F$. Since the exponent of the Boltzmann factor in \eqref{eq:EK:Z*Gauss(HS)} is linear with respect to each link variable, the Cabibbo-Marinari pseudo-heatbath can be used to update them. The update is performed with respect to the probability distribution of individual links \eqref{eq:F:Distrib:Single},
where $U_{\mu,x} \equiv U_\mu$, and $V_{\mu,x} \equiv V_\mu$ is analogous to the sum of `staples' \eqref{eq:F:Staples}:
\bea
\label{eq:EK:Staples}
V_\mu
&=& 
\sum_{{\nu=1}\atop{(\nu\neq\mu)}}^d 
\l(
U_\nu^\dag Q_{\mu\nu} + 
Q_{\mu\nu} U_\nu^\dag
\r)
\eea
Note that $V_\mu$ does not depend on $U_\mu$. In essence, the change of variables in \eqref{eq:EK:ChangeVars} is an example of a Hubbard-Stratonovich transformation \cite{Stratonovich, Hubbard}.

Despite the apparent coupling between the gauge and auxiliary degrees of freedom in \eqref{eq:EK:Z*Gauss(HS)}, the gauge dynamics is completely unaffected by the presence of the auxiliary fields: the partition function \eqref{eq:EK:Z*Gauss(HS)} can always be transformed back to its original form \eqref{eq:EK:Z*Gauss} via the inverse of the non-singular transformations \eqref{eq:EK:ChangeVars}. Another apparent paradox resides in the fact that we are actually increasing the number if degrees of freedom that need to be updated in numerical simulations. However, all components $(\wt Q_{\mu\nu})_{ab}$ of the auxiliary fields are independent normally-distributed complex numbers. These can be generated very fast with a known algorithm, like the Box-M\"uller transform.\footnote{
The Box-M\"uller transform \cite{BoxMuller} is a method for generating pairs of independent random real numbers $(g_1, g_2)$ with normal distribution $dP(g1,g2) = dg_1 dg_2~ (2\pi\sigma^2)^{-1} \exp(-(g_1^2 + g_2^2)/2)$ from a pair of random real numbers $(u_1, u_2)$ uniformly distributed over the unit interval (0,1]. The pair $(g_1, g_2)$ can also be understood as a random complex number $z = g_1 + i g_2$ with normal distribution over the complex plane $dP(z) = dz^\ast dz~ (2\pi\sigma^2)^{-1} \exp(-|z|^2/2)$. The Gaussian complex number $z$ is generated from
the pair $(u_1, u_2)$ via the Box-M\"uller transform: 
\bea
\label{eq:BoxMuller} 
z = \l(- 2 \ln u_1 \r)^{\frac{1}{2}} \exp\l(i 2 \pi u_2\r)
\eea 
} 
Therefore, the time needed to update the auxiliary variables is  negligible as compared with the time needed to update the link variables. Consequently, the Fabricius-Haan method for the EK model is actually faster than its Metropolis alternative, given in \cite{OkawaMetropolis}.

\subsection
{Partially reduced lattices
\label{sec:red:partial}}

The Fabricius-Haan method for the EK model can easily be generalized to lattices with any number of reduced directions.\footnote{
An example of the Fabricius-Haan method applied to partially reduced lattices is used in \cite{BietenholzInstability}.} 
In such lattices, only those plaquettes that are parallel to at least one reduced direction are quadratic with respect to the link variables. We call them reduced plaquettes. The remaining unreduced plaquettes are linear, hence the respective action terms do not need to be replaced. It then suffices to introduce one normally-distributed $N \times N$ complex matrix per reduced plaquette, i.e. $\wt Q_{\mu\nu,x} \equiv \wt Q_{\nu\mu,x}$ such that $N_\mu = 1$ or $N_\nu = 1$. The partition function \eqref{eq:F:Z} of the partially reduced lattice gauge theory is then multiplied by the Gaussian integral over those auxiliary variables, $\int \mu_1[\wt Q]$, which again amounts to multiply the partition function by `1'. With the following change of variables:
\bea
\label{eq:red:ChangeVars}
\wt Q_{\mu\nu,x}
&=& 
\l( \frac{\beta_F}{N} \r)^{\frac{1}{2}} 
\l(
Q_{\mu\nu,x} - U_{\mu,x} U_{\nu,x + \hat\mu} - U_{\nu,x} U_{\mu,x+\hat\nu} 
\r)
\eea
all action terms involving reduced plaquettes are eliminated, and the partition function \eqref{eq:F:Z} becomes:
\bea
\label{eq:red:Z*Gauss(HS)}
Z_F(\beta_F) &=&
\int \mu_{\rm H} [U]~ \mu_{\sigma}[Q]~ \exp\l(
-S_F'(\beta_F;[Q,U])
\r)
\eea
where $ \sigma^2 = N/\beta_F$, and $S_F'$ is the `linearised' action:
\bea
\label{eq:red:Action(HS)}
S_F'(\beta_F;[Q,U]) 
&=& 
-\frac{\beta_F}{N} \sum_{x\in\Lambda} \sum_{\mu<\nu\atop{(L_\mu,L_\nu > 1)}}^d \Re\Tr\{U_{\mu\nu, x}\} 
\nn\\
&&-\frac{\beta_F}{N} \sum_{x\in\Lambda} \sum_{\mu<\nu\atop{(L_\mu=1 \vee L_\nu = 1)}}^d \Re\Tr\{ Q_{\mu\nu,x}^\dag (U_{\mu,x} U_{\nu,x + \hat\mu} + U_{\nu,x} U_{\mu,x+\hat\nu}) \} 
\eea

The contribution from the unreduced plaquettes is unchanged, because those terms are already linear in the link variables. On the other hand, the quadratic contributions coming from the reduced plaquettes are replaced by the linear terms involving auxiliary variables. The Cabibbo-Marinari pseudo-heatbath can then be used to update the link variables. The updates are performed with respect to the probability distribution \eqref{eq:F:Distrib:Single} of individual links, where the sum of `staples' $V_{\mu,x}$ is given by:
\bea
\label{eq:red:Staples}
V_{\mu,x} 
&=& 
\frac{\beta_F}{N} \!\!\! \sum_{\nu=1\atop (\nu \neq \mu, L_\nu > 1)}^d 
\l(
U_{\nu,x} U_{\mu,x+\hat\nu} U_{\nu,x+\hat\mu}^\dag +
U_{\nu,x-\hat\nu}^\dag U_{\mu,x-\hat\nu} U_{\nu,x-\hat\nu+\hat\mu}
\r)
\nn\\
&+& \frac{\beta_F}{N} \!\!\! \sum_{\nu=1\atop (\nu \neq \mu, L_\nu = 1)}^d 
\l(
Q_{\mu\nu,x} U_{\nu,x+\hat\mu}^\dag +
U_{\nu,x}^\dag Q_{\mu\nu,x} 
\r)
\eea

The Fabricius-Haan algorithm for partially reduced lattices is summarised in the Appendix \ref{app:MC:red}. The Fabricius-Haan algorithm for the EK model is contained as the special case for which the lattice is fully reduced.

\section
{Mixed actions
\label{sec:mix}}

A suitable action for a $SU(N)$ lattice gauge theory must be gauge-invariant and converge to the $SU(N)$ Yang-Mills action in the continuum limit. It is well known \cite{MunsterMontvay} that class functions\footnote{
A {\em class function} is a function $f$ defined on a group $G$ that is constant on the conjugacy classes of $G$, i.e. $f(ghg^{-1})=f(h), \forall g,h \in G$. Here, $g$ corresponds to a gauge transformation at a lattice site, and $h$ corresponds to a lattice operator. Therefore, the definition of class function corresponds to the statement that $f$ is gauge-invariant.
} 
on $SU(N)$ can be used to construct lattice actions that satisfy the conditions above. Examples of class functions are the characters of the irreducible representations $\R$ of $SU(N)$. These are used to define an important class of lattice actions, namely Wilson actions whose plaquettes are in different representations of $SU(N)$:
\bea 
\label{eq:R:Action}
S_\R(\beta_\R; [U])
&=& 
-\frac{\beta_\R}{d_\R} \sum_p \Re\chi_\R \l\{ U_p \r\}
\eea
where $\chi_\R$ is the character of the representation $\R$, $d_\R = d_\R \equiv \chi_\R(\1)$ is its dimension, and $\beta_\R$ is its associated bare lattice coupling. Equally suitable are the lattice actions consisting of arbitrary linear combinations of irreducible characters, known as mixed Wilson actions:
\bea
\label{eq:mix:Action}
S_{\rm mix}(\vec\beta; [U]) 
&=& 
\sum_\R S_\R(\beta_\R; [U])
\eea
where $\vec\beta$ denotes the set of independent lattice couplings $\beta_\R$.

All irreducible characters of $SU(N)$ can be expressed in terms of the character of the fundamental representation, $\chi_F \equiv \Tr$. In this paper we only consider explicitly those representations of  $SU(N)$ with $N$--ality $k \leq 3$, whose characters are given by:
\bea
    \label{eq:F:chi}
    \chi_F(U) &=& \Tr\l\{ U \r\}
    \\\nn\\
    \label{eq:A:chi}
    \chi_{A}(U) &=& \l| \Tr\l\{ U \r\} \r|^2 - 1
    \\\nn\\
    \label{eq:2:chi}
    \chi_{(2)}(U) &=& \frac{1}{2} \l( \Tr\l\{ U \r\}^2 + \Tr\l\{ U^2 \r\} \r)
    \\\nn\\
    \label{eq:11:chi}
    \chi_{(1,1)}(U) &=& \frac{1}{2} \l( \Tr\l\{ U \r\}^2 - \Tr\l\{ U^2 \r\} \r)
    \\\nn\\
    \label{eq:3:chi}
    \chi_{(3)}(U) &=& \frac{1}{6} \l( \Tr\l\{ U \r\}^3 + 3 \Tr\l\{ U \r\} \Tr\l\{ U^2 \r\} + 2 \Tr\l\{ U^3 \r\} \r)
    \\\nn\\
    \label{eq:111:chi}
    \chi_{(1,1,1)}(U) &=& \frac{1}{6} \l( \Tr\l\{ U \r\}^3 - 3 \Tr\l\{ U \r\} \Tr\l\{ U^2 \r\} + 2 \Tr\l\{ U^3 \r\} \r)
    \\\nn\\
    \label{eq:21:chi}
    \chi_{(2,1)}(U) &=& \frac{1}{3} \l( \Tr\l\{ U \r\}^3 - \Tr\l\{ U^3 \r\} \r)
\eea
Here $U$ is a generic element of $SU(N)$, and the label in the characters denotes the Young tableau of a representation (e.g. $(2) \equiv \Yvcentermath1\Yboxdim6pt\yng(2)$, $(1,1) \equiv \Yvcentermath1\Yboxdim6pt\yng(1,1)$, etc.); $F$ denotes the fundamental representation, and $A$ denotes the adjoint representation.  

It is clear that only the fundamental Wilson action is linear with respect to link variables. Consequently, the efficient updating algorithms discussed in Section \ref{sec:intro} cannot be used directly in different representations. For this reason, all numerical studies of mixed Wilson actions have been performed using algorithms based on Metropolis, like the Cabibbo-Marinari-Metropolis and the overrelaxation-Metropolis algorithms \cite{NeccoHasenbusch}, or the biased Metropolis algorithm for the $SU(2)$ fundamental/adjoint action \cite{BazavovBergHeller}. However, it is possible to imagine that a generalisation of the Fabricius-Haan method could also be applied to mixed Wilson actions, and in that way evade the nonlinear terms coming from the $SU(N)$ characters.  

The philosophy behind the Fabricius-Haan method consists in adding a minimal (but sufficient) number of normally-distributed auxiliary fields, together with an appropriate change of variables, in order to eliminate all the nonlinear terms in the lattice action. These are then replaced by linear terms on the link variables and Gaussian terms for the auxiliary fields. The possibility of using the efficient updating algorithms of Section \ref{sec:intro} follows immediately. We argue that, by choosing a fair number of auxiliary variables and the correct transformation rules, it is ideally possible to eliminate all nonlinear plaquette terms in arbitrary mixed Wilson actions (and eventually in any lattice action whose dependence on the plaquettes is polynomial). Some of the auxiliary fields may eventually generate new nonlinear terms that also need to be eliminated. But as long as these are `less nonlinear' than the terms they replace (e.g. by depending on smaller powers of the plaquette), the total amount of `nonlinearity' is reduced, the process eventually stops, and all nonlinear terms are eliminated after a sufficient number of auxiliary fields is introduced. The objective is then to keep the number of necessary auxiliary fields at a minimum. We will make these statements more precise below.

\subsection
{Linearisation of arbitrary characters
\label{sec:mix:linearisation}}

For each representation $\R$ of $SU(N)$, we wish to linearise the corresponding Wilson action \eqref{eq:R:Action} with respect to the plaquette operator, $U_p$. By `linearising' the lattice action we mean replacing its nonlinear terms with linear and Gaussian terms, after adding a sufficient number $n_\R$ of normally-distributed auxiliary variables. Effectively, this linearisation corresponds to a Hubbard-Stratonovich transformation of the corresponding partition function. 

Let us associate to each positively oriented plaquette on the lattice $n_\R$ complex $N \times N$ matrix variables $\wt Q_{\mu\nu,x}^{(i)}$ ($i=1,\ldots,n_\R$, and $\mu < \nu$) with the normal distribution $\mu_1(\wt Q_{\mu\nu,x}^{(i)})$. For the negatively oriented plaquettes we define $\wt Q_{\nu\mu,x}^{(i)} \equiv \wt Q_{\mu\nu,x}^{(i)\dag}$. First, we multiply the partition function of the mixed lattice gauge theory by $\int \mu_1[\wt Q] = 1$, where $\mu_1[\wt Q]$ is the product of Gaussian measures of all auxiliary variables. We then make a change of variables of the form:
\bea
\label{eq:R:ChangeVars}
\wt Q_p^{(i)} 
&=& 
\sqrt{2 \beta_\R '}
\l( Q_p^{(i)} - h_{p}^{(i)} \r)
\eea
Here $p \equiv (\mu\nu,x)$ labels plaquettes, $\beta_\R ' = \beta_\R N^\alpha/d_\R$ is a redefinition of the lattice coupling,\footnote{
$\alpha$ is an integer often equalling the degree of $d_\R$ as a polynomial in $N$ minus one. See Table \ref{tab:Summary.Mix} for the definitions of $\beta_\R$ for each particular representation.
}
$Q_p^{(i)}$ are the transformed auxiliary variables, and $h_{p}^{(i)} \equiv h^{(i)}( Q_p^{(j)}; U_p )$ are functions of the plaquette $U_p$ and of a single auxiliary variable $Q_p^{(j)}$ with $j<i$. The condition $j<i$, whose origin becomes clear below, ensures that the change of variables $( \wt Q, U ) \mapsto ( Q, U )$ is non-singular, hence invertible. In fact, its Jacobian is just a non-zero constant. In other words, the integration measure of the HS-transformed partition function is invariant under such a change of variables, up to a multiplicative constant. After the change of variables \eqref{eq:R:ChangeVars}, all nonlinear terms coming from the Gaussian measure $\mu_1[\wt Q]$ must have been exactly cancelled by the nonlinear terms in the Wilson action \eqref{eq:R:Action}. In the end, only the linear terms on $U_p$ and the Gaussian terms for $Q^{(i)}_p$ must survive.  

The way we choose suitable $h$'s is the most straightforward possible. Basically, we associate one auxiliary variable $\wt Q_p^{(j)}$ to each nonlinear term of $\chi_\R$. In order to eliminate them, we perform a change of variables $\wt Q_p^{(j)} \to Q_p^{(j)}$ using a $h^{(j)}_p$ that only depends on $U_p$, i.e. $h_{p}^{(j)} \equiv h^{(j)}(U_p)$. If the $h$'s themselves generate new nonlinear terms, these must be eliminated by adding even more auxiliary variables (again, one auxiliary variable per nonlinear term). Each of these secondary nonlinear terms depends both on $U_p$ and on the auxiliary variable $Q_p^{(j)}$ that generated it. Hence the new $h$'s that we need to eliminate them must also depend on $U_p$ and on $Q_p^{(j)}$, i.e. $h_{p}^{(i)} \equiv h^{(i)} (Q_p^{(j)}, U_p)$ with $j < i$. If the secondary nonlinear terms have a `lower $N$--ality' than the terms they replace (i.e. if the original and secondary nonlinear terms, respectively ${\cal O}_1(U_p)$ and ${\cal O}_2(U_p)$, transforms under $U_p \to z U_p$ as $z^{k_1} {\cal O}_1(U_p)$ and $z^{k_2} {\cal O}_2(U_p)$, with $z \in Z_N$ and $k_1 > k_2$), the process may be repeated until all nonlinear terms are eliminated and a linearised Wilson action emerges.

Let us split $h^{(i)}$ into its linear and nonlinear parts:
\bea
h^{(i)} (Q_p^{(j)}, U_p) 
&=& 
\bar h^{(i)} (Q_p^{(j)}, U_p) 
~+~
h^{(i)}_{\rm nlin}(U_p)
\eea
The linear piece $\bar h^{(i)}_p$, in its most general form, is given by:
\bea
\label{eq:R:hlin}
\bar h^{(i)} (Q_p^{(j)}, U_p) 
&=& 
A_1 U_p + A_2 U_p^\dag + A_3
\Tr\l\{ A_4 U_p \r\} + A_5 \Tr\l\{ A_6 U_p^\dag \r\}
\eea
where the $A_i \equiv A_i(Q_p^{(j)})$ are complex $N \times N$ matrices depending solely on one $Q_p^{(j)}$, with $j<i$. The change of variables \eqref{eq:R:ChangeVars} has the following effect on the Gaussian measures $\mu_1(\wt Q_p^{(i)})$:
\bea
\label{eq:R:Gauss:Terms:Expand}
-\frac{1}{2}~ \Tr\l\{ \wt Q_p^{(i)\dag} \wt Q_p^{(i)} \r\} 
&\stackrel{ \eqref{eq:R:ChangeVars} }{=}& 
- \beta_\R ' \Tr\l\{ Q_p^{(i)\dag} Q_p^{(i)} \r\} 
+2\beta_\R ' \Re\Tr\l\{ Q_p^{(i)\dag} h_{p}^{(i)} \r\} 
- \beta_\R ' \Tr\l\{ h_{p}^{(i)\dag} h_{p}^{(i)} \r\}
\nn\\
&=& 
- \beta_\R ' \Tr\l\{ Q_p^{(i)\dag} Q_p^{(i)} \r\} 
+2\beta_\R ' \Re\Tr\l\{ Q_p^{(i)\dag} \bar h_{p}^{(i)} \r\} 
+ S_{\rm nlin}(U_p)
\eea
In the expression above, $S_{\rm nlin}$ collects all the nonlinear terms generated by $h^{(i)}$. We hypothesise that $S_{\rm nlin}$ is either the symmetric of the sum of some nonlinear terms of $\chi_\R$, or they are secondary nonlinear terms that can be cancelled out with the addition of more auxiliary variables. In other words, we assume that a linearisation of the Wilson action is possible. Given \eqref{eq:R:hlin}, the linear term on the r.h.s. of \eqref{eq:R:Gauss:Terms:Expand} can be rearranged as follows:
\bea
\label{eq:R:Gauss:Terms:Reorder}
2\beta_\R ' \Re\Tr\l\{ 
Q_p^{(i)\dag} \bar h_{p}^{(i)} 
\r\} 
&\stackrel{\eqref{eq:R:hlin}}{=}& 
2\beta_\R ' \Re\Tr\l\{ 
g_{p}^{(i)\dag} U_p 
\r\}
\eea
where $g_{p}^{(i)}$ are complex $N \times N$ matrices that only depend on the auxiliary fields:
\bea
\label{eq:R:g}
g^{(i)}(Q_p^{(j)}) 
&=& 
A_1^\dag Q_p^{(i)} + Q_p^{(i)\dag} A_2 +
A_4^\dag \Tr \{ A_3^\dag Q_p^{(i)} \} + A_6 \Tr \{ A_5 Q_p^{(i)\dag}
\}
\eea
and $A_i \equiv A_i(Q_p^{(j)})$, $j < i$. After all $n_\R$ auxiliary variables are added and the change of variables \eqref{eq:R:ChangeVars} is performed, the exponent in the Gaussian measure $\mu_1[\wt Q]$ becomes:

\bea
\label{eq:R:Gauss:Terms(HS)}
-\frac{1}{2} \sum_{i=1}^{n_\R} \sum_p \Tr\l\{ \wt Q_p^{(i)\dag} \wt Q_p^{(i)} \r\} 
&\stackrel{ \eqref{eq:R:ChangeVars} }{=}&
\nn\\
\comm{Gaussian}
&&-~ \beta_\R ' \sum_{i=1}^{n_\R} \sum_p \Tr \{ Q_p^{(i)\dag} Q_p^{(i)} \} 
\nn\\
\comm{linear}
&&+~ 2\beta_\R ' \sum_p \Re\Tr \{ f_p^{\R\dag} U_p \} 
\nn\\
\comm{cancels out}
&& -~ \beta_\R ' \sum_p \Re\chi_\R\l\{ U_p \r\} 
\eea
where $f^\R_p$ is a complex $N \times N$ matrix that only depends on the auxiliary variables:
\bea
\label{eq:R:f}
f^\R_p \equiv \sum_{i=1}^{n_\R} g_{p}^{(i)}
\eea
In sum, $f^\R_p$ encodes the nonlinear dependence of $\chi_\R$ on $U_p$.

Once the Wilson action \eqref{eq:R:Action} is linearised, the efficient updating algorithms discussed in Section \ref{sec:intro} can be applied directly. The updates are performed with respect to the probability distribution of individual links \eqref{eq:F:Distrib:Single}, 
where $V_{\mu,x} \equiv V_{\mu,x}^\R$ is the sum of `staples':
\bea
\label{eq:R:Staples}
V_{\mu,x}^\R 
&=&
2 \beta_\R ' \sum_{{\nu=1}\atop{(\nu\neq\mu)}}^{d} 
\l(
f^{\R}_{\mu\nu,x} U_{\nu,x} U_{\mu,x+\hat\nu} U_{\nu,x+\hat\mu}^\dag 
~+~ 
U_{\nu,x-\hat\nu}^\dag f^{\R}_{\nu\mu,x-\hat\nu} U_{\mu,x-\hat\nu} U_{\nu,x-\hat\nu+\hat\mu}
\r)
\eea
and $f^{\R}_{\nu\mu,x} \equiv f^{\R\dag}_{\mu\nu,x}$, $\forall \mu < \nu$. It must be noted that the order in which $f_p^\R$ appears in the `staples' depends on the particular choice for the initial point of the plaquette operator \eqref{eq:F:Plaquette}. Changing the initial point of the plaquette to any other vertex would simply result in a relocation of $f^\R_p$ within the `staples'. For the particular case when $\R = F$ we have $f^F_p = \1$, so we have recovered the original algorithm.

In the next Sections we linearise explicitly the Wilson action for each representation of $SU(N)$ with $N$--ality $k \leq 3$; Table \ref{tab:Summary.Mix} summarises the changes of variables necessary for each linearisation. We then discuss the case of mixed actions, with and without reduced directions. In the end, we perform numerical tests on some of these algorithms against Metropolis, in order to check their accuracy and performance.

\subsection
{Adjoint representation
\label{sec:mix:A}}

The best studied example of a mixed Wilson action is the one involving both fundamental and adjoint plaquette terms,
\bea
\label{eq:F/A:Action}
S_{F+A}(\beta_F, \beta_A; [U])
&=&
S_{F}(\beta_F; [U]) +
S_{A}(\beta_A; [U])
\eea
The parameter space of $(\beta_F, \beta_A)$ lattice couplings is known as the fundamental/adjoint plane. This action is important, for example, in the study of the role of centre of the gauge group $Z_N$ in colour confinement, because the adjoint character $\chi_A(U_p)$ is invariant under centre shifts, $U_p \mapsto z U_p$, $z \in Z_N$.

Due to the quadratic nature of $\chi_A$ \eqref{eq:A:chi}, it is not possible to use the efficient updating algorithms of the fundamental Wilson action, discussed in Section \ref{sec:intro}. A Metropolis algorithm is often used instead. In particular, combining a Cabibbo-Marinari-Metropolis algorithm with overrelaxation-Metropolis sweeps results in a rather good performance, at least for $SU(3)$ \cite{NeccoHasenbusch}. Recently, a biased Metropolis algorithm with heatbath efficiency has been constructed for the $SU(2)$ case \cite{BazavovBergHeller}. In this Section we construct a pseudo-heatbath algorithm for the $SU(N)$ adjoint Wilson action, 
\bea
\label{eq:A:Action}
S_A(\beta_A; [U]) &=& -\frac{\beta_A}{N^2-1} {\sum}_p |\Tr \{ U_p \}|^2 
\eea
for all $N$, using the method of auxiliary variables. But because the double-trace term is always positive, the cases for positive and negative adjoint coupling $\beta_A$ must be considered separately.

\subsubsection{\texorpdfstring
{$\beta_A > 0$}
{beta(A) > 0}
\label{sec:mix:A:+}}

When $\beta_A$ is positive, the adjoint Wilson action \eqref{eq:A:Action} is always negative. In order to cancel out the double-trace terms, it suffices to introduce one complex number per plaquette $\wt z_p$ with the normal distribution $\mu_1(\wt z_p)$. We then multiply the partition function of the adjoint Wilson theory $Z_A(\beta_A)$ by the Gaussian integral $\int \mu_1[\wt z] = 1$. Consider the change of variables:
\bea
\label{eq:A:+:ChangeVars}
\wt z_p 
&=& 
\sqrt{2\beta_A '} 
\l( z_p - \frac{1}{N}\Tr\{ U_p \} \r)
\eea
where $\beta_A ' = \beta_A N^2 / (N^2-1)$. The effect of \eqref{eq:A:+:ChangeVars} on the Gaussian exponent of $\mu_1[\wt z]$ is:
\bea
\label{eq:A:+:Gauss:Terms(HS)}
-\frac{1}{2}~ {\sum}_p 
\l| \wt z_p \r|^2
&\stackrel{ \eqref{eq:A:+:ChangeVars} }{=}&
\nn\\ 
\comm{Gaussian} 
&&-\beta_A '~ 
{\sum}_p 
\l| z_p \r|^2
\nn\\
\comm{linear} 
&&+\frac{2\beta_A '}{N}~ 
{\sum}_p 
\Re\Tr\l\{ z_p^\ast U_p \r\} 
\nn\\
\comm{cancels out} 
&&-\frac{\beta_A '}{N^2}~ 
{\sum}_p 
|\Tr \{ U_p \}|^2 
\eea
The last term in the r.h.s. of \eqref{eq:A:+:Gauss:Terms(HS)} cancels out all the adjoint terms of \eqref{eq:A:Action} exactly, as long as $\beta_A$ is positive. It comes from conjugating $h_p^A = \Tr\{U_p\}/N$ with itself. If $\beta_A < 0$, then it is not possible to do the same with $z$--variables only. We discuss this case later. The partition function of the adjoint Wilson theory then becomes:
\bea
\label{eq:A:+:Z*Gauss(HS)}
Z_A(\beta_A) 
&=& 
\int \mu_{\rm H} [U]~ \mu_\sigma[z]~ 
\exp\l( 
\frac{2\beta_A'}{N}~ {\sum}_p \Re\Tr\l\{z_p^\ast U_p \r\}
\r)
\eea
where $\mu_\sigma[z] \equiv \prod_p \mu_{\sigma}(z_p)$ and $\sigma^2 = 1/\beta_A'$. Since the exponent of the Boltzmann factor is now linear, the link variables can be updated using the efficient algorithms of Section \ref{sec:intro}. The updates are performed with respect to the probability distribution of individual links \eqref{eq:F:Distrib:Single}, where $V_{\mu,x} \equiv V^A_{\mu,x}$ is given by \eqref{eq:R:Staples} with $f_p^A = z_p / N$. And because $z_p$ and $f_p^A$ are just complex numbers, the cost of updating the auxiliary variables in this case is negligible as compared with link updates.

\subsubsection{\texorpdfstring
{$\beta_A < 0$}
{beta(A) < 0}
\label{sec:mix:A:-}}

In the case of negative $\beta_A$, the adjoint Wilson action \eqref{eq:A:Action} is positive.
Therefore, the double-trace terms cannot be eliminated using \eqref{eq:A:+:ChangeVars}. However, it is possible to linearise the action if we introduce different auxiliary variables, namely one complex $N \times N$ matrix per plaquette $\wt Q_p$ with the normal distribution $\mu_1(\wt Q_p)$. As usual, we multiply the partition function $Z_A(-|\beta_A|)$ with the Gaussian integral $\int \mu_1[\wt Q] = 1$. Then we consider the change of variables:
\bea
\label{eq:A:-:ChangeVars}
\wt Q_p &=& \sqrt{2\beta_A '} \l( Q_p - \l( U_p - \frac{1}{N} \Tr\l\{ U_p \r\} \1 \r) \r) 
\eea
where $\beta_A' = |\beta_A| N / (N^2-1)$. The effect of \eqref{eq:A:-:ChangeVars} on the Gaussian exponent of $\mu_1[\wt Q]$ is:
\bea
\label{eq:A:-:Gauss:Terms(HS)}
-\frac{1}{2}~ {\sum}_p \Tr\l\{ {\wt Q_{p}}^\dag \wt Q_{p} \r\}
&\stackrel{ \eqref{eq:A:-:ChangeVars} }{=}&
\nn\\ 
\comm{Gaussian} 
&& -\beta_A'
{\sum}_p \Tr\l\{ Q_{p}^\dag Q_{p} \r\} 
\nn\\
\comm{linear} 
&& +2\beta_A' {\sum}_p \Re\Tr\l\{ \l( Q_{p} - \frac{1}{N} \Tr\l\{ Q_p \r\} \1 \r)^\dag U_p \r\} 
\nn\\
\comm{cancels out} 
&& +\frac{\beta_A'}{N} {\sum}_p \l|\Tr\l\{ U_p \r\}\r|^2
\eea
The last term in the r.h.s. of \eqref{eq:A:-:Gauss:Terms(HS)} carries the correct sign to cancel out the double-trace terms with negative $\beta_A$ from \eqref{eq:A:Action}. The partition function of this theory then becomes:
\bea
\label{eq:A:-:Z*Gauss(HS)}
Z_A(\beta_A) 
&=& 
\int \mu_{\rm H}[U]~ \mu_\sigma[Q]~ \exp\l(2\beta_A' {\sum}_p \Re\Tr\l\{ \l( Q_{p} - \frac{1}{N} \Tr\l\{ Q_p \r\} \1 \r)^\dag U_p \r\}\r)
\eea
where $\mu_\sigma[Q] \equiv \prod_p \mu_{\sigma}(Q_p)$, and $\sigma^2 = 1/\beta_A'$. Individual links are then updated with respect to the probability distribution \eqref{eq:F:Distrib:Single}, with $V_{\mu,x} \equiv V^A_{\mu,x}$ given by \eqref{eq:R:Staples}, and $f_p^A$ is the matrix factor:
\bea
\label{eq:A:-:f}
f^A_p = Q_{p} - \frac{1}{N} \Tr\l\{ Q_p \r\} \1
\eea
Since $Q_p$ and $f_p^A$ are $N \times N$ matrices, and not just complex numbers, it is expected for the $\beta_A < 0$ case to be less efficient than the $\beta_A > 0$ case.

\subsection{\texorpdfstring
{Higher $N$--ality representations}
{Higher N-ality representations}
\label{sec:mix:Nality}}

The case of higher $N$--ality representations is more complicated, because there are more nonlinear terms to be eliminated, and these depend on higher powers of the link variables. This means that a larger number of auxiliary variables is needed in order to eliminate all nonlinear terms.  In the following, we explicitly linearise the Wilson actions for two- and three-index irreducible representations of $SU(N)$ using the method of auxiliary variables.

\subsubsection{\texorpdfstring
{$\R=2\pm$}
{R=2+/-}
\label{sec:mix:Nality:k=2,+/-}}

Here we consider Wilson actions with plaquettes in the symmetric ($+$) or antisymmetric ($-$) two-index representations of $SU(N)$, respectively $\R=(2)$ or $\R=(1,1)$ : 
\bea
\label{eq:2,11:Action}
S_\pm (\beta_\pm;[U]) &=& 
-\frac{\beta_\pm}{N(N \pm 1)} \sum_p \Re\l( \Tr \{U_p\}^2 \pm \Tr\{U_p^2\} \r)
\eea
We consider both situations of positive or negative lattice coupling $\beta_\pm$, whose sign we denote by $\sigma \equiv {\rm sgn}(\beta_\pm)$. One way of linearising \eqref{eq:2,11:Action} requires the addition of three auxiliary complex matrix variables per plaquette $\wt Q_p^{(i)}$, $i = 1,2,3$, and the corresponding change of variables:
\bea
\label{eq:2,11:ChangeVars:v1:1}
\wt Q_p^{(1)} 
&=& 
\sqrt{2\beta_\pm '} \l( Q_p^{(1)} - \frac{1}{2} \l( U_p + \frac{\sigma}{N} \Tr\{ U_p^\dag \} \1 \r) \r) 
\\\nn\\
\label{eq:2,11:ChangeVars:v1:2}
\wt Q_p^{(2)} 
&=& 
\sqrt{2\beta_\pm '} \l( Q_p^{(2)} - \frac{1}{2} \l( U_p \pm \frac{\sigma}{N} U_p^\dag \r) \r) 
\\\nn\\
\label{eq:2,11:ChangeVars:v1:3}
\wt Q_p^{(3)} 
&=& 
\sqrt{2\beta_\pm '} \l( Q_p^{(3)} - \frac{1}{2} \l( U_p - \frac{1}{N} \Tr\{ U_p \} \1 \r) \r) 
\eea
where $\beta_\pm ' = 2|\beta_\pm|/(N\pm 1)$. The variable $Q_p^{(1)}$ is responsible for the elimination of the quadratic term $\Tr\{U_p\}^2$ in the action, and its replacement by a linear term. However, it also generates a secondary nonlinear term of the form $|\Tr\{ U_p\}|^2$. The variable $Q_p^{(2)}$ eliminates the quadratic term $\Tr\{U_p^2\}$ in the action and replaces it with a linear term, with no other side effects. Finally, the variable $Q_p^{(3)}$ eliminates the term $|\Tr\{U_p\}|^2$ generated by $Q_p^{(1)}$. This is achieved using the transformation \eqref{eq:2,11:ChangeVars:v1:3}, which is similar to the one used in the elimination of negative-coupling adjoint plaquettes \eqref{eq:A:-:ChangeVars}. The remaining linear terms contribute to the sum of `staples' \eqref{eq:R:Staples} with the matrix factor:
\bea
\label{eq:2,11:f:v1}
f_p^\pm 
&=& 
\frac{1}{2} \l( Q_p^{(1)} + \frac{\sigma}{N} \Tr\{  Q_p^{(1)\dag} \}  + Q_p^{(2)} \pm \frac{\sigma}{N} Q_p^{(2)\dag} + Q_p^{(3)} - \frac{1}{N} \Tr\{Q_p^{(3)} \} \1 \r)
\eea

However, the above choice is not unique. Is is also possible to linearise the Wilson action \eqref{eq:2,11:Action} by adding only two auxiliary complex matrix variables per plaquette. The corresponding change of variables is:
\bea
\label{eq:2,11:ChangeVars:v2:1}
\wt Q_p^{(1)} 
&=& 
\sqrt{2\beta_\pm '} \l( 
Q_p^{(1)} - \frac{1}{2} \l( \sigma U_p^\dag \pm  \frac{1}{N} U_p + \frac{1}{N} \Tr\{ U_p \} \1 \r) \r) 
\\\nn\\
\label{eq:2,11:ChangeVars:v2:2}
\wt Q_p^{(2)} 
&=& 
\sqrt{2\beta_\pm '} \l( Q_p^{(2)} - \l(\frac{N\pm 2}{4N}\r)^{\frac{1}{2}} \l( U_p - \frac{1}{N} \Tr\{ U_p \} \1 \r) \r) 
\eea
The variable $Q_p^{(1)}$ is responsible for the simultaneous elimination of both nonlinear terms from $\chi_\pm$. However, it also generates a $|\Tr\{U_p\}|^2$ term, which is promptly eliminated by  variable $Q_p^{(2)}$. After some algebra, we find the contribution of this change of variables to the sum of `staples' \eqref{eq:R:Staples} to be:
\bea
\label{eq:2,11:f:v2}
f_p^\pm 
&=& 
\frac{1}{2} \l( 
\sigma Q_p^{(1)\dag} \pm \frac{1}{N} Q_p^{(1)} 
+ \frac{1}{N} \Tr\{ Q_p^{(1)} \} \1 
\r) 
+ \l(\frac{N\pm 2}{4N}\r)^{\frac{1}{2}} 
\l( Q_p^{(2) } - \frac{1}{N} \Tr\{ Q_p^{(2)} \} \1 \r) 
\eea

Albeit different, both situations describe the same theories. However, it is clear that the second method, with only two auxiliary variables, is superior to the first method: the smaller the number of auxiliary variables, the more efficient we expect the corresponding MC algorithm to be.

\subsubsection{\texorpdfstring
{$\R=3\pm$}
{R=3+/-}
\label{sec:mix:Nality:k=3,+/-}}

Here we consider Wilson actions with plaquettes in the symmetric ($+$) or antisymmetric ($-$) three-index representations of $SU(N)$, respectively $\R=(3)$ or $\R=(1,1,1)$ : 
\bea
\label{eq:3,111:Action}
S_\pm(\beta_\pm;U) 
&=&
-\frac{\beta_\pm}{N(N^2 \pm 3N + 2)} \sum_p \Re\l( \Tr \l\{ U_p \r\}^3 \pm 3 \Tr\l\{ U_p \r\} \Tr\l\{ U_p^2 \r\} + 2 \Tr\l\{ U_p^3 \r\} \r)~~~
\eea
We again consider both situations of positive or negative lattice coupling $\beta_\pm$, whose sign we denote by $\sigma \equiv {\rm sgn}(\beta_\pm)$. One way of linearising \eqref{eq:3,111:Action} requires the addition of seven auxiliary complex matrix variables per plaquette $\wt Q_p^{(i)}$, $i = 1,\ldots,7$, and the corresponding change of variables:

\bea
\label{eq:3,111:ChangeVars:v1:1}
\wt Q_p^{(1)} &=& \sqrt{2\beta_\pm '} \l( Q_p^{(1)} - \frac{1}{\sqrt{12}N} \l( U_p \Tr\{ U_p \} + \sigma \Tr\{ U_p^\dag \} \1 \r) \r) 
\\\nn\\
\label{eq:3,111:ChangeVars:v1:2}
\wt Q_p^{(2)} &=& \sqrt{2\beta_\pm '} \l( Q_p^{(2)} - \frac{1}{2N} \l( U_p \Tr\{ U_p \} \pm \sigma U_p^\dag \r) \r) 
\\\nn\\
\label{eq:3,111:ChangeVars:v1:3}
\wt Q_p^{(3)} &=& \sqrt{2\beta_\pm '} \l( Q_p^{(3)} - \frac{1}{\sqrt{6}N} \l( U_p^2 + \sigma U_p^\dag \r) \r) 
\\\nn\\
\label{eq:3,111:ChangeVars:v1:4}
\wt Q_p^{(4)} &=& \sqrt{2\beta_\pm '} \l( Q_p^{(4)} - \frac{1}{\sqrt{12}} Q_p^{(1)} U_p^\dag - \frac{1}{N} \Tr\{ U_p \} \1 \r)
\\\nn\\
\label{eq:3,111:ChangeVars:v1:5}
\wt Q_p^{(5)} &=& \sqrt{2\beta_\pm '} \l( Q_p^{(5)} - \frac{1}{2} Q_p^{(2)} U_p^\dag - \frac{1}{N} \Tr\{ U_p \} \1 \r) 
\\\nn\\
\label{eq:3,111:ChangeVars:v1:6}
\wt Q_p^{(6)} &=& \sqrt{2\beta_\pm '} \l( Q_p^{(6)} - \frac{1}{\sqrt{6}} Q_p^{(3)} U_p^\dag - \frac{1}{N} U_p \r) 
\\\nn\\
\label{eq:3,111:ChangeVars:v1:7}
\wt Q_p^{(7)} &=& \sqrt{2\beta_\pm '} \l( Q_p^{(7)} - \l(\frac{29}{12}\r)^{\frac{1}{2}} 
\l( U_p - \frac{1}{N} \Tr\{ U_p \} \1 \r) \r) 
\eea
where $\beta_\pm ' = 6|\beta_\pm|N / (N^2 \pm 3N + 2)$.  
The variable $Q_p^{(1)}$ eliminates the cubic term $\Tr\{U_p\}^3$ from the action, but it also generates nonlinear terms of the form $\Tr\{Q_p^{(1)\dag} U_p\} \Tr\{ U_p\}$ and $|\Tr\{ U_p \}|^2$. The variable $Q_p^{(2)}$ eliminates the cubic term $\Tr\{U_p^2\}\Tr\{U_p\}$ and generates similar nonlinear terms, namely $\Tr\{Q_p^{(2)\dag} U_p\} \Tr\{ U_p\}$ and $|\Tr\{U_p\}|^2$. The variable $Q_p^{(3)}$ eliminates the cubic term $\Tr\{U_p^3\}$ and generates one nonlinear term of the form $\Tr\{Q_p^{(3)\dag} U_p^2\}$. The variables $Q_p^{(4)}$, $Q_p^{(5)}$ and $Q_p^{(6)}$ eliminate the secondary nonlinear terms generated by $Q_p^{(1)}$, $Q_p^{(2)}$ and $Q_p^{(3)}$, respectively, except the term of the form $|\Tr\{U_p\}|^2$; in fact, the variables $Q_p^{(4)}$ and $Q_p^{(5)}$ also produce such a term. The sum of these double-traces is eliminated by the variable $Q_p^{(7)}$. All the seven auxiliary variables generate linear terms, whose contribution to the sum of `staples' \eqref{eq:R:Staples} is given by:
\bea
\label{eq:3,111:v1:f}
f_p^\pm &=& 
\frac{\sigma}{\sqrt{12}N} \Tr\{ Q_p^{(1)\dag}\} \1
\pm \frac{\sigma}{2N} Q_p^{(2)\dag} 
+ \frac{\sigma}{\sqrt{6}N} Q_p^{(3)\dag} 
+ \frac{1}{N} \Tr\{Q_p^{(4)}\} \1  
+ \frac{1}{\sqrt{6}} Q_p^{(4)\dag} Q_p^{(1)}
\nn\\\nn\\
&&
+ \frac{1}{N} \Tr\{Q_p^{(5)}\} \1  
+ \frac{1}{\sqrt{2}} Q_p^{(5)\dag} Q_p^{(2)}
+ \frac{1}{N} Q_p^{(6)} 
+ \frac{1}{\sqrt{3}} Q_p^{(6)\dag} Q_p^{(3)}
\nn\\\nn\\
&&
+ \l(\frac{29}{12}\r)^{\frac{1}{2}}
\l( Q_p^{(7)} - \frac{1}{N} \Tr\{ Q_p^{(7)} \} \1 \r) 
\eea

Just like in the previous case of two-index representations, the choice above is not unique. In fact, there is a cheaper way to eliminate all nonlinear terms, involving only five auxiliary variables. The corresponding change of variables is:

\bea
\label{eq:3,111:ChangeVars:v2:1}
\wt Q_p^{(1)} &=& \sqrt{2\beta_\pm '} \l( Q_p^{(1)} - \frac{1}{\sqrt{12}N} \l( U_p \Tr\{ U_p \} + \sigma \Tr\{ U_p^\dag \} \1 \pm 3\sigma U_p^\dag \r) \r) 
\\\nn\\
\label{eq:3,111:ChangeVars:v2:2}
\wt Q_p^{(2)} &=& \sqrt{2\beta_\pm '} \l( Q_p^{(2)} - \frac{1}{\sqrt{6}N} \l( U_p^2 + \sigma U_p^\dag \r) \r) 
\\\nn\\
\label{eq:3,111:ChangeVars:v2:3}
\wt Q_p^{(3)} &=& \sqrt{2\beta_\pm '} \l( Q_p^{(3)} - \frac{1}{\sqrt{12}} Q_p^{(1)} U_p^\dag - \frac{1}{N} \Tr\{ U_p \} \1 \r) 
\\\nn\\
\label{eq:3,111:ChangeVars:v2:4}
\wt Q_p^{(4)} &=& \sqrt{2\beta_\pm '} \l( Q_p^{(4)} - \frac{1}{\sqrt{6}} Q_p^{(2)} U_p^\dag - \frac{1}{N} U_p \r) 
\\\nn\\
\label{eq:3,111:ChangeVars:v2:5}
\wt Q_p^{(5)} &=& 
\sqrt{2\beta_\pm '} \l( Q_p^{(5)} 
- \l(\frac{7N \pm 3}{6N}\r)^{\frac{1}{2}} 
\l( U_p - \frac{1}{N} \Tr\{ U_p \} \1 \r) \r) 
\eea
The variable $Q_p^{(1)}$ eliminates simultaneously both the $\Tr\{U_p\}^3$ and $\Tr\{U_p^2\}\Tr\{U_p\}$ terms in the action, but it also generates nonlinear terms of the form $\Tr\{Q_p^{(1)\dag} U_p\} \Tr\{ U_p\}$ and $|\Tr\{ U_p \}|^2$. The variable $Q_p^{(2)}$ eliminates the cubic term $\Tr\{U_p^3\}$ and generates one nonlinear term of the form $\Tr\{Q_p^{(2)\dag} U_p^2\}$. The variables $Q_p^{(3)}$ and $Q_p^{(4)}$ eliminate the nonlinear terms generated by $Q_p^{(1)}$ and $Q_p^{(2)}$, respectively, except the term of the form $|\Tr\{U_p\}|^2$ (which is also generated by $Q_p^{(3)}$). The sum of these double-traces is eliminated by $Q_p^{(5)}$. In the end, the contribution of the remaining linear terms to the sum of `staples' \eqref{eq:R:Staples} is given by:
\bea
f_p^\pm &=& 
\label{eq:3,111:v2:f}
\frac{\sigma}{\sqrt{12}N} \l( \Tr\{ Q_p^{(1)}\} \1
\pm 3 Q_p^{(1)} \r)^\dag 
+ \frac{\sigma}{\sqrt{6}N} Q_p^{(2)\dag} 
+ \frac{1}{\sqrt{12}} Q_p^{(3)\dag} Q_p^{(1)}
+ \frac{1}{N} \Tr\{ Q_p^{(3)}\} \1
\nn\\\nn\\
&&+ \frac{1}{\sqrt{6}} Q_p^{(4)\dag} Q_p^{(2)}
+ \frac{1}{N} Q_p^{(4)} 
+ \l(\frac{7N \pm 3}{6N}\r)^{\frac{1}{2}} 
\l( Q_p^{(5)} - \frac{1}{N} \Tr\{ Q_p^{(5)} \} \1 \r)
\eea

\subsubsection{\texorpdfstring
{$\R = (2,1)$}
{R=(2,1)}
\label{sec:mix:Nality:k=3,(2,1)}}

Finally, we consider Wilson actions with plaquettes in the representation $\R = (2,1)$:
\bea
\label{eq:21:Action}
S_{(2,1)} (\beta_{(2,1)}; U)
&=& 
-\frac{\beta_{(2,1)}}{N(N^2 - 1)} \sum_p \Re\l( \Tr\l\{ U_p \r\}^3 - \Tr\l\{ U_p^3 \r\} \r) 
\eea
The sign of $\beta_{(2,1)}$ is denoted by $\sigma \equiv {\rm sgn}(\beta_\pm)$. One way of linearising \eqref{eq:3,111:Action} requires the addition of five auxiliary complex matrix variables per plaquette $\wt Q_p^{(i)}$, $i = 1,\ldots,5$, and the corresponding change of variables:

\bea
\label{eq:21:ChangeVars:v1:1}
\wt Q_p^{(1)} &=& \sqrt{2\beta_{(2,1)}'} \l( Q_p^{(1)} - \frac{1}{\sqrt{6}N} \l( U_p \Tr\{ U_p \} + \sigma \Tr\{ U_p^\dag \} \1 \r) \r) 
\\\nn\\
\label{eq:21:ChangeVars:v1:2}
\wt Q_p^{(2)} &=& \sqrt{2\beta_{(2,1)}'} \l( Q_p^{(2)} - \frac{1}{\sqrt{6}N} \l( U_p^2 - \sigma U_p^\dag \r) \r) 
\\\nn\\
\label{eq:21:ChangeVars:v1:3}
\wt Q_p^{(3)} &=& \sqrt{2\beta_{(2,1)}'} \l( Q_p^{(3)} - \frac{1}{\sqrt{6}} Q_p^{(1)} U_p^\dag - \frac{1}{N} \Tr\{ U_p \} \1 \r) 
\\\nn\\
\label{eq:21:ChangeVars:v1:4}
\wt Q_p^{(4)} &=& \sqrt{2\beta_{(2,1)}'} \l( Q_p^{(4)} - \frac{1}{\sqrt{6}} Q_p^{(2)} U_p^\dag - \frac{1}{N} U_p \r) 
\\\nn\\
\label{eq:21:ChangeVars:v1:5}
\wt Q_p^{(5)} &=& \sqrt{2\beta_{(2,1)}'} \l( Q_p^{(5)} - \frac{2}{\sqrt{3}} \l( U_p - \frac{1}{N} \Tr\{ U_p \} \1 \r) \r) 
\eea
where $\beta_{(2,1)}' = 3|\beta_{(2,1)}|N / (N^2 - 1)$.
The variable $Q_p^{(1)}$ eliminates the cubic term $\Tr\{U_p\}^3$ in the action, and it generates nonlinear terms of the form $\Tr\{Q_p^{(1)\dag} U_p\} \Tr\{ U_p\}$ and $|\Tr\{ U_p \}|^2$. The variable $Q_p^{(2)}$ eliminates the cubic term $\Tr\{U_p^3\}$ and generates one nonlinear term of the form $\Tr\{Q_p^{(2)\dag} U_p^2\}$. The variables $Q_p^{(3)}$ and $Q_p^{(4)}$ eliminate the secondary nonlinear terms generated by $Q_p^{(1)}$ and $Q_p^{(2)}$, respectively, except the term of the form $|\Tr\{U_p\}|^2$ (which is also generated by $Q_p^{(3)}$). The sum of these quadratic terms is eliminated by the fifth variable. In the end, the contribution of the remaining linear terms to the sum of `staples' \eqref{eq:R:Staples} is given by:
\bea
\label{eq:21:v1:f}
f_p^{(2,1)} &=& 
\frac{\sigma}{\sqrt{6}N} \Tr\{ Q_p^{(1)\dag} \} \1
-\frac{\sigma}{\sqrt{6}N} Q_p^{(2)\dag} 
+\frac{1}{N} \Tr\{ Q_p^{(3)} \} \1
+ \frac{1}{\sqrt{6}} Q_p^{(3)\dag} Q_p^{(1)} 
\nn\\\nn\\
&&
+ \frac{1}{N} Q_p^{(4)} 
+ \frac{1}{\sqrt{6}} Q_p^{(4)\dag} Q_p^{(2)}
+ \frac{2}{\sqrt{3}} \l( Q_p^{(5)} 
- \frac{1}{N} \Tr\{ Q_p^{(5)} \} \1 \r) 
\eea

Also in this case it is possible to find a smaller set of auxiliary variables that does the same job. In fact, only three auxiliary variables suffice to eliminate all the nonlinear terms from the action \eqref{eq:21:Action}. The corresponding change of variables is:
\bea
\label{eq:21:ChangeVars:v2:1}
\wt Q_p^{(1)} &=& \sqrt{2\beta_{(2,1)}'} \l( Q_p^{(1)} - \frac{1}{\sqrt{6}N} \l( U_p \l(\Tr\{U_p\} \1 - U_p \r) + \sigma \l( U_p + \Tr \{U_p\} \1 \r)^\dag \r) \r) 
\\\nn\\
\label{eq:21:ChangeVars:v2:2}
\wt Q_p^{(2)} &=& \sqrt{2\beta_{(2,1)}'} \l( Q_p^{(2)} - \frac{1}{\sqrt{6}}~ Q_p^{(1)} U_p^\dag - \frac{1}{N} \l(U_p - \Tr\{U_p\} \1 \r) \r) 
\\\nn\\
\label{eq:21:ChangeVars:v2:3}
\wt Q_p^{(3)} &=& \sqrt{2\beta_{(2,1)}'} \l( Q_p^{(3)} - \l( \frac{4N-6}{3N} \r)^\frac{1}{2} \l( U_p - \frac{1}{N} \Tr\{U_p\} \1 \r) \r) 
\eea
The variable $Q_p^{(1)}$ eliminates both nonlinear terms in the action, but it also generates nonlinear terms of the form $\Tr\{Q_p^{(1)\dag} U_p\} \Tr\{ U_p\}$, $\Tr\{Q_p^{(1)\dag} U_p^2\}$ and $|\Tr\{ U_p \}|^2$. The variable $Q_p^{(2)}$ eliminates all secondary nonlinear terms generated by  $Q_p^{(1)}$, except the term $|\Tr\{U_p\}|^2$, which is also generated by it. The sum of these double-traces is eliminated by  $Q_p^{(3)}$. In the end, the contribution of the remaining linear terms to the sum of `staples' \eqref{eq:R:Staples} is given by:
\bea
\label{eq:21:v2:f}
f_p^{(2,1)} &=& 
\frac{\sigma}{\sqrt{6}N} \l( Q^{(1)}_p + \Tr\{Q^{(1)}_p\} \1 \r)^\dag 
- \frac{1}{N} \l( Q_p^{(2)} - \Tr\{Q_p^{(2)}\} \1 \r)
+\frac{1}{\sqrt{6}} Q_p^{(1)} Q_p^{(2)} 
\nn\\\nn\\
&&+\l( \frac{4N-6}{3N} \r)^\frac{1}{2} \l( Q_p^{(3)} - \frac{1}{N} \Tr\{Q_p^{(3)}\} \1 \r)
\eea

\subsection
{Mixed Wilson actions
\label{sec:mix:mixed}}

The generalisation to the case of mixed Wilson actions is trivial, because a linearised mixed Wilson action is simply the sum of linearised Wilson actions for each representation:
\bea
\label{eq:mix:Action:Single}
S_{\rm mix}' (\vec\beta; U_{\mu,x}) 
&=& 
\sum_\R S_{\R}' (\beta_\R; U_{\mu,x}) 
\eea
Therefore, the probability distribution of individual links associated with a generic mixed Wilson action is given by \eqref{eq:F:Distrib:Single},
where $V_{\mu,x}$ is the sum of `staples':
\bea
\label{eq:mix:Staples}
V_{\mu,x} 
&=& 
\sum_{{\nu=1}\atop{(\nu\neq\mu)}}^d
\l(
f_{\mu\nu,x} 
U_{\nu,x} 
U_{\mu,x+\hat\nu} 
U_{\nu,x+\hat\mu}^\dag
~+~
U_{\nu,x-\hat\nu}^\dag 
f_{\nu\mu,x-\hat\nu}
U_{\mu,x-\hat\nu}
U_{\nu,x-\hat\nu+\hat\mu}
\r)
\eea
and $f_{\mu\nu,x} \equiv f_p$ encodes the information about all representations involved:
\bea
\label{eq:mix:f}
f_p = 2 \sum_\R \beta_\R ' f^\R_p
\eea
The possibility of using the efficient MC algorithms discussed in Section \ref{sec:intro} follows immediately. The MC algorithm for a general mixed Wilson action is summarised in Appendix \ref{app:MC:mix}.

This straightforward approach to mixed Wilson actions may not be the most efficient, however. In this approach, each representation is treated independently, and the total number of auxiliary variables is $n_{\rm mix} = \sum_\R n_\R$. But in some situations it is possible to reduce $n_{\rm mix}$. For example, consider the mixed `$A + (2)$' Wilson action. During the linearisation of $\chi_{(2)}$, a secondary nonlinear term of the form $| \Tr\{ U_p \} |^2$ is generated, and it can be eliminated in simultaneous with the double-trace terms of $\chi_A(U_p)$. Therefore, only two auxiliary variables per plaquette are needed, instead of the naive $n_A + n_{(2)} = 3$. This is the same ambiguity that exists in the case of individual representations, as shown above. In sum, the process of linearisation is not unique, and may be improved by a wise choice of auxiliary variables and respective change of variables. In the end, one must always choose the set with the smallest number of auxiliary variables.

\subsection
{A note on reduced mixed models
\label{sec:mix:red}}

A special note must be taken in the case of mixed Wilson actions on lattices with reduced directions. These may be useful to study the (non-)universality of the symmetry breaking transitions that invalidate the large $N$ equivalence in TEK models \cite{ArroyoPrivate}.

As usual, the construction of an efficient updating algorithm for a mixed reduced model consists in adding enough auxiliary variables in order to eliminate all the nonlinear terms in the action. In reduced models, however, the nonlinearities have two origins: the quadratic nature of the reduced plaquette operator, and the nonlinear nature of the $SU(N)$ characters. Clearly, the nonlinearities associated with $SU(N)$ characters need to be dealt with first. For that end, the linearisation of the mixed lattice action proceeds exactly as described in Sections \ref{sec:mix:linearisation}--\ref{sec:mix:mixed}. 

The problem occurs in the final step, when dealing with the reduced plaquette operator. One would be tempted to solve the problem with the original Fabricius-Haan treatment for the EK model. The nonlinear terms to be eliminated are of the form:
\bea
\label{eq:mixred:Plaquette}
-2\beta_\R '~
\Re\Tr\l\{ 
f_{\mu\nu, x}^{\R\dag}
U_{\mu, x} U_{\nu, x + \hat\mu}
U_{\mu, x}^\dag  U_{\nu, x}^\dag
\r\}
\eea
which differ from the EK plaquette terms by the matrix factor $f_{\mu\nu,x}^\R$. If one adds an auxiliary variable $\wt R_{\mu\nu,x}$ and perform the change of variables
\bea
\label{eq:mixred:ChangeVars}
\wt R_{\mu\nu,x} 
&=& 
\sqrt{2 \beta_\R '}
\l(
R_{\mu\nu,x} 
-f_{\mu\nu,x}^{\R\dag}
U_{\mu,x} U_{\nu,x + \hat\mu} 
-U_{\nu,x} U_{\mu,x}
\r)
\eea
the linearisation with respect to the link variables is promptly achieved:
\bea
\label{eq:mixred:Gauss:Terms}
-\frac{1}{2}~ 
\Tr\l\{ {\wt R_{\mu\nu,x}}^\dag \wt R_{\mu\nu,x} \r\}
&\stackrel{ \eqref{eq:mixred:ChangeVars} }{=}&
\nn\\\nn\\ 
\comm{Gaussian} 
&& -\beta_\R '
\Tr\l\{ R_{\mu\nu,x}^\dag R_{\mu\nu,x} \r\} 
\nn\\\nn\\ 
\comm{non-Gaussian} 
&& -\beta_\R '
\Tr\l\{ f_{\mu\nu,x}^{\R\dag} f_{\mu\nu,x}^{\R} \r\} 
\nn\\\nn\\
\comm{linear} 
&& +2\beta_\R ' \Re\Tr\l\{ R_{\mu\nu,x}^\dag \l( f_{\mu\nu,x}^{\R\dag} U_{\mu,x} U_{\nu,x + \hat\mu} + U_{\nu,x} U_{\mu,x} \r)\r\} 
\nn\\\nn\\
\comm{cancels out} 
&& -2\beta_\R ' \Re\Tr\l\{ f_{\mu\nu,x}^{\R\dag} U_{\mu,x} U_{\nu,x+\hat\mu} U_{\mu,x}^{\dag} U_{\nu,x}^{\dag}  \r\}
\eea
However, new terms of the form $\Tr\l\{ f_{\mu\nu,x}^{\R\dag} f_{\mu\nu,x}^{\R} \r\}$ appear in the linearised action. Even though they do not depend on the link to be updated, they give a non-Gaussian weight to the auxiliary variables. Therefore, the efficient algorithms discussed in Section \ref{sec:intro} cannot be applied, unless such terms are eliminated too. The elimination of the term $\Tr\l\{ f_{\mu\nu,x}^{\R\dag} f_{\mu\nu,x}^{\R} \r\}$ is not easy. In general, $f_{\mu\nu,x}^\R$ has a rather complicated dependence on the auxiliary variables, as can be seen in the examples derived in Sections \ref{sec:mix:A}--\ref{sec:mix:mixed}. The only situation for which we have an easy solution to the problem (or, at least, a solution that does not undermine the efficiency of the resulting algorithm) is the case of the adjoint representation.

For positive $\beta_A$, the solution is trivial. Given that $f_{\mu\nu,x}^A = z_{\mu\nu,x}/N$, the non-Gaussian term is in fact Gaussian: 
\bea
-\beta_A '
\Tr\l\{ f_{\mu\nu,x}^{A\dag} f_{\mu\nu,x}^{A} \r\} 
&=&
-\frac{\beta_A '}{N}
\l| z_{\mu\nu,x} \r|^2 
\eea
and so the problem is solved. The update of the link variables for a partially reduced adjoint Wilson action is performed with respect to the probability distribution of individual links \eqref{eq:F:Distrib:Single}, with $V_{\mu,x} \equiv V_{\mu,x}^{A}$ given by:
\bea
V_{\mu,x}^{A} 
&=& 
\frac{2\beta_A '}{N}\!\!\! \sum_{\nu=1\atop (\nu \neq \mu, L_\nu > 1)}^d \l(
z_{\mu\nu,x} U_{\nu,x} U_{\mu,x+\hat\nu} U_{\nu,x+\hat\mu}^\dag ~+~ 
z^{\ast}_{\mu\nu,x-\hat\nu} U_{\nu,x-\hat\nu}^\dag U_{\mu,x-\hat\nu} U_{\nu,x-\hat\nu+\hat\mu} \r)
\nn\\
&&+~ 2\beta_A '\!\!\! \sum_{\nu=1\atop (\nu \neq \mu, L_\nu = 1)}^d 
\l( \frac{1}{N} z_{\mu\nu,x} R_{\mu\nu,x} U_{\nu,x+\hat\mu}^\dag ~+~
U_{\nu,x}^\dag R_{\mu\nu,x} \r)
\eea
The first term in the r.h.s. of the equation above is the contribution from the unreduced plaquettes, and the second term is the contribution from the reduced plaquettes.

For negative $\beta_A$, $f_{\mu\nu,x}^A$ is a matrix. In order to eliminate this term, we first expand it in terms of the $Q$--variables:
\bea
\label{eq:mixred:nonGauss:Expand}
-\beta_A '
\Tr\l\{ f_{\mu\nu,x}^{A\dag} f_{\mu\nu,x}^{A} \r\} 
&\stackrel{ \eqref{eq:A:-:f} }{=}&
\nn\\\nn\\
\comm{Gaussian}
&&
-\beta_A ' \Tr\l\{ 
Q_{\mu\nu,x}^\dag Q_{\mu\nu,x} 
\r\} 
\nn\\\nn\\
\comm{nonlinear}
&&
+\frac{\beta_A '}{N} \l|\Tr\l\{ Q_{\mu\nu,x} \r\}\r|^2
\eea
This term is clearly non-Gaussian. However, it is the sum of a Gaussian term and a double-trace term that can easily be eliminated. For that end, we introduce yet another auxiliary variable $\wt M_{\mu\nu,x}$ with the normal distribution $\mu_1(\wt M_{\mu\nu,x})$, associated with positively-oriented reduced plaquettes (for negatively-oriented plaquettes we define $\wt M_{\nu\mu,x} \equiv \wt M_{\mu\nu,x}^\dag$). Let us perform the change of variables:
\bea
\label{eq:mixred:ChangeVars:M}
\wt M_{\mu\nu,x} 
&=&
\sqrt{2\beta_A '}
\l( M_{\mu\nu,x}  - \frac{1}{N} \Tr\l\{ Q_{\mu\nu,x} \r\} U_{\mu,x+\hat\nu} U_{\nu,x+\hat\mu}^\dag \r)
\eea
The effect of \eqref{eq:mixred:ChangeVars:M} on the Gaussian exponent in $\mu_1(\wt M_{\mu\nu,x})$ generates the following terms:

\bea
-\frac{1}{2} 
\Tr\l\{ 
\wt M_{\mu\nu,x}^{\dag} 
\wt M_{\mu\nu,x} 
\r\} 
&\stackrel{\eqref{eq:mixred:ChangeVars:M}}{=}&
\nn\\\nn\\
\comm{Gaussian}
&&
-\beta_A '
\Tr\l\{ 
M_{\mu\nu,x}^{\dag} 
M_{\mu\nu,x} 
\r\}
\nn\\\nn\\
\comm{linear}
&&
+\frac{2\beta_A '}{N}
\Re\Tr\l\{ 
\Tr\l\{Q_{\mu\nu,x}\r\} M_{\mu\nu,x}^\dag 
U_{\mu,x+\hat\nu} U_{\nu,x+\hat\mu}^\dag
\r\} 
\nn\\\nn\\
\comm{cancels out}
&&
-\frac{\beta_A '}{N}
\l|\Tr\l\{ Q_{\mu\nu,x} \r\}\r|^2
\eea
the last of which cancels out the nonlinear term coming from the non-Gaussian piece, thus solving the problem for negative $\beta_A$. The sum of `staples' for the probability distribution of individual links \eqref{eq:R:Staples} is then given by:
\bea
V_{\mu,x}^{A} 
&=& 
2\beta_A ' \!\!\! \sum_{\nu=1\atop (\nu \neq \mu, L_\nu > 1)}^d 
\l(
f^{A}_{\mu\nu,x} U_{\nu,x} U_{\mu,x+\hat\nu} U_{\nu,x+\hat\mu}^\dag 
~+~ 
U_{\nu,x-\hat\nu}^\dag f^{A\dag}_{\mu\nu,x-\hat\nu} U_{\mu,x-\hat\nu} U_{\nu,x-\hat\nu+\hat\mu}
\r)
\nn\\
&&+~ 2\beta_A ' \!\!\! \sum_{\nu=1\atop (\nu \neq \mu, L_\nu = 1)}^d 
\l( 
f_{\mu\nu,x}^{A} R_{\mu\nu,x} U_{\nu,x+\hat\mu}^\dag 
~+~
U_{\nu,x}^\dag R_{\mu\nu,x} 
~+~
\frac{1}{N}\Tr\l\{Q_{\mu\nu,x}^\dag\r\} M_{\mu\nu,x} U_{\nu,x+\hat\mu}
\r)~~~
\eea
where $f_{\mu\nu,x}^{A}$ is defined as in \eqref{eq:A:-:f}.

For higher representations, the analogous of the expansion \eqref{eq:mixred:nonGauss:Expand} of the non-Gaussian term results in a higher number of complicated nonlinear terms. In order to eliminate them, many more auxiliary terms would have to be introduced. This would certainly render the resulting algorithms inefficient and useless. Although unlikely, it is not a priori impossible that a clever choice of a smaller set of auxiliary variables would result in viable updating algorithms for higher-representation reduced models.

\subsection
{Numerical tests
\label{sec:mix:numerical}}

We simulated some of the new algorithms proposed above, with the purpose of comparing them with the Metropolis algorithm. We considered mixed `$F + \R$' Wilson actions in $d=4$, i.e. the Wilson action \eqref{eq:R:Action} with plaquettes in the fundamental representation, $F$, and in another representation $\R \neq F$ of $SU(N)$. We simulated each theory with both a Metropolis algorithm and the new MC algorithm proposed in Section \ref{sec:mix:mixed}. We tested both algorithms for their compatibility, by checking if the expectation values of gauge-invariant observables coincide in both cases. We also tested them for their relative efficiency, by comparing the magnitude of the autocorrelations in the Markov chains they generate. 

For the thermal Metropolis updates we used a variant of the (1-hit) Cabibbo-Marinari-Metropolis (CMM) algorithm described in \cite{NeccoHasenbusch}, and appropriately adapted to an arbitrary representation $\R$. In our CMM algorithm, new link proposals are generated via Cabibbo-Marinari updates with respect to the fundamental part of the action only, in which $\beta_F$ is replaced with the tuning parameter of the Metropolis algorithm $\beta_{\rm M}$ ($\beta_{\rm M} < \beta_F$). The link proposal is then accepted or rejected a la Metropolis with respect to the full `$F + \R$' action. The acceptance rates are tuned to stay in the range 40-60\%. For the overrelaxation updates, we adapted the overrelaxation-Metropolis algorithm also described in \cite{NeccoHasenbusch}. In this algorithm, $SU(2)$-- or $SU(N)$--overrelaxed link variables are accepted or rejected with respect to the $S_\R$ part of the action only. Acceptance rates for the overrelaxation-Metropolis algorithm stayed well above 85\%.

For both the Metropolis and the new algorithms, each configuration update consisted of one thermal update followed by 5 overrelaxation updates. For each configuration, we evaluated the characters $\chi_F$ and $\chi_\R$ of the plaquette, in order to estimate their expectation values. In each simulation we performed $O(10^5)$ measurements, after discarding the initial 2,000 configurations for equilibration. The simulation parameters and measured observables, together with their naive confidence intervals, are shown in Table \ref{tab:Plaquette.Mix}.

The parameters of the simulations of the `$F + A$' action were chosen to coincide with those of \cite{NeccoHasenbusch}. In this way, we could compare our results with the literature. We considered two particular values for the adjoint lattice coupling, one positive and one negative, that were also considered in that article. We obtained compatible results for both plaquette characters, which is good evidence that our new algorithm reproduces accurately the `$F + A$' lattice gauge theory. For higher representations, we performed simulations of the `$F + \R$' theories on a $8^4$ lattice. In each of these cases, the observables calculated for a particular value of the lattice coupling also matched in both algorithms. There is, however, some discrepancies in the last significant digits of each observable. This is probably due to the fact that the confidence intervals shown in Table \ref{tab:Plaquette.Mix} do not take autocorrelations into account, i.e. they are underestimated. 

\begin{figure}[p]
\begin{center}
\leavevmode
\resizebox{11.5cm}{!}{\includegraphics
{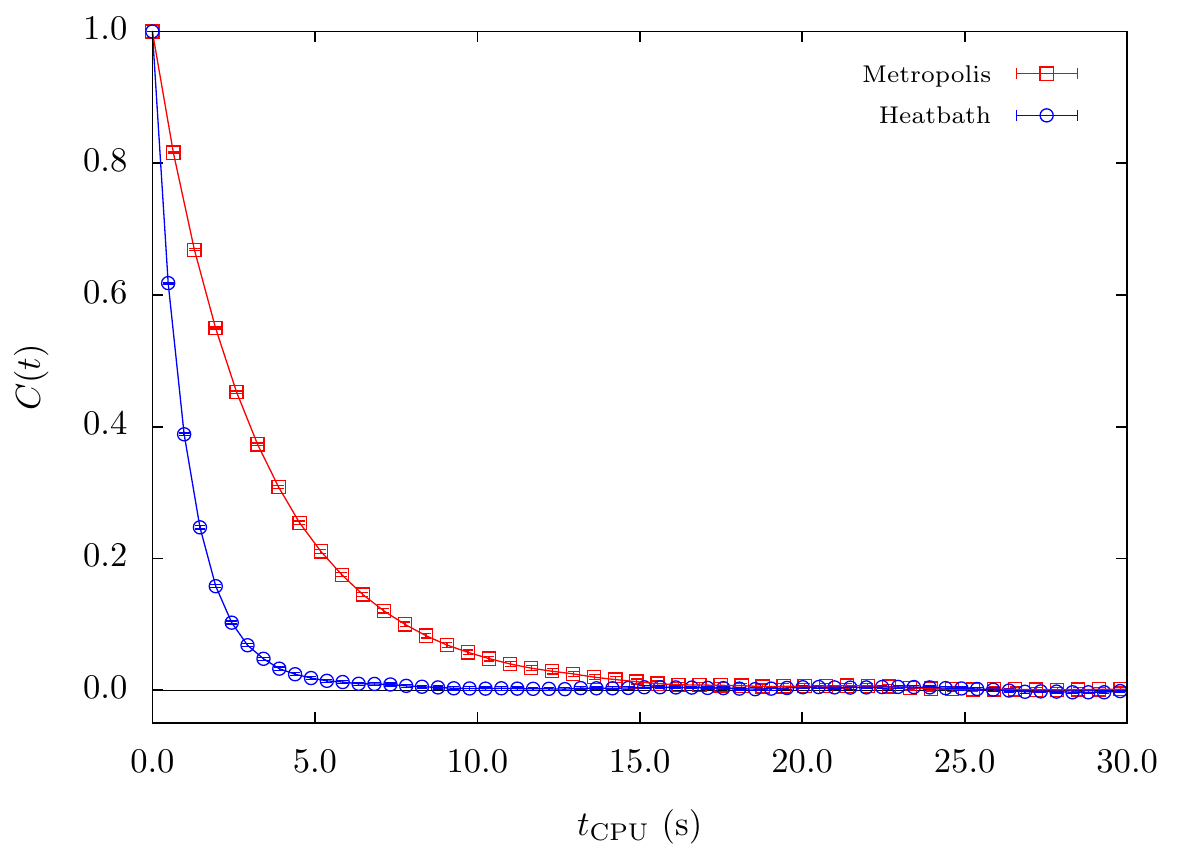}
}
\end{center}
\vskip -4mm
\caption{Estimator of the autocorrelation function, $C(t)$, vs. the CPU time, $t_{\rm CPU}$, for the $SU(3)$ `$F + A$' mixed Wilson action, simulated on a $12^3 8$ lattice with $(\beta_F, \beta_A) = (4.00,2.00)$, via a Cabibbo-Marinari-Metropolis algorithm (squares) and a Fabricius-Haan-type pseudo-heatbath algorithm (circles).}
\label{fig:A+:C}
\end{figure}

\begin{figure}[p]
\begin{center}
\leavevmode
\resizebox{11.5cm}{!}{\includegraphics
{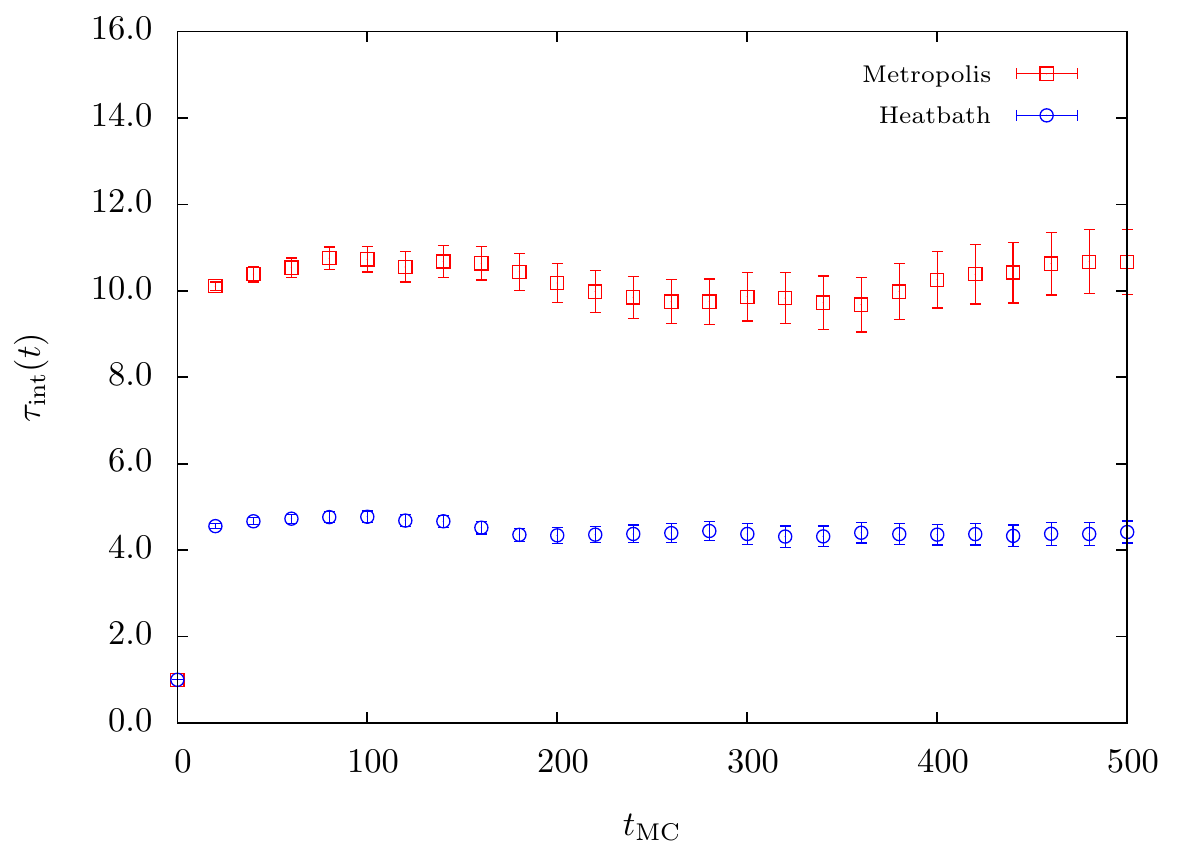}
}
\end{center}
\vskip -4mm
\caption{Estimator of the $t$-dependent integrated autocorrelation time, $\tau_{\rm int}(t)$, vs. the MC time, $t_{\rm MC}$, for the $SU(3)$ `$F + A$' mixed Wilson action, simulated on a $12^3 8$ lattice with $(\beta_F, \beta_A) = (4.00,2.00)$, via a Cabibbo-Marinari-Metropolis algorithm (squares) and a Fabricius-Haan-type pseudo-heatbath algorithm (circles).}
\label{fig:A+:tau}
\end{figure}

\begin{figure}[p]
\begin{center}
\leavevmode
\resizebox{11.5cm}{!}{\includegraphics
{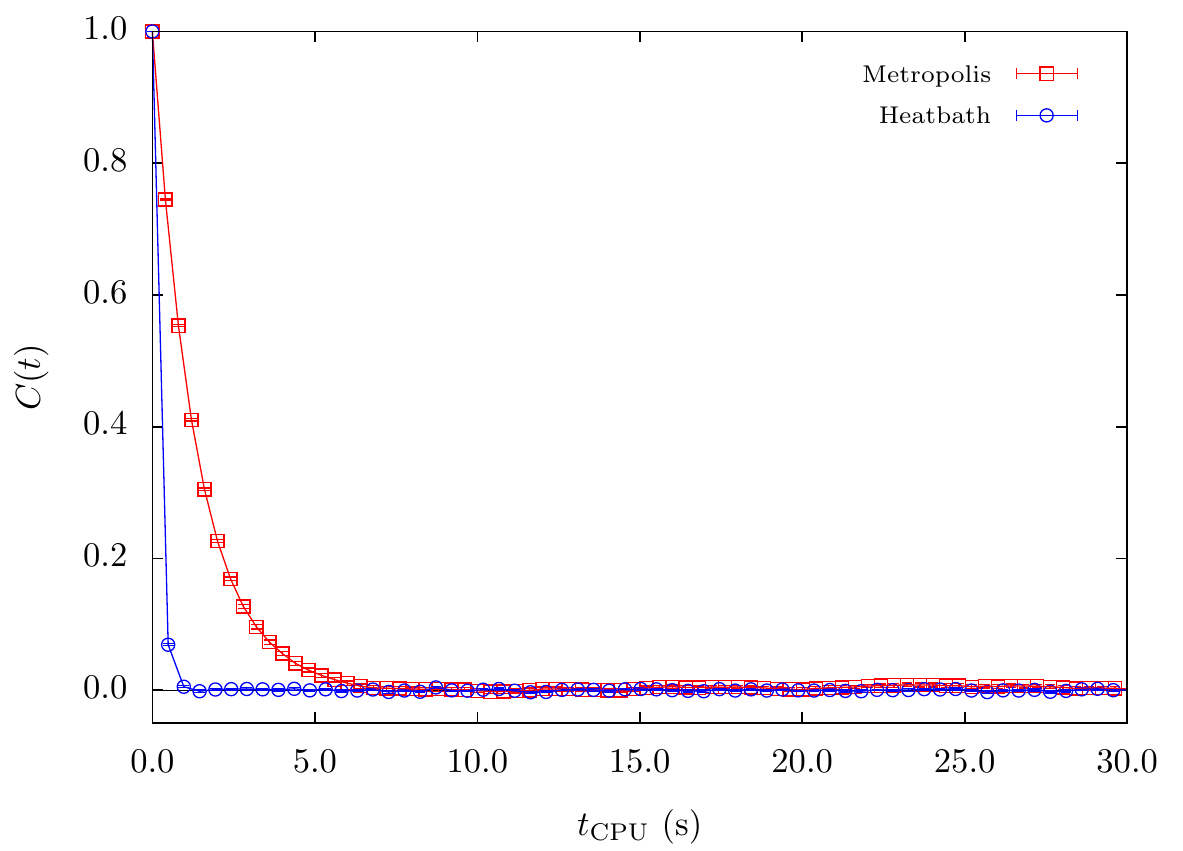}
}
\end{center}
\vskip -4mm
\caption{Estimator of the autocorrelation function, $C(t)$, vs. the CPU time, $t_{\rm CPU}$, for the $SU(4)$ `$F + (2)$' mixed Wilson action, simulated on a $8^4$ lattice with $(\beta_F, \beta_{(2)}) = (10.665,0.3556)$, via a Cabibbo-Marinari-Metropolis algorithm (squares) and a Fabricius-Haan-type pseudo-heatbath algorithm (circles).}
\label{fig:2:C}
\end{figure}

\begin{figure}[p]
\begin{center}
\leavevmode
\resizebox{11.5cm}{!}{\includegraphics
{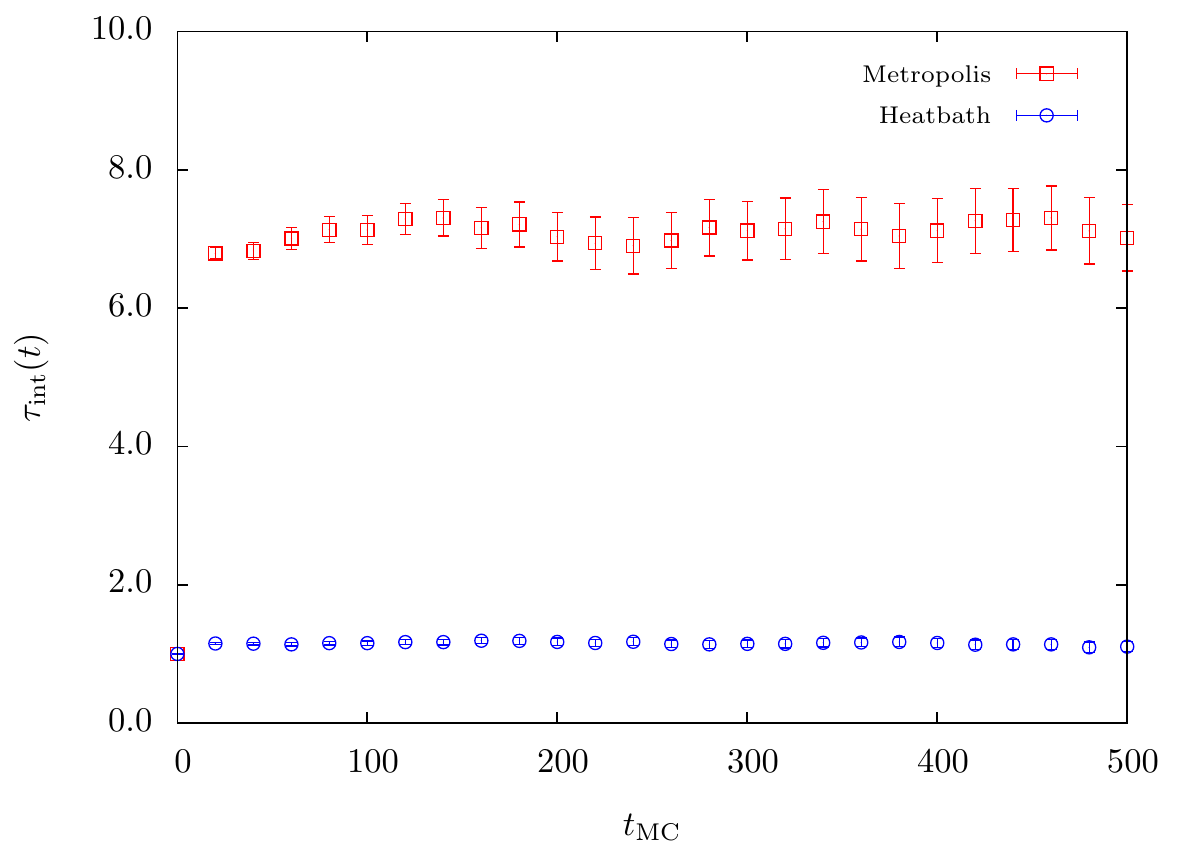}
}
\end{center}
\vskip -4mm
\caption{Estimator of the $t$-dependent integrated autocorrelation time, $\tau_{\rm int}(t)$, vs. the MC time, $t_{\rm MC}$, for the $SU(4)$ `$F + (2)$' mixed Wilson action, simulated on a $8^4$ lattice with $(\beta_F, \beta_{(2)}) = (10.665,0.3556)$, via a Cabibbo-Marinari-Metropolis algorithm (squares) and a Fabricius-Haan-type pseudo-heatbath algorithm (circles).}
\label{fig:2:tau}
\end{figure}

\begin{figure}[p]
\begin{center}
\leavevmode
\resizebox{11.5cm}{!}{\includegraphics
{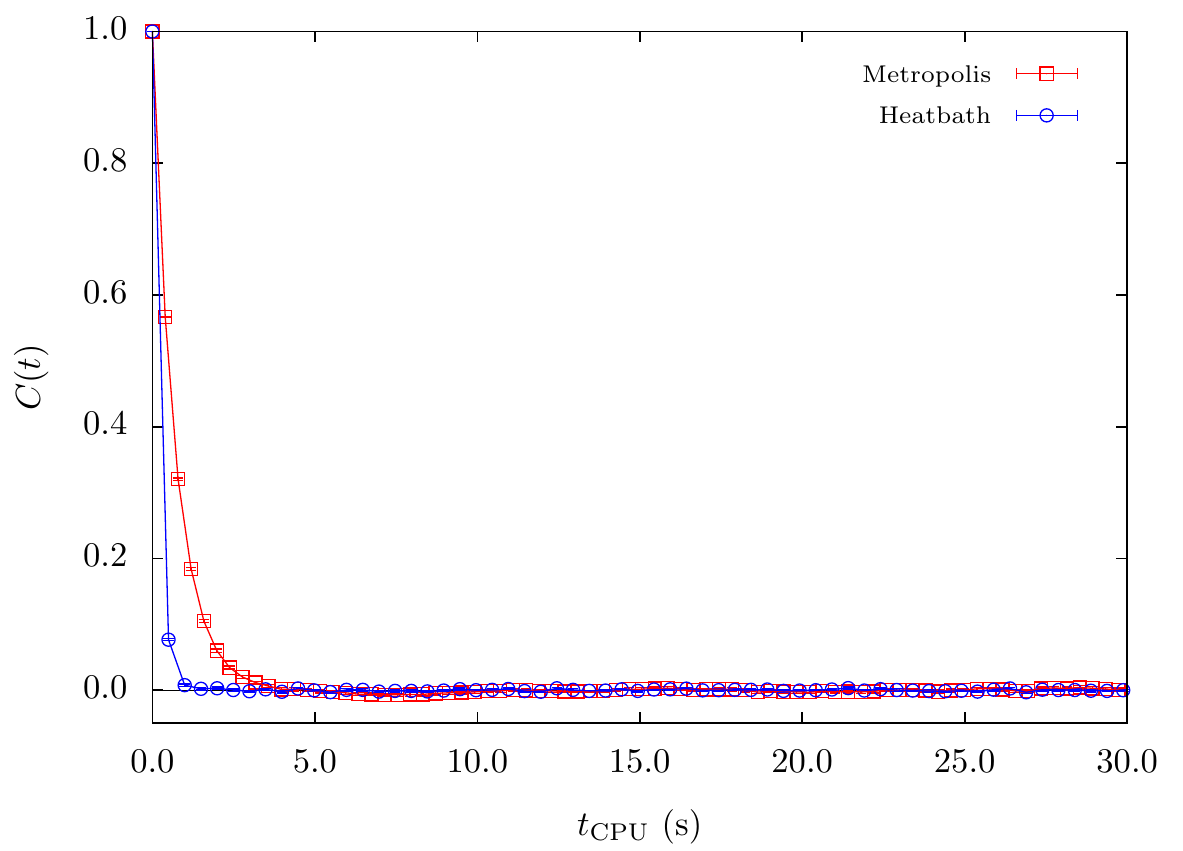}
}
\end{center}
\vskip -4mm
\caption{Estimator of the autocorrelation function, $C(t)$, vs. the CPU time, $t_{\rm CPU}$, for the $SU(4)$ `$F + (1,1)$' mixed Wilson action, simulated on a $8^4$ lattice with $(\beta_F, \beta_{(1,1)}) = (10.665,1.0668)$, via a Cabibbo-Marinari-Metropolis algorithm (squares) and a Fabricius-Haan-type pseudo-heatbath algorithm (circles).}
\label{fig:11:C}
\end{figure}

\begin{figure}[p]
\begin{center}
\leavevmode
\resizebox{11.5cm}{!}{\includegraphics
{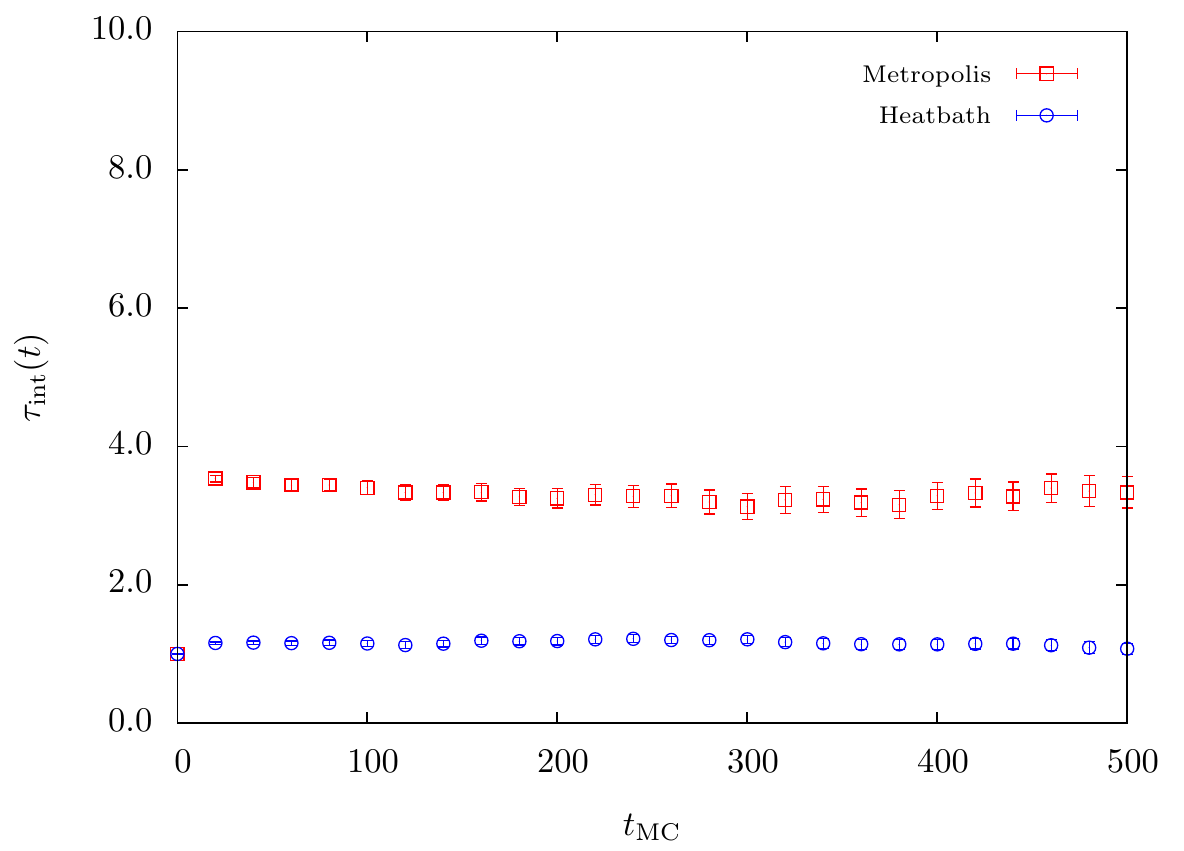}
}
\end{center}
\vskip -4mm
\caption{Estimator of the $t$-dependent integrated autocorrelation time, $\tau_{\rm int}(t)$, vs. the MC time, $t_{\rm MC}$, for the $SU(4)$ `$F + (1,1)$' mixed Wilson action, simulated on a $8^4$ lattice with $(\beta_F, \beta_{(1,1)}) = (10.665,1.0668)$, via a Cabibbo-Marinari-Metropolis algorithm (squares) and a Fabricius-Haan-type pseudo-heatbath algorithm (circles).}
\label{fig:11:tau}
\end{figure}

\begin{figure}[p]
\begin{center}
\leavevmode
\resizebox{11.5cm}{!}{\includegraphics
{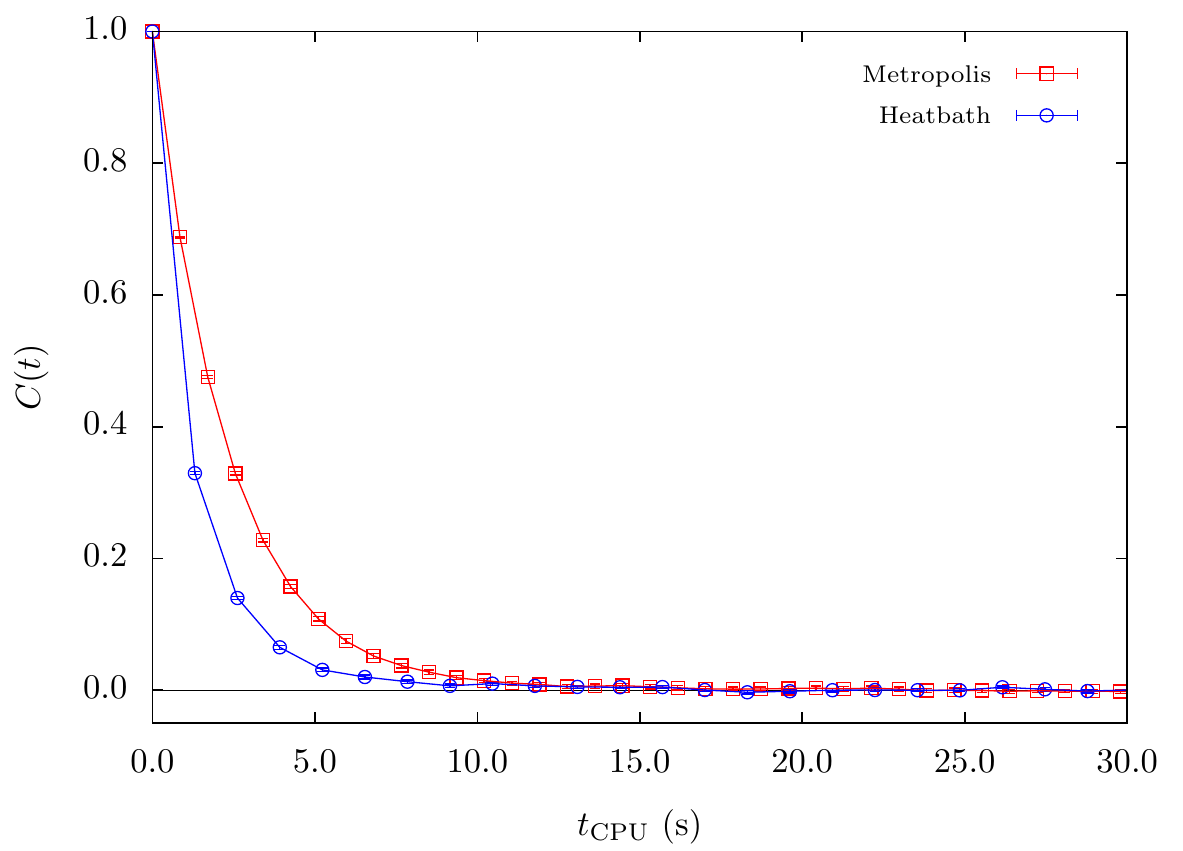}
}
\end{center}
\vskip -4mm
\caption{Estimator of the autocorrelation function, $C(t)$, vs. the CPU time, $t_{\rm CPU}$, for the $SU(5)$ `$F + (3)$' mixed Wilson action, simulated on a $8^4$ lattice with $(\beta_F, \beta_{(3)}) = (16.665,4.1487)$, via a Cabibbo-Marinari-Metropolis algorithm (squares) and a Fabricius-Haan-type pseudo-heatbath algorithm (circles).}
\label{fig:3:C}
\end{figure}

\begin{figure}[p]
\begin{center}
\leavevmode
\resizebox{11.5cm}{!}{\includegraphics
{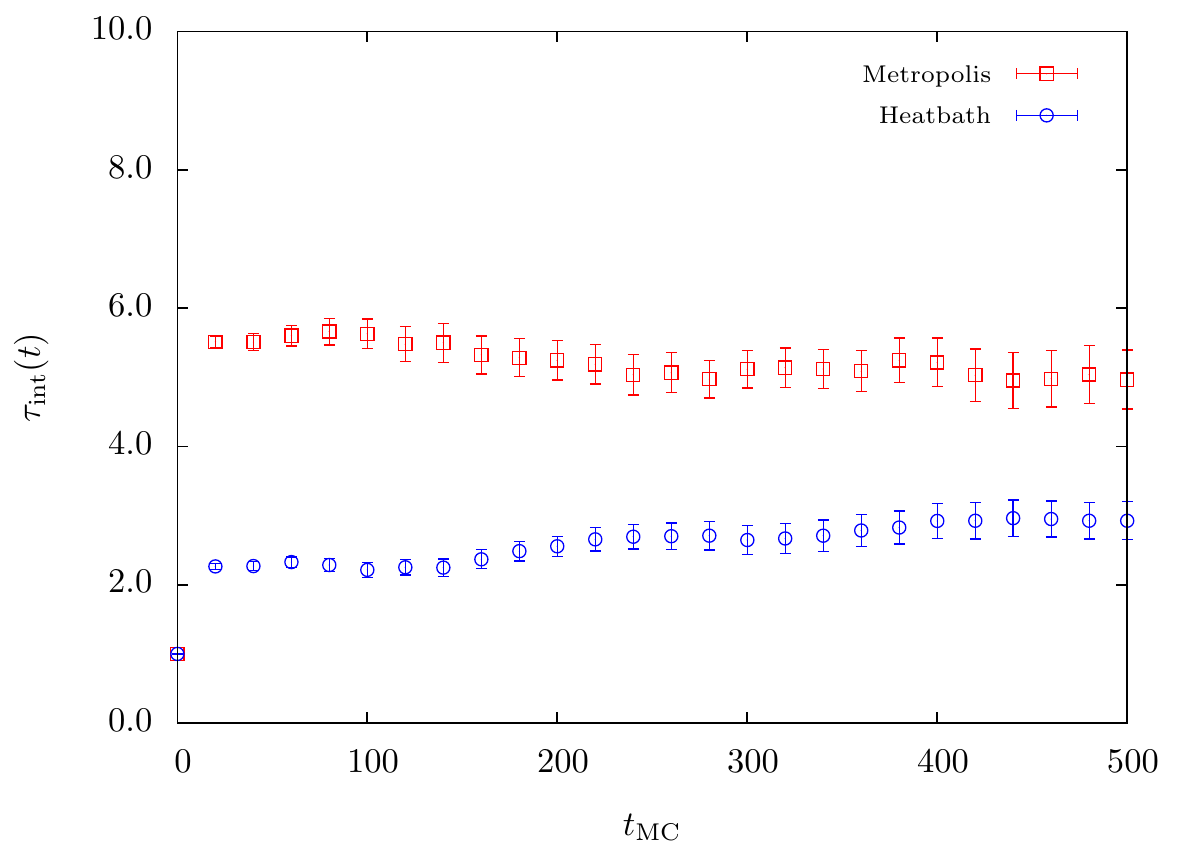}
}
\end{center}
\vskip -4mm
\caption{Estimator of the $t$-dependent integrated autocorrelation time, $\tau_{\rm int}(t)$, vs. the MC time, $t_{\rm MC}$, for the $SU(5)$ `$F + (3)$' mixed Wilson action, simulated on a $8^4$ lattice with $(\beta_F, \beta_{(3)}) = (16.665,4.1487)$, via a Cabibbo-Marinari-Metropolis algorithm (squares) and a Fabricius-Haan-type pseudo-heatbath algorithm (circles).}
\label{fig:3:tau}
\end{figure}

\begin{figure}[p]
\begin{center}
\leavevmode
\resizebox{11.5cm}{!}{\includegraphics
{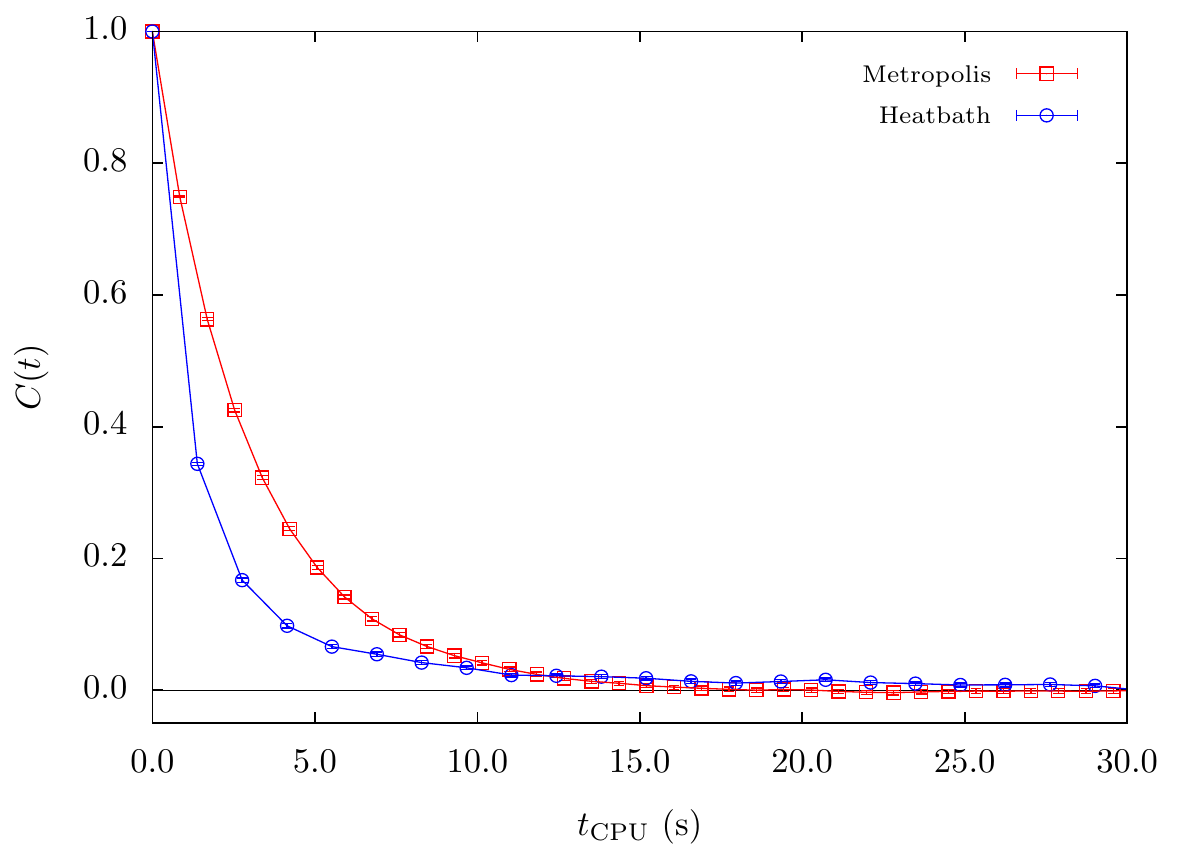}
}
\end{center}
\vskip -4mm
\caption{Estimator of the autocorrelation function, $C(t)$, vs. the CPU time, $t_{\rm CPU}$, for the $SU(5)$ `$F + (1,1,1)$' mixed Wilson action, simulated on a $8^4$ lattice with $(\beta_F, \beta_{(1,1,1)}) = (16.665,1.1853)$, via a Cabibbo-Marinari-Metropolis algorithm (squares) and a Fabricius-Haan-type pseudo-heatbath algorithm (circles).}
\label{fig:111:C}
\end{figure}

\begin{figure}[p]
\begin{center}
\leavevmode
\resizebox{11.5cm}{!}{\includegraphics
{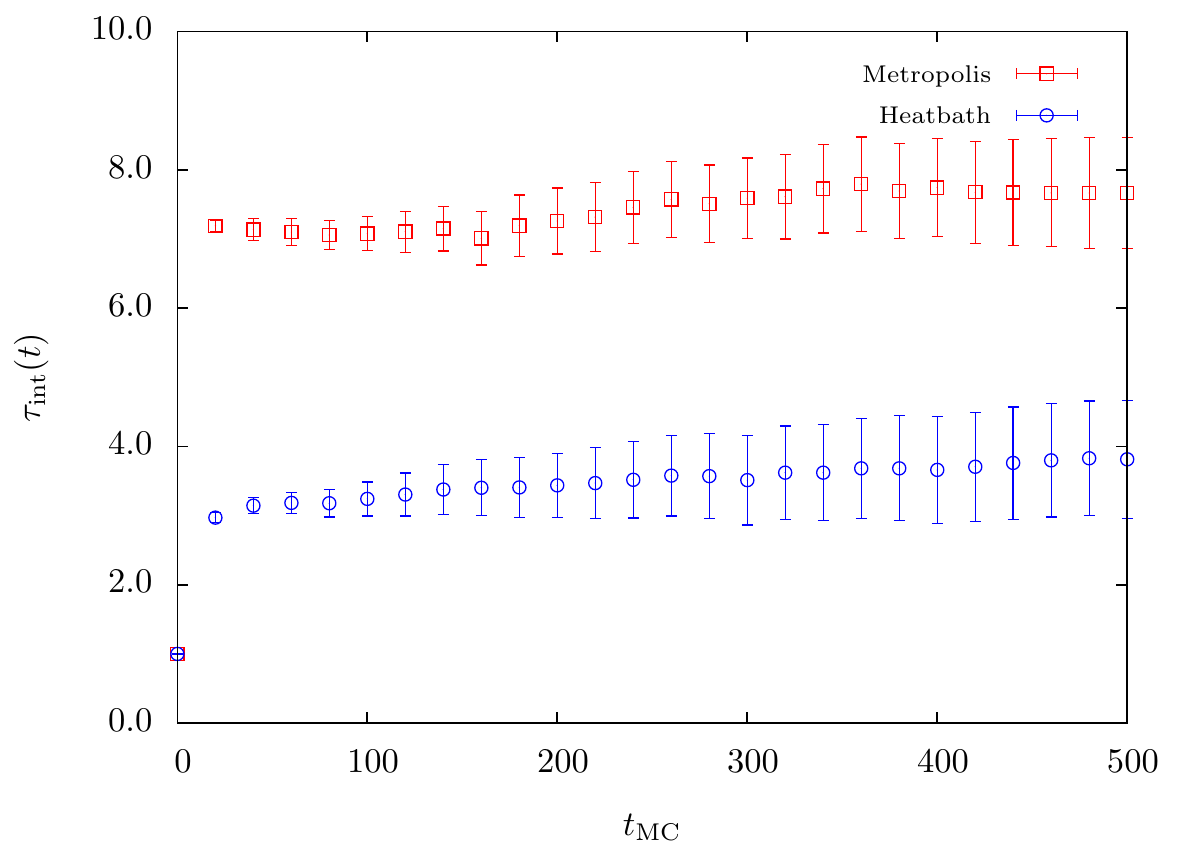}
}
\end{center}
\vskip -4mm
\caption{Estimator of the $t$-dependent integrated autocorrelation time, $\tau_{\rm int}(t)$, vs. the MC time, $t_{\rm MC}$, for the $SU(5)$ `$F + (1,1,1)$' mixed Wilson action, simulated on a $8^4$ lattice with $(\beta_F, \beta_{(1,1,1)}) = (16.665,1.1853)$, via a Cabibbo-Marinari-Metropolis algorithm (squares) and a Fabricius-Haan-type pseudo-heatbath algorithm (circles).}
\label{fig:111:tau}
\end{figure}

\begin{figure}[p]
\begin{center}
\leavevmode
\resizebox{11.5cm}{!}{\includegraphics
{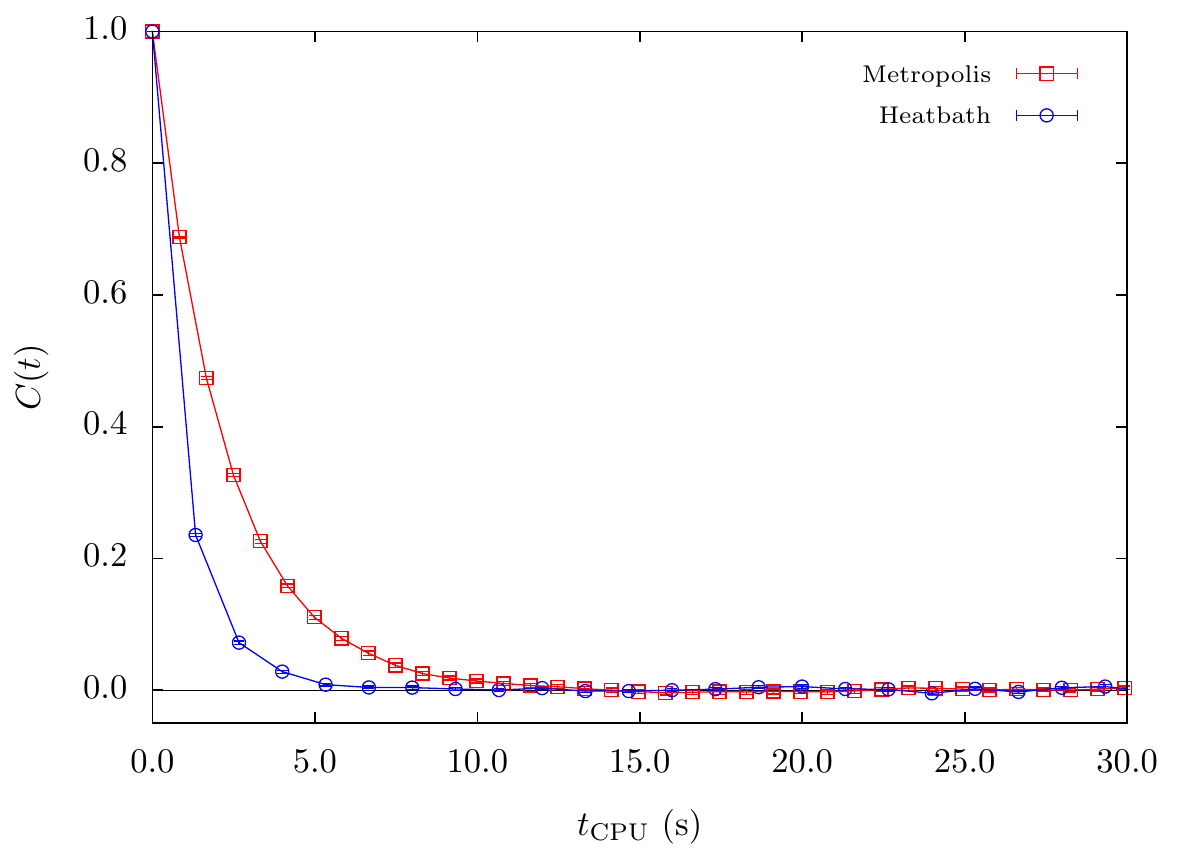}
}
\end{center}
\vskip -4mm
\caption{Estimator of the autocorrelation function, $C(t)$, vs. the CPU time, $t_{\rm CPU}$, for the $SU(5)$ `$F + (2,1)$' mixed Wilson action, simulated on a $8^4$ lattice with $(\beta_F, \beta_{(2,1)}) = (16.665,4.7413)$, via a Cabibbo-Marinari-Metropolis algorithm (squares) and a Fabricius-Haan-type pseudo-heatbath algorithm (circles).}
\label{fig:21:C}
\end{figure}

\begin{figure}[p]
\begin{center}
\leavevmode
\resizebox{11.5cm}{!}{\includegraphics
{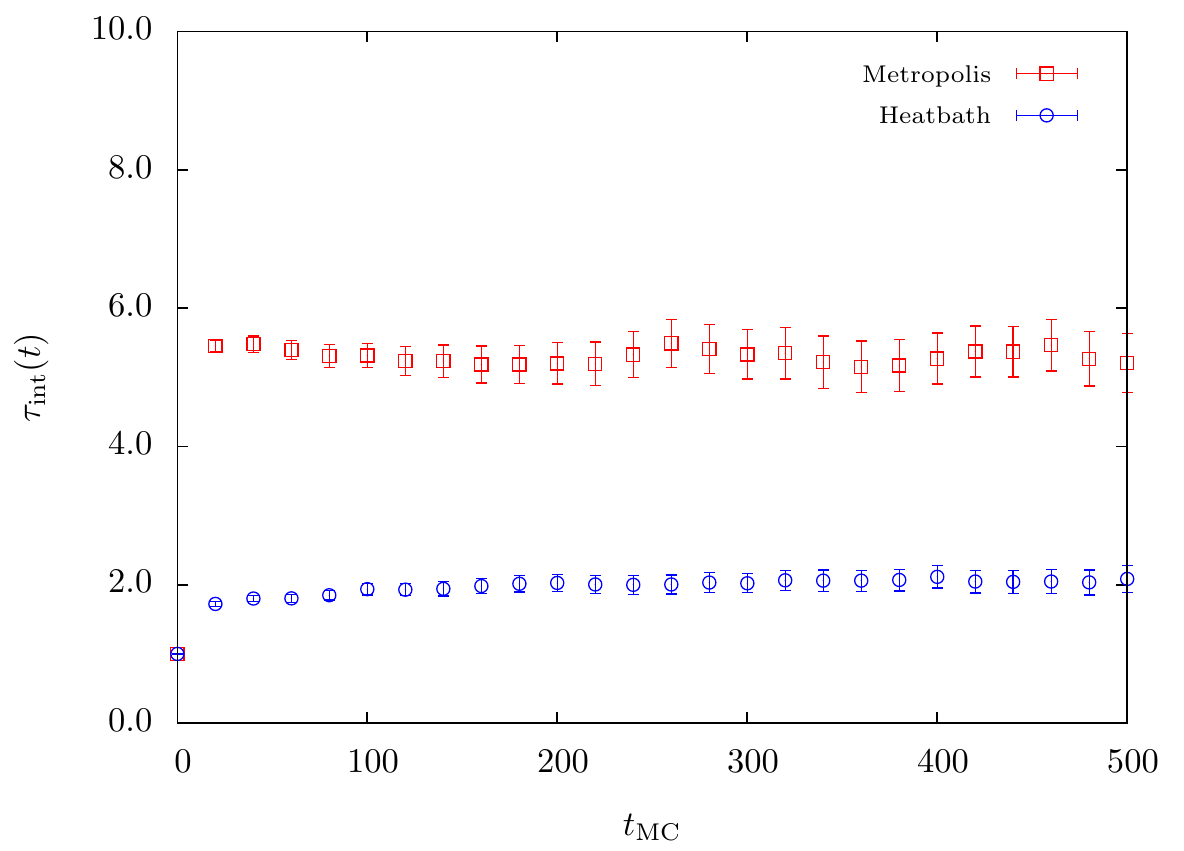}
}
\end{center}
\vskip -4mm
\caption{Estimator of the $t$-dependent integrated autocorrelation time, $\tau_{\rm int}(t)$, vs. the MC time, $t_{\rm MC}$, for the $SU(5)$ `$F + (2,1)$' mixed Wilson action, simulated on a $8^4$ lattice with $(\beta_F, \beta_{(2,1)}) = (16.665,4.7413)$, via a Cabibbo-Marinari-Metropolis algorithm (squares) and a Fabricius-Haan-type pseudo-heatbath algorithm (circles).}
\label{fig:21:tau}
\end{figure}

We also measured the autocorrelations of the Markov chains generated by both types of algorithms. We calculated the values of the fundamental trace of the plaquette $u_i$ as a function of the MC time $i$, and then used them to estimate the normalised autocorrelation function:
\bea
\label{eq:C(t)}
C(t)
&=&
\frac{
\langle u_{i + t} u_i \rangle - \langle u_i \rangle^2
}{
\langle u_i^2 \rangle - \langle u_i \rangle^2
}
\eea
From $C(t)$ we estimated the time-dependent integrated autocorrelation time:

\bea
\label{eq:tau_int(t)}
\tau_{\rm int}(t) 
&=&
1 + 2 \sum_{i=1}^t C(i)
\eea
The estimator of the integrated autocorrelation time, $\bar\tau_{\rm int}$, is the plateau value of $\tau_{\rm int}(t)$ when $t \to \infty$. In terms of these autocorrelations, the new pseudo-heatbath algorithms performs significantly better than their Metropolis counterparts, as can be seen in  Figs.\ref{fig:A+:C}--\ref{fig:21:tau}. The graphs show that the new algorithms decorrelate faster than Metropolis, for all representations, both in MC time and in effective CPU time.

We also observed that, except for the case of the adjoint representation, one pseudo-heatbath update takes longer to perform (in CPU time) than one Metropolis hit. This is not surprising, taking into account the large number of auxiliary variables that some representations require. Only for the adjoint case, which requires only one auxiliary variable per plaquette, were the pseudo-heatbath updates faster than Metropolis. In particular, this difference was largest for $\beta_A > 0$, as expected. Nevertheless, the fast decorrelation of the new algorithms overshadows the disadvantage of slower updates, making them a superior alternative to Metropolis.

\section
{Double-trace deformations
\label{sec:cym}}

Recently, \"Unsal and Yaffe suggested \cite{UnsalYaffe} a strategy to prevent the spontaneous breaking of the centre symmetry in pure $SU(N)$ Yang-Mills theories on manifolds with small compactified directions, namely $\mathds{R}^{d-k} \times (S^1)^k$. The strategy consists in deforming the Yang-Mills action with a centre-stabilising potential, which contains double-trace operators dependent on the holonomies wrapping the compact directions. This mimics the effective potential contribution of adjoint fermions with periodic boundary conditions in the compact direction, which are known to stabilise the centre symmetry \cite{KovtunUnsalYaffe}. The gauge theories deformed by such double-trace terms are called centre-stabilised Yang-Mills (CYM) theories. 

The breaking of the centre symmetry is the reason why the EK model (and most of its variants) is not equivalent to pure Yang-Mills theories in the large $N$ limit. CYM theories, however, are believed to be completely volume-independent in the large $N$ limit. If this is true, then it would be possible to represent large $N$ Yang-Mills theory defined on $\mathds{R}^d$ by the zero volume limit of the compact directions of CYM theories. Taking the volume of the compactified directions to zero with no consequences for the large $N$ equivalence has clear analytical and numerical advantages.

In this section we consider the lattice regularisation of CYM theories defined on manifolds with a single compactified direction, $\mathds{R}^{d-1} \times S^1$. Due to the double-trace terms, the CYM lattice action is highly nonlinear with respect to the link variables, hence Metropolis seem to be the only possible algorithm to simulate such theories. However, we will show that it is possible to linearise the CYM lattice action using the method of auxiliary variables, and from there obtain a rather efficient MC algorithms.

\subsection{\texorpdfstring
{CYM theory on $\mathds{R}^{d-1} \times S^1$}
{CYM theory on R(d-1) x S(1)}
\label{sec:cym:single}}

The lattice action for the $SU(N)$ CYM theory defined on a manifold with only one compact direction, namely $\mathds{R}^{d-1} \times S^1$, is given by:
\bea
\label{eq:cym:Action}
S_\CYM\l( \vec\alpha, \beta_F; [U] \r) 
&=&
S_F\l(\beta_F; [U] \r) + S_{\rm def} \l( \vec\alpha; [U] \r)
\eea
where $S_F$ is the fundamental Wilson action \eqref{eq:F:Action} and $ \vec\alpha \equiv (\alpha_1, \ldots, \alpha_\Nhalf )$ are free parameters of the CYM model that must be positive. The deformation potential $S_{\rm def}$ is a sum of double-traces:

\bea
\label{eq:cym:Action(def)}
S_\mathrm{def}\l( \vec\alpha; [U] \r) 
&=& 
\frac{1}{N_t^{d-1}} \sum_{\vecx\in \Lambda_\perp} \sum_{n=1}^\Nhalf \alpha_n \l|\Tr\l\{\Omega_{\vecx}^n\r\}\r|^2 
\eea
where $N_t$ is the size of the compact direction ($S^1$) in lattice units, $\vecx$ labels the lattice sites on the $(d-1)$--dimensional lattice $\Lambda_\perp$ (the discretization of $\mathds{R}^{d-1}$), and $\Omega_{\vecx}$ is the holonomy (Polyakov loop) wrapping the compact direction, $\wh d$:
\bea
\label{eq:cym:Polyakov}
\Omega_{\vecx} 
&=& 
{\cal P} \prod_{i=0}^{N_t-1} U_{d, \vecx + i \wh d} 
~=~ 
U_{d, \vecx} U_{d, \vecx + \wh d} \cdots U_{d, \vecx + (N_t  - 1) \wh d}
\eea

The CYM action \eqref{eq:cym:Action(def)} is highly nonlinear with respect to the link variables, especially for large $N$ (because it contains terms that depend up to the $\lfloor N/2 \rfloor$--th power of the link variable). In addition, $S_{\rm def}$ contains $O(N^1)$ terms, which makes the theory even harder to simulate in that limit. From these facts, we should expect to use a relatively large number of auxiliary variables in order to linearise the whole action. Let us consider the following set: 
\bea
\wt R_{n,\vecx}    ~, && 1 \leq n \leq K
\nn\\
\wt Q^{(m)}_{n,\vecx} ~, && 1 \leq m < n \leq K
\nn\\
\wt Q_{n,\vecx}^{(K)} ~, && 2 \leq n \leq K
\nn
\eea
where $K = \Nhalf$. We then multiply the partition function of the CYM theory by the Gaussian integrals $\int\mu_1[\wt R,\wt Q] = 1$. We now show that the following change of variables linearises the CYM action:
\bea
\label{eq:cym:ChangeVars:R}
\wt R_{n,\vecx} &=&
\l(\frac{2N\alpha_n}{N_t^{d-1}}\r)^{\frac{1}{2}} 
\l( R_{n,\vecx} - \l( \Omega_{\vecx}^n - \frac{1}{N}\Tr\{\Omega_{\vecx}^n\}\1 \r) \r)
\\\nn\\
\label{eq:cym:ChangeVars:Q}
\wt Q^{(m)}_{n,\vecx} &=&
\l(\frac{2N\alpha_n}{N_t^{d-1}}\r)^{\frac{1}{2}} 
\l( Q^{(m)}_{n,\vecx} - \l( Q_{n,\vecx}^{(m-1)} \Omega_{\vecx}^\dag + \Omega_{\vecx}^{n-m} \r) \r)
\\\nn\\
\label{eq:cym:ChangeVars:Qlast}
\wt Q_{n,\vecx}^{(K)} &=&
\l(\frac{2N\alpha_n}{N_t^{d-1}}\r)^{\frac{1}{2}} 
\l( Q_{n,\vecx}^{(K)} - \frac{1}{N} \Tr\{ R_{n,\vecx} \} \Omega_{\vecx}^\dag \r)
\eea
where we define $Q^{(0)}_{n,\vecx}$ to be the traceless part of $R_{n,\vecx}$:
\bea
\label{eq:cym:ChangeVars:Q0}
Q^{(0)}_{n,\vecx} 
&\equiv&
R_{n,\vecx} - \frac{1}{N}\Tr\{ R_{n,\vecx} \}\1
\eea
The $\wt R$--variables cancel out the double-trace terms from the CYM action. We can see this when we apply the change of variables \eqref{eq:cym:ChangeVars:R} on the exponent of the Gaussian measure $\mu_1[\wt R]$:

\bea
\label{eq:cym:Gauss:Terms:R}
-\frac{1}{2} \sum_{n=1}^K 
\Tr\l\{\wt R_{n,\vecx}^\dag \wt R_{n,\vecx} \r\} 
&\stackrel{\eqref{eq:cym:ChangeVars:R}}{=}& 
\nn\\
\comm{Gaussian}
&&
-\frac{N}{N_t^{d-1}} \sum_{n=1}^K \alpha_n \Tr\l\{ R_{n,\vecx}^\dag R_{n,\vecx} \r\}
\nn\\
\comm{linear}
&&
+\frac{2N}{N_t^{d-1}} ~\alpha_1 \Re\Tr\l\{ Q_{1,\vecx}^{(0)\dag} \Omega_{\vecx} \r\}
\nn\\
\comm{nonlinear}
&&
+ \frac{2N}{N_t^{d-1}} \sum_{n=2}^K \alpha_n \Re\Tr\l\{ Q_{n,\vecx}^{(0)\dag} \Omega_{\vecx}^n \r\}
\nn\\
\comm{cancels out}
&&
+ \frac{1}{N_t^{d-1}} \sum_{n=1}^K \alpha_n \l|\Tr\l\{ \Omega_{\vecx}^n \r\}\r|^2
\eea
The nonlinear terms in the r.h.s. of the expression above are cancelled by the $\wt Q$--variables:
\bea
\label{eq:cym:Gauss:Terms:Q}
-\frac{1}{2} 
\sum_{n=2}^K
\sum_{m=1}^{n-1} 
\Tr\l\{ 
\wt Q_{n,\vecx}^{(m)\dag} 
\wt Q_{n,\vecx}^{(m)} 
\r\} 
&\stackrel{\eqref{eq:cym:ChangeVars:Q}}{=}&
\nn\\
\comm{Gaussian}
&&
-\frac{2N}{N_t^{d-1}} 
\sum_{n=2}^K \sum_{m=1}^{n-1} 
\alpha_n \Tr\l\{ 
Q_{n,\vecx}^{(m)\dag} Q_{n,\vecx}^{(m)} 
\r\}
\nn\\
\comm{Gaussian}
&&
-\frac{N}{N_t^{d-1}} 
\sum_{n=2}^K
\alpha_n \Tr\l\{ 
Q_{n,\vecx}^{(n-1)\dag} Q_{n,\vecx}^{(n-1)} 
\r\}
\nn\\
\comm{non-Gaussian}
&&
-\frac{N}{N_t^{d-1}} 
\sum_{n=2}^K
\alpha_n \Tr\l\{ 
Q_{n,\vecx}^{(0)\dag} Q_{n,\vecx}^{(0)} 
\r\} 
\nn\\
\comm{linear}
&&
-\frac{2N}{N_t^{d-1}} \sum_{n=2}^K \sum_{m=1}^{n-1} \alpha_n \Re\Tr\l\{ Q_{n,\vecx}^{(m)\dag} Q_{n,\vecx}^{(m-1)} \Omega_{\vecx}^\dag \r\}
\nn\\
\comm{linear}
&&
+ \frac{2N}{N_t^{d-1}} \sum_{n=2}^K \alpha_n \Re\Tr\l\{ Q_{n,\vecx}^{(n-1)\dag} \Omega_{\vecx} \r\} 
\nn\\
\comm{cancels out}
&&
- \frac{2N}{N_t^{d-1}} \sum_{n=2}^K \alpha_n \Re\Tr\l\{ Q_{n,\vecx}^{(0)\dag} \Omega_{\vecx}^n \r\}
\eea
The only remaining pathological terms that need to be eliminated are non-Gaussian. This situation is similar to the case of reduced mixed models with a negative adjoint coupling $\beta_A$ (see Section \ref{sec:mix:red}). Hence let us expand the non-Gaussian piece in terms of the $R$--variables:

\bea
\label{eq:cym:nonGauss:Q0:Expand}
-\frac{N}{N_t^{d-1}} 
\sum_{n=2}^K
\alpha_n \Tr\l\{ 
Q_{n,\vecx}^{(0)\dag} Q_{n,\vecx}^{(0)} 
\r\} 
&\stackrel{\eqref{eq:cym:ChangeVars:Q0}}{=}&
\nn\\
\comm{Gaussian}
&&
-\frac{N}{N_t^{d-1}} 
\sum_{n=2}^K
\alpha_n \Tr\l\{ 
R_{n,\vecx}^\dag R_{n,\vecx} 
\r\} 
\nn\\
\comm{nonlinear}
&&
+\frac{1}{N_t^{d-1}} 
\sum_{n=2}^K
\alpha_n \l|\Tr\l\{ R_{n,\vecx} \r\}\r|^2
\eea
In order to eliminate the nonlinear term in \eqref{eq:cym:nonGauss:Q0:Expand}, we use the auxiliary variables $\wt Q^{(K)}_{n,\vecx}$ :
\bea
\label{eq:cym:Gauss:Terms:Qlast}
-\frac{1}{2} 
\sum_{n=2}^K
\Tr\l\{ 
\wt Q_{n,\vecx}^{(K)\dag} 
\wt Q_{n,\vecx}^{(K)} 
\r\} 
&\stackrel{\eqref{eq:cym:ChangeVars:R}}{=}&
\nn\\
\comm{Gaussian}
&&
-\frac{N}{N_t^{d-1}} \sum_{n=2}^K
\alpha_n \Tr\l\{ 
Q_{n,\vecx}^{(K)\dag} 
Q_{n,\vecx}^{(K)} 
\r\}
\nn\\
\comm{linear}
&&
+\frac{2N}{N_t^{d-1}} 
\sum_{n=2}^K 
\alpha_n \Re\Tr\l\{ 
\frac{1}{N} \Tr\l\{R_{n,\vecx}\r\} 
Q_{n,\vecx}^{(K)\dag} 
\Omega_{\vecx}^\dag 
\r\} 
\nn\\
\comm{cancels out}
&&
-\frac{1}{N_t^{d-1}} 
\sum_{n=2}^K
\alpha_n \l|\Tr\l\{ R_{n,\vecx} \r\}\r|^2
\eea
All nonlinear terms cancel out and are replaced by Gaussian and linear terms. The applicability of the efficient updating algorithms discussed in Section \ref{sec:intro} follows immediately.

With auxiliary variables, the partition function of the CYM theory becomes:
\bea
\label{eq:cym:Gauss(HS)}
Z_{\rm CYM}(\vec\alpha,\beta_F)
&=&
\int \mu_{\rm H} [U]~ \mu[R,Q]~ 
\exp\l(
-S_F(\beta_F;[U])
+\frac{2N}{N_t^3} \sum_{\vecx} \Re\Tr\l\{ 
f_{\vecx}^\dag \Omega_{\vecx} 
\r\}
\r)
\eea
where $\mu_{\rm H}[U]$ is the usual product of $SU(N)$--invariant Haar measures of the link variables, and $\mu[R,Q]$ is the product of Gaussian measures of the auxiliary variables,
\bea
\label{eq:cym:Gauss:mu[R,Q]}
\mu[R,Q] &\equiv& \prod_{\vecx\in\Lambda_\perp} \prod_{n=2}^K \l( \mu_{\sigma_n} (R_{n,\vecx})~ \mu_{\sigma_{K,n}} (Q_{n,\vecx}^{(K)}) \prod_{m=1}^{n-1} \mu_{\sigma_{m,n}} (Q_{n,\vecx}^{(m)}) \r)
\eea
with $\sigma$'s given by:
\bea
\sigma^2_1 = 2 \sigma^2_n = 2 \sigma^2_{m,n} = \sigma^2_{n-1,n} = \sigma^2_{K,n} = \frac{N_t^{d-1}}{2N\alpha_n}, \quad 2 \leq n \leq K, \quad 1 \leq m \leq n-2
\eea
and $f_\vecx$ is a matrix factor that encodes all the information about the deformation terms, which is obtained from the linear terms generated by the auxiliary variables:
\bea
f_{\vecx} &=& 
  \alpha_1 Q_{1,\vecx}^{(0)}
+ \sum_{n=2}^K \alpha_n \l( 
Q_{n,\vecx}^{(n-1)}
+ \sum_{m=1}^{n-1} Q_{n,\vecx}^{(m)\dag} Q_{n,\vecx}^{(m-1)} 
+ \frac{1}{N} \Tr\l\{ R_{n,\vecx} \r\} Q_{n,\vecx}^{(K)\dag} 
\r)
\eea

The exponent of the Boltzmann factor in the partition function is now fully linear. Hence each link variable can be updated with respect to a probability distribution \eqref{eq:F:Distrib:Single},
where $V_{\mu,x}$ is analogous to the typical sum of `staples'. If the link variable to be updated is parallel to a non-compact direction, then $V_{\mu,x}$ contains only the contribution $V_{\mu,x}^F$ from the neighboring plaquettes; if the link variable is parallel to the compact direction, then it will also contain the contribution $V_{x}^{\rm def}$ from the deformation terms:
\bea
V_{\mu,x} &=& 
\begin{cases}
\frac{\beta_F}{N}~ V_{\mu,x}^F ~,
& {\rm if}~~ \mu = 1,\ldots,d-1
\\
\\
\frac{\beta_F}{N}~ V_{\mu,x}^F ~+~ \frac{2N}{N_t^{d-1}}~ V_{x}^{\rm def} ~,
& {\rm if}~~ \mu = d
\end{cases}
\eea
The plaquette contribution $V_{\mu,x}^F$ is given by \eqref{eq:red:Staples}, which already takes into account the possibility of a fully reduced compact direction (i.e. $N_t = 1$). On the other hand, the contribution from the double-trace terms is given by:
\bea
V_{x}^{\rm def} &=& 
\l( {\cal P} \prod_{i=0}^{t-1} U_{d, \vecx + i \wh d} \r)^\dag
\cdot f_{\vecx} \cdot
\l( {\cal P} \prod_{i=t+1}^{L_d-1} U_{d, \vecx + i \wh d} \r)^\dag
\eea
This term resembles the Hermitian conjugate of the Polyakov loop wrapping the compact direction and starting at $x = (\vecx,t)$, except that the link variable $U_{d,x}$ is replaced with $f_\vecx^\dag$. When implementing the algorithm, one may find convenient to redefine the Polyakov loop $\Omega_\vecx$ to start at $x$ before the link $U_{d,x}$ is updated,
\bea
\Omega_\vecx &\to& 
\mathcal{P} \prod_{i=0}^{N_t-1} U_{d, \vecx + i' \wh d} 
~=~
U_{d, \vecx + t\wh d} \cdots U_{d, \vecx + (N_t-1) \wh d} U_{d, \vecx}  U_{d, \vecx + \wh d} \cdots U_{d, \vecx + (t - 1) \wh d}
\eea
where $i' \equiv i + t~ ({\rm mod}~ N_t)$. The `staple' contribution coming from the deformation terms is then given by:
\bea
V_{x}^\mathrm{def} 
&=& 
f_{\vecx} \cdot \l( \mathcal{P} \prod_{i=1}^{N_t-1} U_{d, \vecx + i' \hat d} \r)^\dag
\eea
If the compact direction is fully reduced, then $V_x^{\rm def}$ is simply given by:
\bea
V_{\vecx}^\mathrm{def} &=& f_{\vecx}
\eea

The MC algorithm for lattice CYM theories on $\mathds{R}^{d-1} \times S^1$ is summarised in Appendix \ref{app:MC:cym.single}.

\subsection
{Numerical tests
\label{sec:cym:single:numerical}}

We performed numerical simulations of the MC algorithm described above and compared it with Metropolis. We considered the lattice regularised version of $SU(5)$ CYM theory defined on $\mathds{R}^3 \times S^1$. We simulated the theory on two different lattices, namely on $10^3 1$ and $10^3 3$ lattices. The purpose was to test the new algorithm in both situations of an unreduced and a fully reduced compact direction.

For the gauge group $SU(5)$, the CYM action has $\lfloor 5/2 \rfloor = 2$ distinct deformation terms. The $\alpha_n$ parameters attached to them supposedly interpolate between different phases of the Yang-Mills theory at fixed coupling. In particular, for large values of $\beta_F$, the CYM theory should interpolate between the deconfining regime for small $\alpha_n$ (where the centre symmetry associated with the compact direction is spontaneously broken) and a confining regime for large $\alpha_n$ (where the centre symmetry is intact). This behaviour is a  result of the competition between the $S_F$ and the $S_{\rm def}$ terms in the CYM action. For this reason we chose to perform our simulations at $\beta_F = 25.0$, which is  located in the deconfining ($Z_N$--broken) regime of the fundamental Wilson action. For the $10^3 1$ lattice, we chose values for $\alpha_n$ that would put the CYM theory on two different phases. Specifically, we chose a `small' $\vec\alpha = (0.20,0.05)$ and a `large' $\vec\alpha = (0.60,0.10)$. For the $10^3 3$ lattice, we only chose one value, $\vec\alpha = (2.80,0.20)$.

For the thermal Metropolis updates, we used a variant of the (1-hit) Cabibbo-Marinari-Metropolis (CMM) algorithm described in \cite{NeccoHasenbusch} and adapted to CYM theories. In our CMM algorithm, new link proposals are generated using an appropriately tuned $S_F$ action, and are then subjected to an accept/reject step with respect to the full CYM action. The acceptance rates are tuned to stay within a range of 40-60\%. 

For both the Metropolis and the new algorithm, each configuration update consists of only one thermal update (no overrelaxation updates were performed). For each configuration we evaluated the CYM action $S_{\rm CYM}$, which was then used to estimate its expectation value $\langle S_{\rm CYM} \rangle$. We also evaluated plaquette and Polyakov loop traces. In each simulation we performed 398,000 measurements, after discarding the initial 2,000 configurations for equilibration. The simulation parameters and measured observables, together with their naive confidence intervals are shown in Table \ref{tab:Plaquette.CYM}.

\begin{figure}[p]
\begin{center}
\leavevmode
\resizebox{11.5cm}{!}{\includegraphics
{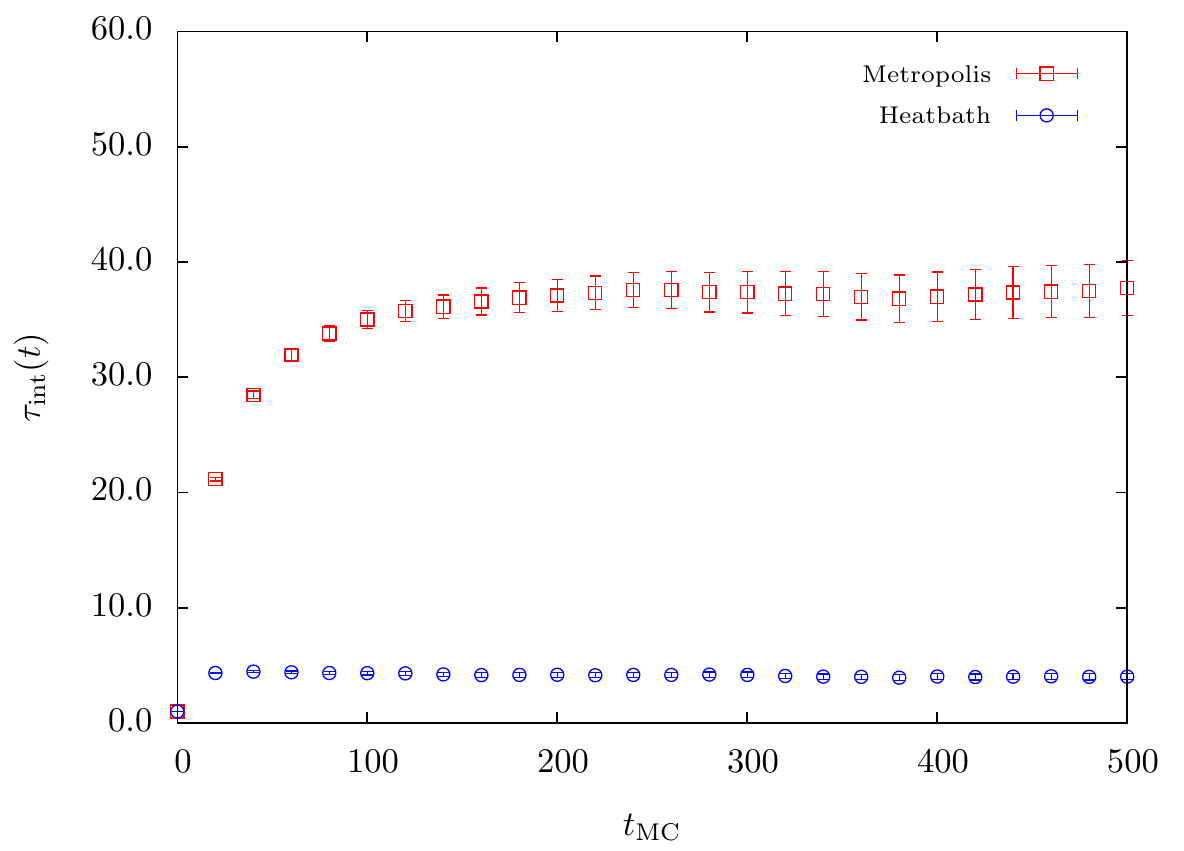}
}
\end{center}
\vskip -4mm
\caption{Estimator of the $t$-dependent integrated autocorrelation time, $\tau_{\rm int}(t)$, vs. the MC time, $t_{\rm MC}$, for the $SU(5)$ CYM theory, simulated on a $10^3 1$ lattice with $\beta_F = 25.0$ and $(\alpha_1,\alpha_2) = (0.20,0.05)$, via a Cabibbo-Marinari-Metropolis algorithm (squares) and a Fabricius-Haan-type pseudo-heatbath algorithm (circles).}
\label{fig:CYM:red(decf):tau}
\end{figure}

Both algorithms agree in terms of the measured values of $\langle S_{\rm CYM} \rangle$, as can be seen in the first and third rows of Table \ref{tab:Plaquette.CYM}. In the second row there is a clear discrepancy in the last significant digits, but this is very likely due to the fact that autocorrelations were not taken into account in the evaluation of the confidence intervals. From Fig.\ref{fig:CYM:red(decf):tau} it is possible to see that $\bar \tau_{\rm int}$ is very large in this example, which means that the confidence intervals on the second row of Table \ref{tab:Plaquette.CYM} are highly underestimated. 

\begin{figure}[p]
\begin{center}
\leavevmode
\resizebox{11.5cm}{!}{\includegraphics
{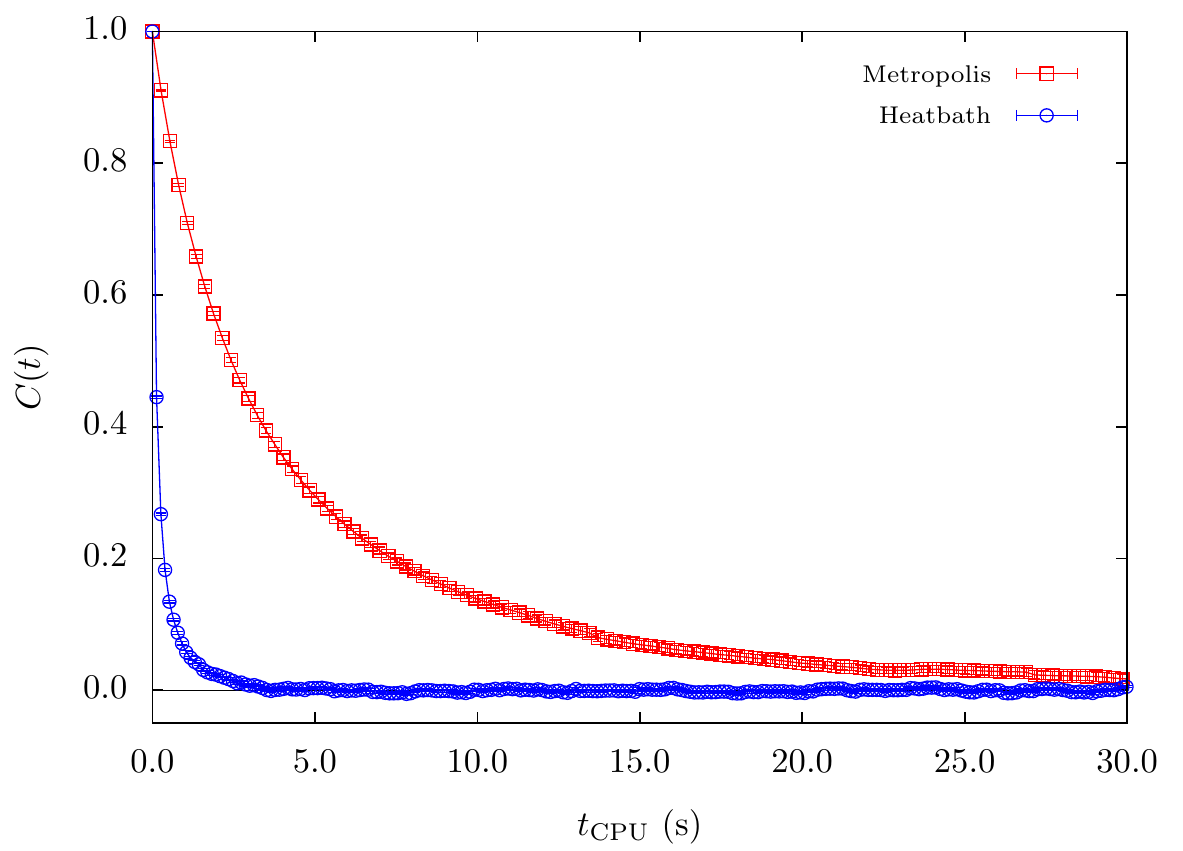}
}
\end{center}
\vskip -4mm
\caption{Estimator of the autocorrelation function, $C(t)$, vs. the CPU time, $t_{\rm CPU}$, for the $SU(5)$ CYM theory, simulated on a $10^3 1$ lattice with $\beta_F = 25.0$ and $(\alpha_1,\alpha_2) = (0.20,0.05)$, via a Cabibbo-Marinari-Metropolis algorithm (squares) and a Fabricius-Haan-type pseudo-heatbath algorithm (circles).}
\label{fig:CYM:red(decf):C}
\end{figure}

\begin{figure}[p]
\begin{center}
\leavevmode
\resizebox{11.5cm}{!}{\includegraphics
{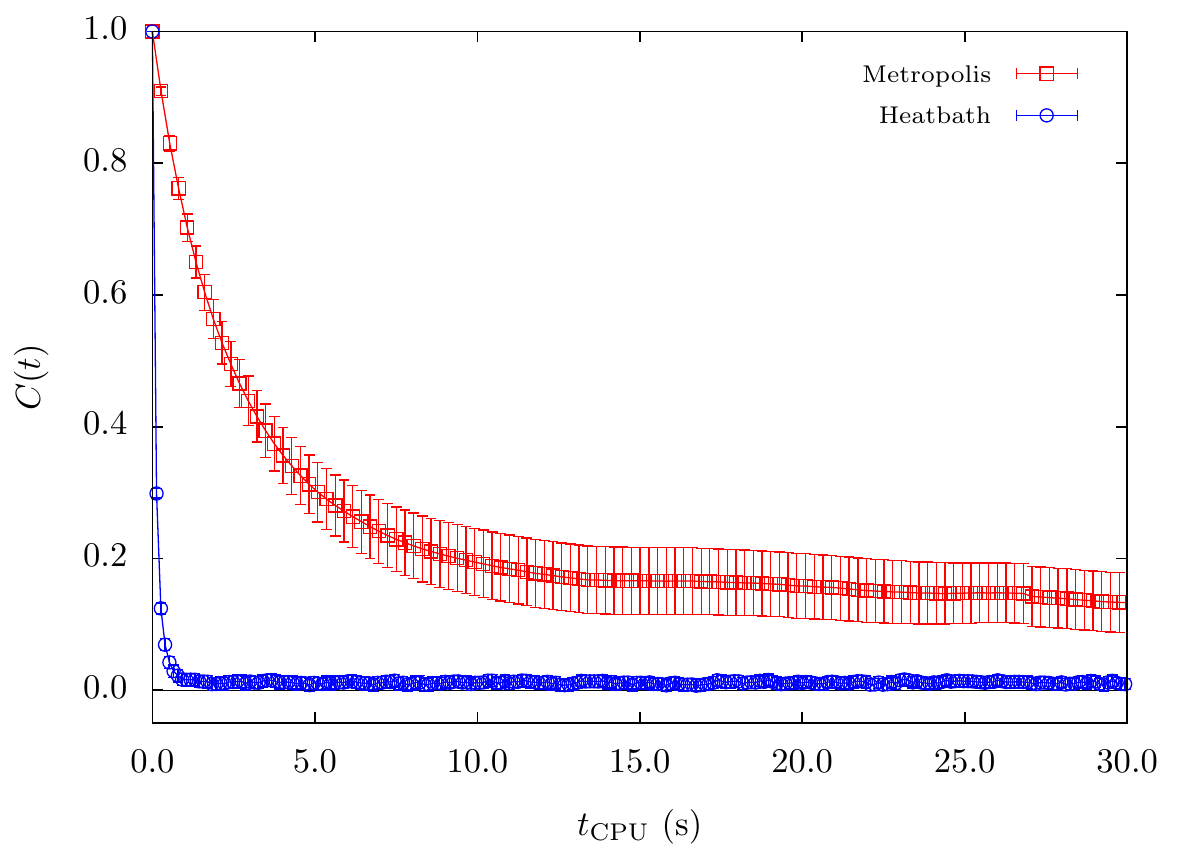}
}
\end{center}
\vskip -4mm
\caption{Estimator of the autocorrelation function, $C(t)$, vs. the CPU time, $t_{\rm CPU}$, for the $SU(5)$ CYM theory, simulated on a $10^3 1$ lattice with $\beta_F = 25.0$ and $(\alpha_1,\alpha_2) = (0.60,0.10)$, via a Cabibbo-Marinari-Metropolis algorithm (squares) and a Fabricius-Haan-type pseudo-heatbath algorithm (circles).}
\label{fig:CYM:red(conf):C}
\end{figure}

\begin{figure}[p]
\begin{center}
\leavevmode
\resizebox{11.5cm}{!}{\includegraphics
{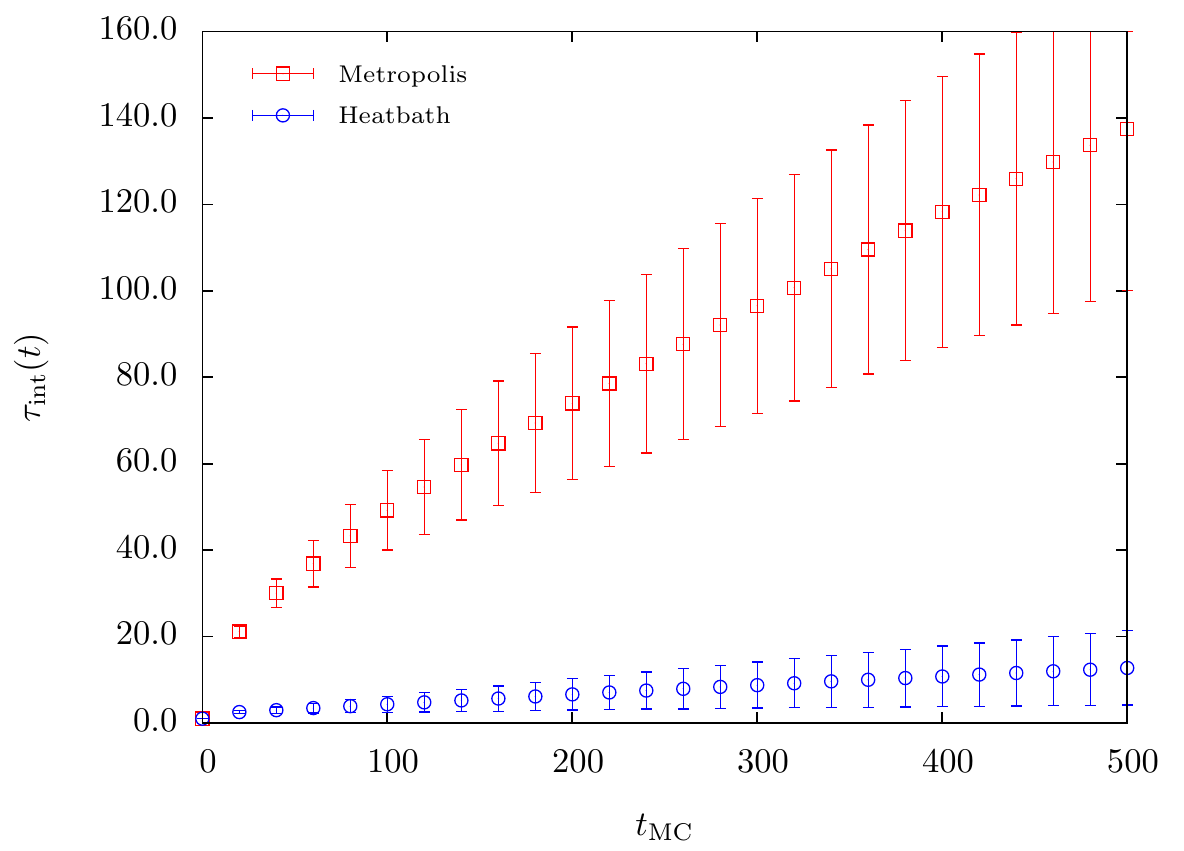}
}
\end{center}
\vskip -4mm
\caption{Estimator of the $t$-dependent integrated autocorrelation time, $\tau_{\rm int}(t)$, vs. the MC time, $t_{\rm MC}$, for the $SU(5)$ CYM theory, simulated on a $10^3 1$ lattice with $\beta_F = 25.0$ and $(\alpha_1,\alpha_2) = (0.60,0.10)$, via a Cabibbo-Marinari-Metropolis algorithm (squares) and a Fabricius-Haan-type pseudo-heatbath algorithm (circles).}
\label{fig:CYM:red(conf):tau}
\end{figure}

\begin{figure}[p]
\begin{center}
\leavevmode
\resizebox{11.5cm}{!}{\includegraphics
{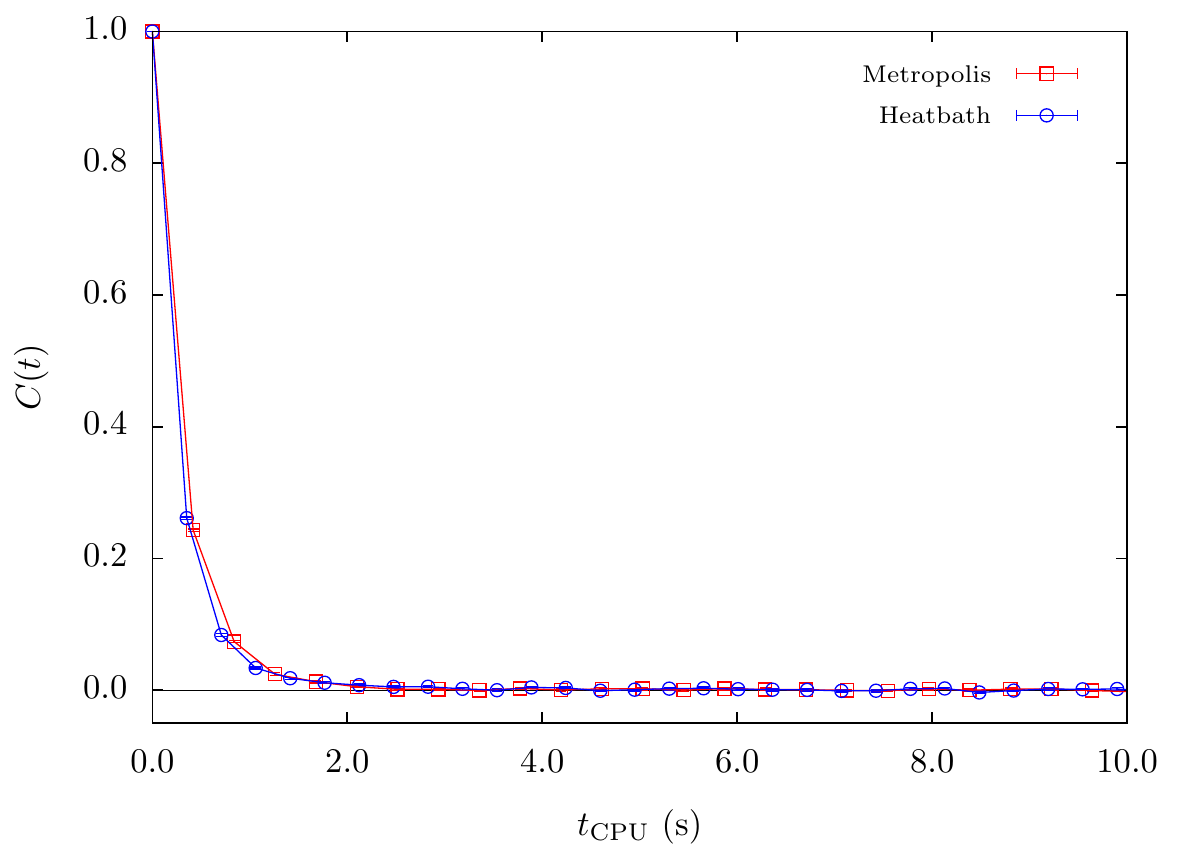}
}
\end{center}
\vskip -4mm
\caption{Estimator of the autocorrelation function, $C(t)$, vs. the CPU time, $t_{\rm CPU}$, for the $SU(5)$ CYM theory, simulated on a $10^3 3$ lattice with $\beta_F = 25.0$ and $(\alpha_1,\alpha_2) = (2.80,0.20)$, via a Cabibbo-Marinari-Metropolis algorithm (squares) and a Fabricius-Haan-type pseudo-heatbath algorithm (circles).}
\label{fig:CYM:unr:C}
\end{figure}

\begin{figure}[p]
\begin{center}
\leavevmode
\resizebox{11.5cm}{!}{\includegraphics
{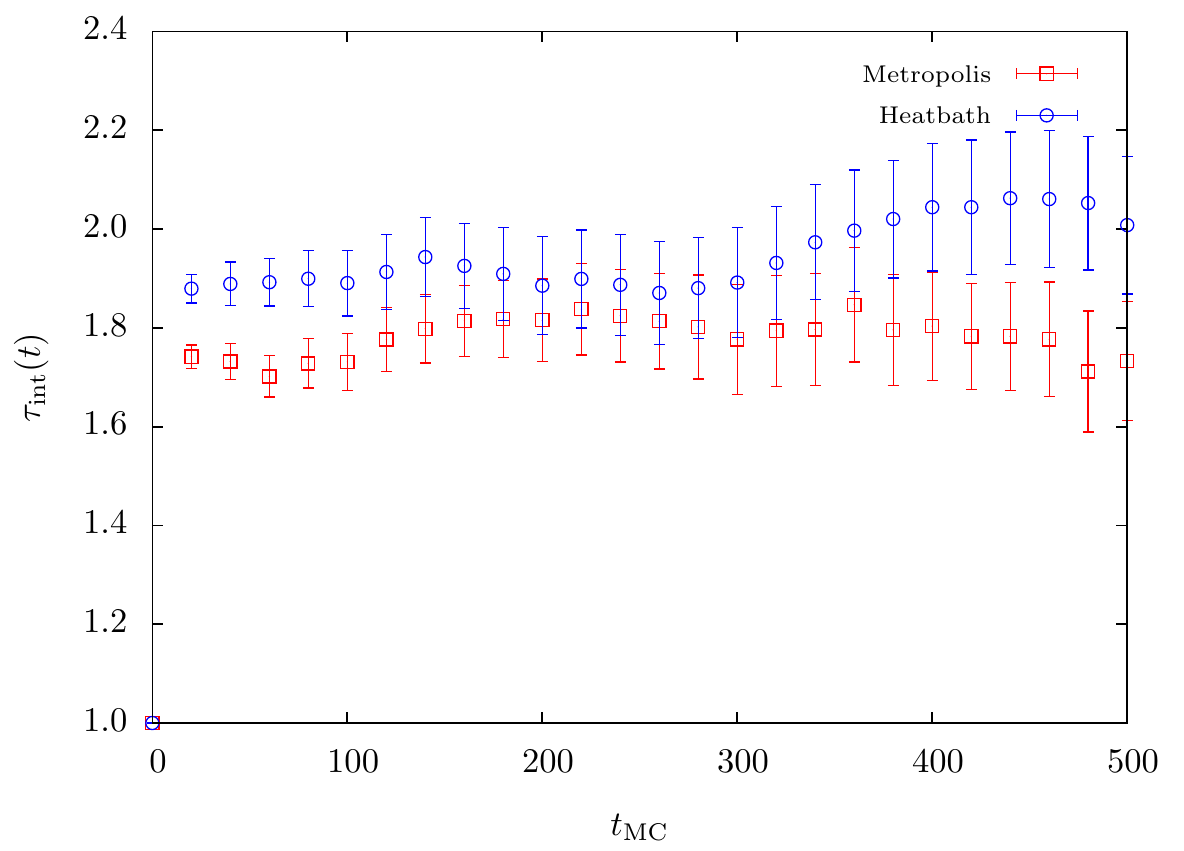}
}
\end{center}
\vskip -4mm
\caption{Estimator of the $t$-dependent integrated autocorrelation time, $\tau_{\rm int}(t)$, vs. the MC time, $t_{\rm MC}$, for the $SU(5)$ CYM theory, simulated on a $10^3 3$ lattice with $\beta_F = 25.0$ and $(\alpha_1,\alpha_2) = (2.80,0.20)$, via a Cabibbo-Marinari-Metropolis algorithm (squares) and a Fabricius-Haan-type pseudo-heatbath algorithm (circles).}
\label{fig:CYM:unr:tau}
\end{figure}

In terms of autocorrelations, the pseudo-heatbath algorithm for the $N_t = 1$ CYM theory performs much better than the (optimally tuned) Metropolis algorithm (see Figs.\ref{fig:CYM:red(decf):C}--\ref{fig:CYM:red(conf):tau}). However, in the case $N_t=3$, the pseudo-heatbath algorithm does not show an improvement over Metropolis (see Figs.\ref{fig:CYM:unr:C}--\ref{fig:CYM:unr:tau}). A possible reason for this behaviour could be an excessive number of auxiliary variables in the $N_t=3$ CYM theory, whose update could easily undermine the efficiency of the pseudo-heatbath algorithm. The smaller the number of auxiliary variables, the more efficient and faster the algorithm is. In sum, the new updating algorithm for CYM theories on $\mathds{R}^3 \times S^1$ are most efficient when the compact direction is fully reduced. Fortunately, this is also the most interesting case for a CYM theory, as long as large $N$ volume-independence holds nonperturbatively for any volume.

\begin{figure}[p]
\centering
\mbox{
\subfigure{
\includegraphics[width=8cm]{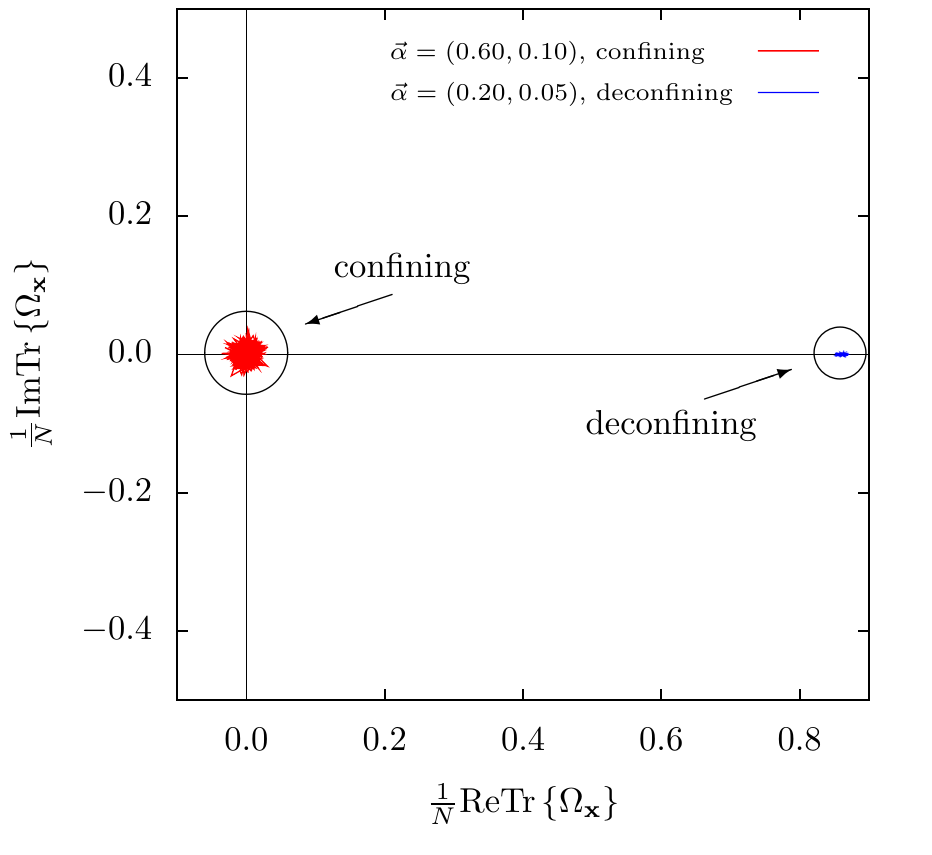}}
\subfigure{
\includegraphics[width=8cm]{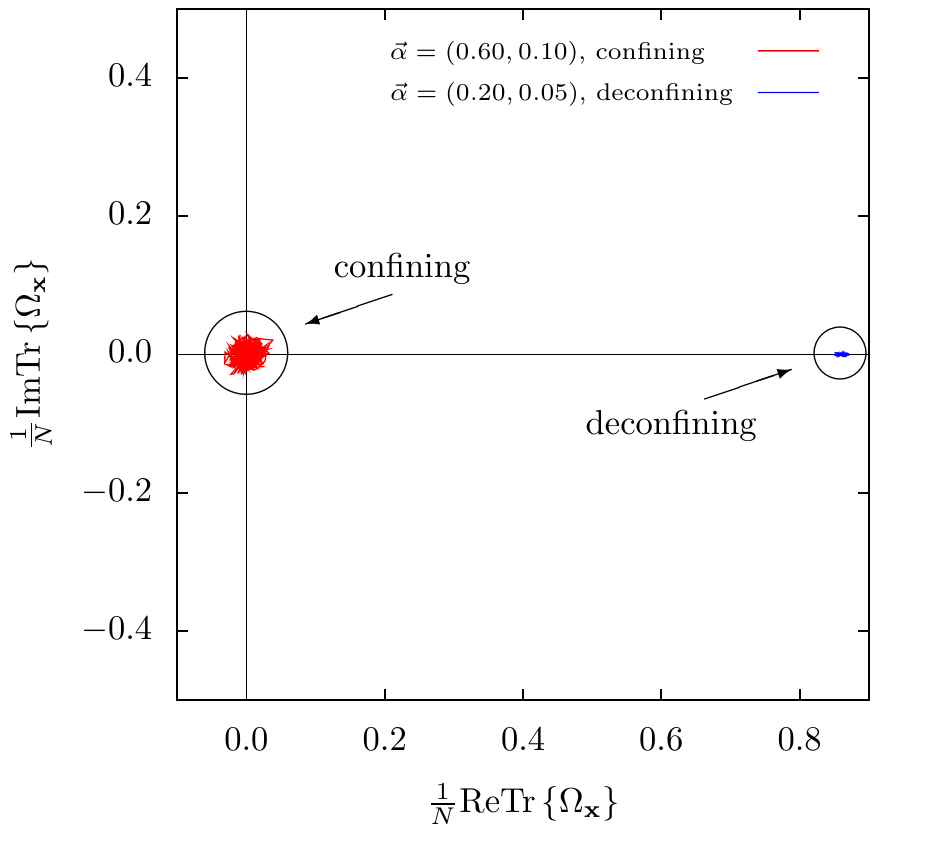}}
}
\caption{Trace of the holonomy, $\Omega_\vecx$, around the compact direction of $\mathds{R}^3 \times S^1$, in simulations of the $SU(5)$ CYM theory on a $10^3 1$ lattice with $\beta_F = 25.0$, using either a Fabricius-Haan-type pseudo-heatbath algorithm (left) or a Cabibbo-Marinari-Metropolis algorithm (right).}
\label{fig:CYM:h.m}
\end{figure}

Finally, Fig.\ref{fig:CYM:h.m} provides a qualitative check for the compatibility of the results from the Metropolis and the pseudo-heatbath algorithms. The graphs show the MC histories of the complex trace of the Polyakov loop, for Metropolis and pseudo-heatbath simulations of the $SU(5)$ CYM theory on a $10^3 1$ lattice. The simulation with a `small' deformation parameter $\vec\alpha = (0.20,0.05)$ results in a deconfining vacuum ($\langle\Tr\{\Omega_\vecx\}\rangle \neq 0$), as expected for a CYM theory whose action is dominated by plaquette terms with a large $\beta_F$. The simulation with `large' deformation parameters $\vec\alpha = (0.60,0.10)$ results in a confining vacuum ($\langle\Tr\{\Omega_\vecx\}\rangle = 0$), as expected for a CYM theory whose action is dominated by the centre-stabilising double-trace terms. Both situations were correctly captured by the Metropolis and the pseudo-heatbath algorithm, which reinforces the validity of the latter.

\section
{Discussion
\label{sec:Discussion}}

In this paper we constructed new algorithms for the update of the link variables for two classes of pure lattice gauge theories with nonlinear actions. The theories under consideration were (i) pure $SU(N)$ lattice gauge theories with a generic mixed Wilson action, and (ii) the lattice regularisation of centre-stabilised $SU(N)$ Yang-Mills (CYM) theories defined on $\mathds{R}^{d-1} \times S^1$. 

We used a generalisation of the Fabricius-Haan method of auxiliary variables in order to construct such algorithms. By adding enough extra degrees of freedom to the lattice gauge theory in question, we were able to get rid of the nonlinear terms in its action, and replace them with linear terms on the link variables. In this way it is possible to perform pseudo-heatbath, overrelaxation or cooling updates on the links of the nonlinear theories, just like in the standard $SU(N)$ lattice gauge theory with the fundamental Wilson action.

As a test for the accuracy of the new algorithms, we evaluated numerically the expectation values of some gauge-invariant observables and compared them with a Metropolis evaluation. Both quantitative and qualitative results showed a match between the outputs of the new pseudo-heatbath and Metropolis algorithms (modulo the autocorrelation correction of the confidence levels).

We also showed numerically that the new algorithms are more efficient than Metropolis. Despite the new updates, in general, being slower than Metropolis hits in CPU time (because of the large number of auxiliary variables that some lattice theories require), they also decorrelate the gauge configurations very fast. Therefore, when taking autocorrelations into account, the new algorithms perform much better than Metropolis.

The method of auxiliary variables is rather general and may be applied to other lattice theories with polynomial dependence on the link variables, for example the 3D Georgi-Glashow model,\footnote{
We thank G. Bali for this suggestion.
}
the TEK--reduced 2D principal chiral model, etc.. In relation to the CYM theories discussed in this paper, the method can also be extended to the more complicated case of centre-stabilised Yang-Mills theories defined on manifolds with multiple compactified directions, namely $\mathds{R}^{d-k} \times (S^1)^k$. In particular, it could be used to construct an efficient MC algorithm for the zero-volume limit of CYM theories compactified on a $d$--torus, also known as deformed Eguchi-Kawai (DEK) models. An efficient algorithm for the DEK model would be very helpful in establishing nonperturbatively (via numerical simulations) if its centre symmetry is indeed intact, as claimed by \"Unsal and Yaffe \cite{UnsalYaffe}, thus providing the first problem-free matrix model representation of the planar sector of pure $SU(N)$ Yang-Mills theories on $\mathbb{R}^d$. We leave the construction of appropriate algorithms and the study of these theories to later publications.


\vskip 5mm
\section*{Acknowledgements}

We are very grateful to Mike Teper and Barak Bringoltz for useful discussions and feedback in the early stage of this project. We are also grateful for the hospitality at CERN, where part of this work was done. 

Our lattice calculations were carried out on the Milipeia cluster at the University of Coimbra, and on PCs equipped with Intel${}^\circledR$ Pentium IV and Core Duo processors.

HV is supported by FCT (Portugal) under the grant
SFRH/BPD/37949/2007.


\clearpage
\appendix

\section
{Monte Carlo algorithms
\label{app:MC}}

Here we summarise the MC algorithms for the different $SU(N)$ lattice gauge theories discussed in the paper.

\subsection
{Reduced lattices
\label{app:MC:red}}

MC algorithm for the fundamental Wilson action \eqref{eq:F:Action} on partially reduced lattices:

\begin{small}
\begin{enumerate}
\item For each reduced plaquette $(\mu\nu,x)$: 
\begin{enumerate}
\item Generate one random complex $N \times N$ matrix, $\wt Q_{\mu\nu,x}$, with normal distribution $\mu_1(\wt Q_{\mu\nu,x})$, using the Box-M\"uller transform \eqref{eq:BoxMuller}.
\item Construct new auxiliary variables:
\bea
Q_{\mu\nu,x} 
&=& 
\l( \frac{N}{\beta_F} \r)^{\frac{1}{2}} 
\wt Q_{\mu\nu,x} 
+ U_{\mu,x} U_{\nu,x + \hat\mu} 
+ U_{\nu,x} U_{\mu,x + \hat\nu}
\eea
\end{enumerate}
\item For each link variable $U_{\mu,x}$ :
\begin{enumerate}
\item Construct the sum of `staples':
\bea
V_{\mu,x} 
&=& 
\frac{\beta_F}{N} \!\!\! \sum_{\nu=1\atop (\nu \neq \mu, L_\nu > 1)}^d 
\l(
U_{\nu,x} U_{\mu,x+\hat\nu} U_{\nu,x+\hat\mu}^\dag ~+~
U_{\nu,x-\hat\nu}^\dag U_{\mu,x-\hat\nu} U_{\nu,x-\hat\nu+\hat\mu}
\r)
\nn\\
&+& \frac{\beta_F}{N} \!\!\! \sum_{\nu=1\atop (\nu \neq \mu, L_\nu = 1)}^d 
\l( Q_{\mu\nu,x} U_{\nu,x+\hat\mu}^\dag ~+~ 
U_{\nu,x}^\dag Q_{\mu\nu,x} \r)
\eea
where $Q_{\mu\nu,x} \equiv Q_{\nu\mu,x}$.
\item Update $U_{\mu,x}$ with respect to the probability distribution \eqref{eq:F:Distrib:Single} 
using the Cabibbo-Marinari pseudo-heatbath algorithm.
\end{enumerate}
\end{enumerate}
\end{small}

\subsection
{Mixed actions
\label{app:MC:mix}}

MC algorithm for generic mixed Wilson actions \eqref{eq:mix:Action} on lattices with unreduced directions. The case of partially reduced directions is discussed in Section \ref{sec:mix:red}.

\begin{small}
\begin{enumerate}
\item For each plaquette $p \equiv (\mu\nu,x)$:
\begin{enumerate}
\item Construct $U_p ~=~ U_{\mu,x} U_{\nu,x + \wh\mu} U_{\mu,x + \wh\nu}^\dag U_{\nu,x}^\dag$
\item For each irreducible representation $\R$ contributing to the mixed Wilson action:
\begin{enumerate}
\item Generate $n_\R$ random complex $N \times N$ matrices $\wt Q^{(i)}_{p}$ with the normal distribution $\mu_1(\wt Q^{(i)}_{p})$, using the Box-M\"uller transform \eqref{eq:BoxMuller}.
\item Construct new auxiliary variables using the appropriate $\beta_\R '$, $h^{(i)}_{p}$ (see Table \ref{tab:Summary.Mix}):
    \bea
    Q^{(i)}_{p} 
    &=& 
    \frac{1}{\sqrt{2\beta_\R '}}
    \wt Q^{(i)}_{p} 
    ~+~ h^{(i)}_{p}
    \eea
\item Construct $f^\R_p$ using the appropriate $g^{(i)}_p$ (see Table \ref{tab:Summary.Mix}):
    \bea
    f^{\R}_p
    &=&
    \sum_{i=1}^{n_\R} g^{(i)}_p
    \eea
\end{enumerate}
\item Construct $f_p$:
    \bea
    f_p
    &=&
    2 \sum_\R \beta_\R ' f^\R_p
    \eea
\end{enumerate}
\item For each link variable $U_{\mu,x}$ :
\begin{enumerate}
\item Construct the sum of `staples':
\bea
V_{\mu,x} 
&=&
\sum_{{\nu=1}\atop{(\nu\neq\mu)}}^{d} 
\l(
f_{\mu\nu,x} U_{\nu,x} U_{\mu,x+\hat\nu} U_{\nu,x+\hat\mu}^\dag 
~+~ 
U_{\nu,x-\hat\nu}^\dag f_{\nu\mu,x-\hat\nu} U_{\mu,x-\hat\nu} U_{\nu,x-\hat\nu+\hat\mu}
\r)
\eea
where $f_{\nu\mu,x} \equiv f_{\mu\nu,x}^\dag \equiv f_p$.
\item Update $U_{\mu,x}$ with respect to the probability distribution \eqref{eq:F:Distrib:Single} 
using any of the following algorithms: 
\begin{itemize}
    \item Cabibbo-Marinari pseudo-heatbath
    \item $SU(2)$ or $SU(N)$ overrelaxation
    \item $SU(2)$ or $SU(N)$ cooling
\end{itemize}
\end{enumerate}
\end{enumerate}
\end{small}

\subsection{Double-trace deformations
\label{app:MC:cym.single}}

MC algorithm for lattice CYM theories on $\mathds{R}^{d-1} \times S^1$, with the possibility of a fully reduced `compact' direction.

\begin{small}
\begin{enumerate}
\item For each $x \in \Lambda$ $(\sim \mathbb{R}^{d-1} \times S^1)$ :
\begin{enumerate}
\item If $N_t = 1$, do step (1) from the algorithm \ref{app:MC:red}
\end{enumerate}

\item For each $\vecx \in \Lambda_\perp$  $(\sim \mathbb{R}^{d-1})$ :
\begin{enumerate}

\item Construct $\Omega_\vecx ~=~ U_{d, \vecx} U_{d, \vecx + \wh d} \cdots U_{d, \vecx + (N_t  - 1) \wh d}$
\item Generate a random complex $N \times N$ matrix, $\wt R_{1,\vecx}$, with normal distribution $\mu_1(\wt R_{1,\vecx})$, using the Box-M\"uller transform \eqref{eq:BoxMuller}.
\item Construct the new auxiliary variables:
\bea
R_{1,\vecx} &=&
\l(\frac{N_t^{d-1}}{2N\alpha_1}\r)^{\frac{1}{2}} 
\wt R_{1,\vecx} ~+~ \Omega_{\vecx} - \frac{1}{N}\Tr\{\Omega_{\vecx}\}\1
\eea
\item Construct $Q^{(0)}_{1,\vecx} ~=~ R_{1,\vecx} - \frac{1}{N}\Tr\l\{ R_{1,\vecx} \r\}\1$.

\item For $n=2,\ldots,K$ :

\begin{enumerate}

\item Construct $\Omega_\vecx^{n}$
\item Generate random complex $N \times N$ matrices $\wt R_{n,\vecx}$ and $\wt Q_{n,\vecx}^{(K)}$ with normal distributions $\mu_1(\wt R_{n,\vecx})$ and $\mu_1(\wt Q_{n,\vecx}^{(K)})$, respectively, using the Box-M\"uller transform \eqref{eq:BoxMuller}.
\item Construct the new auxiliary variables:
\bea
R_{n,\vecx} &=&
\l(\frac{N_t^{d-1}}{2N\alpha_n}\r)^{\frac{1}{2}} 
\wt R_{n,\vecx} ~+~ \Omega_{\vecx}^{n} - \frac{1}{N}\Tr\{\Omega_{\vecx}^{n}\}\1
\\
Q_{n,\vecx}^{(K)} &=&
\l(\frac{N_t^{d-1}}{2N\alpha_n}\r)^{\frac{1}{2}} 
\wt Q_{n,\vecx}^{(K)} ~+~ \frac{1}{N} \Tr\{ R_{n,\vecx} \} \Omega_{\vecx}^\dag
\eea
\item Construct $Q^{(0)}_{n,\vecx} ~=~ R_{n,\vecx} - \frac{1}{N}\Tr\l\{ R_{n,\vecx} \r\}\1$.

\item For $m=1,\ldots,n-1$ :

\begin{enumerate}

\item Generate a random complex $N \times N$ matrix $\wt Q^{(m)}_{n,\vecx}$ with the normal distribution $\mu_1(\wt Q_{n,\vecx}^{(m)})$, using the Box-M\"uller transform \eqref{eq:BoxMuller}.
\item Construct the new auxiliary variables:
\bea
Q^{(m)}_{n,\vecx} 
&=&
\l(\frac{N_t^{d-1}}{2N\alpha_n}\r)^{\frac{1}{2}} 
\wt Q^{(m)}_{n,\vecx} + Q_{n,\vecx}^{(m-1)} \Omega_{\vecx}^\dag + \Omega_{\vecx}^{n-m}
\eea
\end{enumerate}
\end{enumerate}

\item Construct $f_{\vecx}$:
\bea
f_{\vecx} &=& 
  \alpha_1 Q_{1,\vecx}^{(0)}
+ \sum_{n=2}^K \alpha_n \l( 
Q_{n,\vecx}^{(n-1)}
+ \sum_{m=1}^{n-1} Q_{n,\vecx}^{(m)\dag} Q_{n,\vecx}^{(m-1)} 
+ \frac{1}{N} \Tr\l\{ R_{n,\vecx} \r\} Q_{n,\vecx}^{(K)\dag} 
\r)
\eea
\end{enumerate}
\item For each link variable $U_{\mu,x}$ :

\begin{enumerate}

\item Construct the sum of `staples' $V_{\mu,x}$ as in the step (2a) of the algorithm \ref{app:MC:red}, and multiply it by $\beta_F/N$:
\bea
V_{\mu,x} ~\leftarrow~ \frac{\beta_F}{N}V_{\mu,x}
\eea

\item If $\mu = d$, then :

\begin{enumerate}
\item Construct $V_{x}^{\rm def}$ :
\bea
{\rm If}~ N_t > 1,&&
V_{x}^{\rm def} =
\l( {\cal P} \prod_{i=0}^{t-1} U_{d, \vecx + i \wh d} \r)^\dag
\cdot f_{\vecx} \cdot
\l( {\cal P} \prod_{i=t+1}^{N_t-1} U_{d, \vecx + i \wh d} \r)^\dag
\\
{\rm If}~ N_t = 1,&&
V_{x}^{\rm def} = f_{\vecx}
\eea
where $x \equiv (\vecx,t)$.

\item Add  $V_{x}^{\rm def}$ to  $V_{x}$ :
\bea
V_{x} ~\leftarrow~  V_{x} + \frac{2N}{N_t^{d-1}} V_{x}^{\rm def}
\eea

\end{enumerate}

\item Update $U_{\mu,x}$ with respect to the probability distribution \eqref{eq:F:Distrib:Single}
using any of the following algorithms: 
\begin{itemize}
    \item Cabibbo-Marinari pseudo-heatbath
    \item $SU(2)$ or $SU(N)$ overrelaxation
    \item $SU(2)$ or $SU(N)$ cooling
\end{itemize}
\end{enumerate}

\end{enumerate}
\end{small}


\vfill\eject\clearpage
\addcontentsline{toc}{section}{\bibname}
\bibliography{montecarlo}

\begin{thebibliography}{10}

\bibitem{Metropolis}
N.~Metropolis, A.~Rosenbluth, M.~Rosenbluth, A.~Teller and E.~Teller, {\em
  Equation of State Calculations by Fast Computing Machines},
\newblock J. Chem. Phys. \textbf{ 21}, 1087--1092 (1953).

\bibitem{CreutzSU2}
M.~Creutz, {\em Monte Carlo Study Of Quantized $SU(2)$ Gauge Theory},
\newblock Phys. Rev. \textbf{ D21}, 2308--2315 (1980).

\bibitem{FabriciusHaan}
K.~Fabricius and O.~Haan, {\em Heat Bath Method For The Twisted Eguchi-Kawai
  Model},
\newblock Phys. Lett. \textbf{ B143}, 459 (1984).

\bibitem{KennedyPendletonSU2}
A.~D. Kennedy and B.~J. Pendleton, {\em Improved Heat Bath Method For Monte
  Carlo Calculations In Lattice Gauge Theories},
\newblock Phys. Lett. \textbf{ B156}, 393--399 (1985).

\bibitem{BrownWoch}
F.~R. Brown and T.~J. Woch, {\em {Overrelaxed Heat Bath and Metropolis
  Algorithms for Accelerating Pure Gauge Monte Carlo Calculations}},
\newblock Phys. Rev. Lett. \textbf{ 58}, 2394 (1987).

\bibitem{CreutzOverrelaxation}
M.~Creutz, {\em {Overrelaxation and Monte Carlo Simulation}},
\newblock Phys. Rev. \textbf{ D36}, 515 (1987).

\bibitem{TeperCooling}
M.~Teper, {\em {Instantons in the Quantized $SU(2)$ Vacuum: A Lattice Monte
  Carlo Investigation}},
\newblock Phys. Lett. \textbf{ B162}, 357 (1985).

\bibitem{IlgenfritzCooling}
E.-M. Ilgenfritz, M.~L. Laursen, G.~Schierholz, M.~Muller-Preussker and
  H.~Schiller, {\em {First Evidence for the Existence of Instantons in the
  Quantized $SU(2)$ Lattice Vacuum}},
\newblock Nucl. Phys. \textbf{ B268}, 693 (1986).

\bibitem{CabibboMarinari}
N.~Cabibbo and E.~Marinari, {\em A New Method For Updating $SU(N)$ Matrices In
  Computer Simulations Of Gauge Theories},
\newblock Phys. Lett. \textbf{ B119}, 387--390 (1982).

\bibitem{Pietarinen}
E.~Pietarinen, {\em {String Tension in $SU(3)$ Lattice Gauge Theory}},
\newblock Nucl. Phys. \textbf{ B190}, 349 (1981).

\bibitem{KiskisNarayananNeuberger}
J.~Kiskis, R.~Narayanan and H.~Neuberger, {\em {Does The Crossover From
  Perturbative To Nonperturbative Physics In QCD Become A Phase Transition At
  Infinite $N$?}},
\newblock Phys. Lett. \textbf{ B574}, 65--74 (2003), {hep-lat/0308033}.

\bibitem{ForcrandJahn}
P.~de~Forcrand and O.~Jahn, {\em {Monte Carlo Overrelaxation for $SU(N)$ Gauge
  Theories}},
\newblock (2005), {hep-lat/0503041}.

\bibitem{EguchiKawai}
T.~Eguchi and H.~Kawai, {\em {Reduction Of Dynamical Degrees Of Freedom In The
  Large $N$ Gauge Theory}},
\newblock Phys. Rev. Lett. \textbf{ 48}, 1063 (1982).

\bibitem{TEK}
A.~Gonz{\'a}lez-Arroyo and M.~Okawa, {\em {The Twisted Eguchi-Kawai Model: A
  Reduced Model For Large $N$ Lattice Gauge Theory}},
\newblock Phys. Rev. \textbf{ D27}, 2397 (1983).

\bibitem{QEK}
G.~Bhanot, U.~M. Heller and H.~Neuberger, {\em {The Quenched Eguchi-Kawai
  Model}},
\newblock Phys. Lett. \textbf{ B113}, 47 (1982).

\bibitem{UnsalYaffe}
M.~{\"U}nsal and L.~G. Yaffe, {\em {Center-Stabilized Yang-Mills Theory:
  Confinement and Large $N$ Volume Independence}},
\newblock Phys. Rev. \textbf{ D78}, 065035 (2008), {0803.0344}.

\bibitem{Stratonovich}
R.~Stratonovich, {\em {On a Method of Calculating Quantum Distribution
  Functions}},
\newblock Soviet Phys. Doklady \textbf{ 2}, 416 (1958).

\bibitem{Hubbard}
J.~Hubbard, {\em {Calculation of Partition Functions}},
\newblock Phys. Rev. Lett. \textbf{ 3}, 77 (1959).

\bibitem{BoxMuller}
G.~E.~P. Box and M.~E. M{\"u}ller, {\em {A Note on the Generation of Random
  Normal Deviates}},
\newblock Ann. Math. Statist. \textbf{ 29}(2), 610–611 (1958).

\bibitem{OkawaMetropolis}
M.~Okawa, {\em {Monte Carlo Study Of The Eguchi-Kawai Model}},
\newblock Phys. Rev. Lett. \textbf{ 49}, 353 (1982).

\bibitem{BietenholzInstability}
W.~Bietenholz, J.~Nishimura, Y.~Susaki and J.~Volkholz, {\em {A
  Non-Perturbative Study Of 4D $U(1)$ Non-Commutative Gauge Theory: The Fate Of
  One-Loop Instability}},
\newblock JHEP \textbf{ 10}, 042 (2006), {hep-th/0608072}.

\bibitem{MunsterMontvay}
I.~Montvay and G.~M{\"u}nster,
\newblock {\em {Quantum Fields On A Lattice}},
\newblock Cambridge, UK: Univ. Pr. (1994) 491 p. (Cambridge Monographs on
  Mathematical Physics).

\bibitem{NeccoHasenbusch}
M.~Hasenbusch and S.~Necco, {\em {$SU(3)$ Lattice Gauge Theory With A Mixed
  Fundamental And Adjoint Plaquette Action: Lattice Artefacts}},
\newblock JHEP \textbf{ 08}, 005 (2004), {hep-lat/0405012}.

\bibitem{BazavovBergHeller}
A.~Bazavov, B.~A. Berg and U.~M. Heller, {\em {Biased Metropolis-Heat-Bath
  Algorithm For Fundamental-Adjoint $SU(2)$ Lattice Gauge Theory}},
\newblock Phys. Rev. \textbf{ D72}, 117501 (2005), {hep-lat/0510108}.

\bibitem{ArroyoPrivate}
A.~Gonz{\'a}lez-Arroyo,
\newblock {private communication}.

\bibitem{KovtunUnsalYaffe}
P.~Kovtun, M.~Unsal and L.~G. Yaffe, {\em {Volume Independence In Large $N_c$
  QCD-Like Gauge Theories}},
\newblock JHEP \textbf{ 06}, 019 (2007), {hep-th/0702021}.

\end{thebibliography}
\vfill\eject\clearpage


\begin{table}
\begin{center}
\begin{scriptsize}
\begin{tabular}{cccllc} 
\hline
\hline
$k$ & $\R$ & Aux. & $h_p^{(i)}$ & $g_p^{(i)}$ & 
$\beta_\R '$
\\
\hline
\hline
\\
0
& $\beta_A > 0$ 
& $\wt z_p$ 
& $\frac{1}{N} \Tr\{ U_p \}$ 
& $z_p$ 
& $\frac{\beta_A N^2}{N^2-1}$
\\
\\
& $\beta_A < 0$
& $\wt Q_p$ 
& $U_p - \frac{1}{N} \Tr\{ U_p \} \1$ 
& $Q_p - \frac{1}{N} \Tr\{ Q_p \} \1$ 
& $\frac{|\beta_A|N}{N^2-1}$
\\
\\
2
& $\pm$
& $\wt Q_p^{(1)}$ 
& $\frac{1}{2} ( U_p + \frac{\sigma}{N} \Tr\{ U_p^\dag \} \1 )$ 
& $\frac{1}{2} ( Q_p^{(1)} + \frac{\sigma}{N} \Tr\{ Q_p^{(1)\dag} \} \1 )$ 
& $\frac{2|\beta_\pm|}{N \pm 1}$
\\
&
& $\wt Q_p^{(2)}$ 
& $\frac{1}{2} ( U_p \pm \frac{\sigma}{N} U_p^\dag )$ 
& $\frac{1}{2} ( Q_p^{(2)} \pm \frac{\sigma}{N} Q_p^{(2)\dag} )$
\\
&
& $\wt Q_p^{(3)}$ 
& $\frac{1}{2} ( U_p - \frac{1}{N} \Tr\{ U_p \} \1)$ 
& $\frac{1}{2} ( Q_p^{(3)} - \frac{1}{N} \Tr\{ Q_p^{(3)} \} \1 )$
\\
\\
& $\pm$
& $\wt Q_p^{(1)}$ 
& $\frac{1}{2} ( \sigma U_p^\dag \pm \frac{1}{N} U_p + \frac{1}{N} \Tr\{ U_p \} \1 )$ 
& $\frac{1}{2} ( \sigma Q_p^{(1)\dag} \pm \frac{1}{N} Q_p^{(1)} + \frac{1}{N} \Tr\{ Q_p^{(1)} \} \1 )$ 
& $\frac{2|\beta_\pm|}{N \pm 1}$
\\
&
& $\wt Q_p^{(2)}$ 
& $\sqrt{\frac{N \pm 2}{4N}} ( U_p - \frac{1}{N} \Tr\{ U_p \} \1 )$ 
& $\sqrt{\frac{N \pm 2}{4N}} ( Q_p^{(2)} - \frac{1}{N} \Tr\{ Q_p^{(2)} \} \1 )$ 
\\
\\
3 
& $\pm$
& $\wt Q_p^{(1)}$ 
& $\frac{1}{\sqrt{12}N} ( \Tr\{ U_p \} U_p + \sigma \Tr\{ U_p^\dag \} \1 )$ 
& $\frac{\sigma}{\sqrt{12}N} \Tr\{ Q^{(1)\dag}_p \} \1$ 
& $\frac{6 |\beta_\pm| N}{N^2 \pm 3N + 2}$
\\
&
& $\wt Q_p^{(2)}$ 
& $\frac{1}{2N} ( \Tr \{ U_p \} U_p \pm \sigma U_p^\dag )$ 
& $\pm \frac{\sigma}{2N} Q_p^{(2)\dag}$
\\
&
& $\wt Q_p^{(3)}$ 
& $\frac{1}{\sqrt{6}N} ( U_p^2 + \sigma U_p^\dag )$ 
& $\frac{\sigma}{\sqrt{6}N} Q_p^{(3)\dag}$
\\
&
& $\wt Q_p^{(4)}$ 
& $\frac{1}{\sqrt{12}} Q_p^{(1)} U_p^\dag + \frac{1}{N} \Tr\{ U_p \} \1$ 
& $\frac{1}{\sqrt{12}} Q_p^{(4)\dag} Q_p^{(1)} + \frac{1}{N} \Tr\{ Q_p^{(4)} \} \1$
\\
&
& $\wt Q_p^{(5)}$ 
& $\frac{1}{2} Q_p^{(2)} U_p^\dag + \frac{1}{N} \Tr\{ U_p \} \1$ 
& $\frac{1}{2} Q_p^{(5)\dag} Q_p^{(2)} + \frac{1}{N} \Tr\{ Q_p^{(5)} \} \1$
\\
&
& $\wt Q_p^{(6)}$ 
& $\frac{1}{\sqrt{6}} Q_p^{(3)} U_p^\dag + \frac{1}{N} U_p$ 
& $\frac{1}{\sqrt{6}} Q_p^{(6)\dag} Q_p^{(3)} +
\frac{1}{N} Q_p^{(6)}$
\\
&
& $\wt Q_p^{(7)}$ 
& $\sqrt{\frac{29}{12}} ( U_p - \frac{1}{N} \Tr \{ U_p \} \1 )$ 
& $\sqrt{\frac{29}{12}} ( Q_p^{(7)} - \frac{1}{N} \Tr \{ Q_p^{(7)} \} \1 )$
\\
\\
& $\pm$
& $\wt Q_p^{(1)}$ 
& $\frac{1}{\sqrt{12}N} ( \Tr\{U_p\} U_p + \sigma ( \Tr\{U_p^\dag\} \1 \pm 3 U_p^\dag ) )$ 
& $\frac{\sigma}{\sqrt{12}N} ( \Tr\{Q^{(1)\dag}_p\} \1 \pm 3 Q^{(1)\dag}_p )$
& $\frac{6 |\beta_\pm| N}{N^2 \pm 3N + 2}$
\\
&& $\wt Q_p^{(2)}$ & $\frac{1}{\sqrt{6}N} ( U_p^2 + \sigma U_p^\dag )$
& $\frac{\sigma}{\sqrt{6}N} Q_p^{(2)\dag}$
\\
&
& $\wt Q_p^{(3)}$ 
& $\frac{1}{\sqrt{12}} Q_p^{(1)} U_p^\dag + \frac{1}{N} \Tr\{U_p\} \1$ 
& $\frac{1}{\sqrt{12}} Q_p^{(3)\dag} Q_p^{(1)} + \frac{1}{N} \Tr\{Q_p^{(3)}\} \1$
\\
&
& $\wt Q_p^{(4)}$ 
& $\frac{1}{\sqrt{6}} Q_p^{(2)} U_p^\dag + \frac{1}{N} U_p$ 
& $\frac{1}{\sqrt{6}} Q_p^{(4)\dag} Q_p^{(2)} +
\frac{1}{N} Q_p^{(4)}$
\\
&
& $\wt Q_p^{(5)}$ 
& $\sqrt{\frac{7N \pm 3}{6N}} ( U_p - \frac{1}{N} \Tr \{U_p \} \1 )$ 
& $\sqrt{\frac{7N \pm 3}{6N}} ( Q_p^{(5)} - \frac{1}{N} \Tr\{Q_p^{(5)} \} \1 )$ 
\\
\\
& $(2,1)$
& $\wt Q_p^{(1)}$ 
& $\frac{1}{\sqrt{6}N} ( \Tr\{U_p\} U_p + \sigma \Tr \{ U_p^\dag \} \1 )$ 
& $\frac{\sigma}{\sqrt{6}N} \Tr\{Q^{(1)\dag}_p\} \1$ 
& $\frac{3|\beta_{(2,1)}| N}{N^2-1}$
\\
&
& $\wt Q_p^{(2)}$ & $\frac{1}{\sqrt{6}N} ( U_p^2 - \sigma U_p^\dag )$
& $-\frac{\sigma}{\sqrt{6}N} Q_p^{(2)\dag}$
\\
&
& $\wt Q_p^{(3)}$ & $\frac{1}{\sqrt{6}} Q_p^{(1)} U_p^\dag + \frac{1}{N} \Tr\{U_p\} \1$ 
& $\frac{1}{\sqrt{6}} Q_p^{(3)\dag} Q_p^{(1)} + \frac{1}{N} \Tr\{Q_p^{(3)}\} \1$
\\
&
& $\wt Q_p^{(4)}$ 
& $\frac{1}{\sqrt{6}} Q_p^{(2)} U_p^\dag + \frac{1}{N} U_p$ 
& $\frac{1}{\sqrt{6}} Q_p^{(4)\dag} Q_p^{(2)} + \frac{1}{N} Q_p^{(4)}$
\\
&
& $\wt Q_p^{(5)}$ 
& $\frac{2}{\sqrt{3}} ( U_p - \frac{1}{N} \Tr\{U_p\} \1 )$ 
& $\frac{2}{\sqrt{3}} ( Q_p^{(5)} - \frac{1}{N} \Tr\{Q_p^{(5)}\} \1 )$
\\
\\
& $(2,1)$
& $\wt Q_p^{(1)}$ 
& $\frac{1}{\sqrt{6}N} ( U_p (\Tr\{U_p\} \1 - U_p) + \sigma ( U_p + \Tr \{U_p\} \1 )^\dag )$ 
& $\frac{\sigma}{\sqrt{6}N} (Q^{(1)}_p + \Tr\{Q^{(1)}_p\} \1)^\dag$ 
& $\frac{3|\beta_{(2,1)}| N}{N^2-1}$
\\
&
& $\wt Q_p^{(2)}$ & $\frac{1}{\sqrt{6}} Q_p^{(1)} U_p^\dag - \frac{1}{N} (U_p - \Tr\{U_p\} \1)$ 
& $\frac{1}{\sqrt{6}} Q_p^{(1)} Q_p^{(2)} - \frac{1}{N} (Q_p^{(2)} - \Tr\{Q_p^{(2)}\} \1 )$
\\
&
& $\wt Q_p^{(3)}$ 
& $\sqrt{\frac{4N-6}{3N}} ( U_p - \frac{1}{N} \Tr\{U_p\} \1 )$ 
& $\sqrt{\frac{4N-6}{3N}} ( Q_p^{(3)} - \frac{1}{N} \Tr\{Q_p^{(3)}\} \1 )$
\\
\\
\hline
\hline
\end{tabular}
\end{scriptsize}
\end{center}
\caption{List of the expressions for $h_{p}^{(i)}$ and
$g_{p}^{(i)}$, useful in the construction of MC algorithms for mixed Wilson actions with plaquettes in irreducible representations of $SU(N)$ with $N$--ality $k \leq 3$. The symbol `$+$' stands for `symmetric representation', either $(2)$ or $(3)$, while `$-$' stands for `antisymmetric representation', either $(1,1)$ or $(1,1,1)$. In $N$--ality $k \geq 2$, two different algorithms are suggested for each representation; they differ in the number of auxiliary variables. For the adjoint representation, the cases of positive and negative $\beta_A$ are considered separately. Here, $p$ labels plaquettes in the hypercubic lattice, $\sigma$ denotes the sign of the lattice coupling, $\sigma \equiv {\rm sgn}(\beta_\R) = \pm 1$, and $\beta_\R '$ is the redefined lattice coupling.
\label{tab:Summary.Mix}}
\end{table}

\vfill\eject\clearpage

\begin{table}
\begin{small}
\begin{center}
\begin{tabular}{cclrrrrcc} 
\hline
\hline
$N$ & lattice & $F + \R$ & $\beta_F$ & $\beta_{\R}$ & $\l\langle \frac{1}{N}\Re\Tr_F U_p \r\rangle$ & $\l\langle \frac{1}{d_\R} \Re\Tr_{\R} U_p \r\rangle$ & $N_{\rm meas}$ & Algor.
\\
\hline
\hline
3 & $12^3 8$ & $F + A$ & 4.0 & 2.0 & 
0.626587(7) & 0.487024(7) & 398,000 & M
\\
& &&&& 
0.626689(6) & 0.487145(7) & 398,000 & H
\\
& &&&& $\sim 0.6267$ & & & \cite{NeccoHasenbusch}
\\
\hline
3 & $16^3 4$ & $F + A$ & 9.25 & --3.556 & 
0.53828(2) & 0.36850(2) & 398,000 & M
\\
&&&&& 
0.53816(2) & 0.36836(3) & 398,000 & H
\\
&&&&& 
$\sim 0.5383$ & $\sim 0.3685$ & & \cite{NeccoHasenbusch}
\\
\hline
4 & $8^4$ & $F + (2)$ & 10.665 & 0.3556 & 
0.743870(5) & 0.494378(7) & 398,000 & M
\\
&&&&& 
0.743913(2) & 0.494446(3) & 398,000 & H
\\
\hline
4 & $8^4$ & $F + (1,1)$ & 10.665 & 1.0668 & 
0.671709(4) & 0.587947(5) & 398,000 & M
\\
&&&&& 
0.671735(3) & 0.587978(3) & 398,000 & H
\\
\hline
5 & $8^4$ & $F + (3)$ & 16.665 & 4.1487 & 
0.620798(5) & 0.159693(5) & 180,000 & M
\\
&&&&& 0.620836(6) & 0.159729(5) & 180,000 & H
\\
\hline
5 & $8^4$ & $F + (1,1,1)$ & 
16.665 & 1.1853 & 
0.571895(6) & 0.428673(7) & 180,000 & M
\\
&&&&& 
0.571923(7) & 0.428706(8) & 180,000 & H
\\
\hline
5 & $8^4$ & $F + (2,1)$ & 16.665 & 4.7413 & 
0.656146(5) & 0.316318(7) & 180,000 & M
\\
&&&&& 
0.656167(4) & 0.316345(6) & 180,000 & H
\\
\hline
\hline
\end{tabular} 
\end{center}
\end{small}
\vskip 3mm
\caption{Expectation values of plaquette characters in the `$F + \R$' mixed Wilson action, using a Cabibbo-Marinari-Metropolis algorithm (M) and a pseudo-heatbath algorithm proposed in this paper (H). For the mixed fundamental/adjoint Wilson action, we also include plaquette values that were calculated by Hasenbusch and Necco with a Cabibbo-Marinari-Metropolis algorithm \cite{NeccoHasenbusch}. The statistical errors in the table do not take autocorrelations into account, which is likely the reason for some discrepancies in the last significant digits.
\label{tab:Plaquette.Mix}
}
\end{table}


\begin{table}
\begin{center}
\begin{small}
\begin{tabular}{ccrrrrcc} 
\hline
\hline
$N$ & lattice & $\beta_F$ & $\alpha_1$ & $\alpha_2$ & $\langle S_{CYM} \rangle$ & $N_{\rm meas}$ & Algor.
\\
\hline
\hline
$5$ & $10^3 1$ & 25.0 & 0.2 & 0.05 &
--4.14109(9) & 398,000 & M
\\
&&&&&
--4.14113(4) & 398,000 & H
\\
\hline
$5$ & $10^3 1$ & 25.0 & 0.6 & 0.1 &
--4.39116(20) & 398,000 & M
\\
&&&&&
--4.39230(6) & 398,000 & H
\\
\hline
$5$ & $10^3 3$ & 25.0 & 2.8 & 0.2 &
--4.39339(2) & 398,000 & M
\\
&&&&& 
--4.39341(2) & 398,000 & H
\\
\hline
\hline
\end{tabular} 
\end{small}
\vskip 3mm
\caption{
Expectation values of the $SU(5)$ CYM action for different compactification radii and $\alpha_n$, using a Cabibbo-Marinari-Metropolis algorithm (M) and the  pseudo-heatbath algorithm (H) proposed in this paper. The statistical errors in the table do not take autocorrelations into account, which is likely the reason for some discrepancies in the last significant digits. The discrepancy in the second row is due to a very slow equilibration of the Metropolis algorithm due to large correlations (see Fig.\ref{fig:CYM:red(conf):C}).
\label{tab:Plaquette.CYM}
}
\end{center}
\end{table}

\vfill\eject\clearpage

\vfill\eject\clearpage

\vfill\eject\clearpage

\vfill\eject\clearpage

\vfill\eject\clearpage

\vfill\eject\clearpage

\vfill\eject\clearpage

\vfill\eject\clearpage

\vfill\eject\clearpage

\vfill\eject\clearpage

\vfill\eject\clearpage


\end{document}